\documentclass[aps,showpacs,nofootinbib,superscriptaddress]{revtex4}
\usepackage{graphicx}
\usepackage{dcolumn}
\newcommand{\ii}{\'\i}
\newcommand{\nn}{\~n}
\begin{document}

\title{ Study of exclusive semileptonic and nonleptonic decays of
$B_c^-$ in a nonrelativistic quark model.}  \author{ E. Hern\'andez}
\affiliation{Grupo de F\'\i sica Nuclear, Departamento de F\ii sica
Fundamental e IUFFyM, Facultad de Ciencias, E-37008 Salamanca, Spain.}
\author {J. Nieves} \affiliation{Departamento de F\'{\i}sica
At\'omica, Molecular y Nuclear, Universidad de Granada, E-18071
Granada, Spain.}  \author{ J. M. Verde-Velasco} \affiliation{Grupo de
F\'\i sica Nuclear, Departamento de F\ii sica Fundamental e IUFFyM,
Facultad de Ciencias, E-37008 Salamanca, Spain.}
\begin{abstract} 
\rule{0ex}{3ex} 

\end{abstract}


\begin{abstract} 
 We present results for different observables measured in semileptonic
and nonleptonic decays of the $B_c^-$ meson. The calculations have
been done within the framework of a nonrelativistic constituent quark
model.  In order to check the sensitivity of all our results against
the inter-quark interaction we use five different quark--quark
potentials.  We obtain form factors, decay widths and asymmetry
parameters for semileptonic $B_c^-\to c\bar c$ and $B_c^-\to \overline
B$ decays.  In the limit of infinite heavy quark mass our model
reproduces the constraints of heavy quark spin symmetry. For the
actual heavy quark masses we find nonetheless large corrections to
that limiting situation for some form factors.  We also analyze
exclusive nonleptonic two--meson decay channels within the
factorization approximation.

\end{abstract}
\maketitle

\section{Introduction}

Since its discovery at Fermilab by the CDF Collaboration~\cite{cdf98}
the $B_c$ meson has drawn a lot of attention. Unlike other heavy
mesons it is composed of two heavy quarks of different flavor
($b,\,c$) and, being below the $B$--$D$ threshold, it can only decay
through weak interactions making an ideal system to study
weak decays of heavy quarks.

It is known~\cite{jenkins93} that one can not apply heavy quark
symmetry (HQS) to hadrons containing two heavy quarks: the kinetic
energy term, needed in those systems to regulate infrared divergences,
breaks heavy flavor symmetry. Still, there is a symmetry that
survives: heavy quark spin symmetry (HQSS). This symmetry amounts to
the decoupling of the two heavy quark spins since the spin--spin
interaction vanishes for infinite heavy quark masses. Using HQSS
Jenkins {\it et al.}~\cite{jenkins93} were able to obtain, in the
infinite heavy quark mass limit, relations between different form
factors for semileptonic $B_c$ decays into pseudoscalar and vector
mesons. Contrary to the heavy-light meson case where standard HQS
applies, no determination of corrections in inverse powers of the
heavy quark masses has been worked out in this case. So one can only
test any model calculation against HQSS predictions in the infinite
heavy quark mass limit.

With both quarks being heavy, a nonrelativistic treatment of the $B_c$
meson should provide reliable results. Besides a nonrelativistic model
will comply with the constraints imposed by HQSS as the spin--spin
interaction vanish in the infinity heavy quark mass limit. In this
paper we will study, within the framework of a nonrelativistic quark
model, exclusive semileptonic and nonleptonic decays of the $B_c^-$
meson driven by a $b\to c$ or $\bar c\to \bar d,\,\bar s$ transitions
at the quark level. We will not consider semileptonic processes driven
 by the quark $b\to
u$ transition.  Our experience with this kind
of processes, like the analogous $B\to\pi$ semileptonic decay
~\cite{albertus05-2}, shows that the nonrelativistic model without any
improvements underestimates the decay width for two reasons: first at
high $q^2$ transfers one might need to include the exchange of a $B^*$
meson, and second the model underestimates the form factors at low
$q^2$ or high three--momentum transfers.  We will concentrate thus on
semileptonic $B_c^-\to c\bar c$ and $B_c^-\to \overline B$ transitions. As for
two--meson nonleptonic decay
we will only consider  channels with a least a $c\bar c$ or
$ B$ final meson. In the first case we will include channels with final $D$ 
mesons for which there is a
  contribution coming from an effective $b\to d,\,s $ transition. As 
later explained  this is not the main contribution to the decay amplitude and
besides the momentum
transfer in those cases is neither too high nor too low so that  the problems
mentioned above are avoided.

The observables studied here have been analyzed before in the
context of different models like the relativistic constituent quark
model~\cite{ivanov06,ivanov05,ivanov01}, the quasi-potential approach
to the relativistic quark model~\cite{ebert03,ebert03-2}, the
instantaneous nonrelativistic approach to the Bethe--Salpeter
equation~\cite{chang94, chang02,chang01}, the Bethe--Salpeter
equation~\cite{hady00,liu97}, the three point sum rules of QCD and
nonrelativistic
QCD~\cite{kiselev00,kiselev00-2,kiselev02,kiselev02-2}, the QCD
relativistic potential model~\cite{colangelo00}, the relativistic
constituent quark model formulated on the light
front~\cite{anisimov99}, the relativistic quark--meson
model~\cite{nobes00} or in models that use the Isgur, Scora, Grinstein
and Wise wave functions~\cite{isgw} like the calculations in
Refs.~\cite{sanchis95,lopezcastro02,lu95}. We will compare our results
with those obtained in these latter references whenever is
possible. Besides, we will perform an exhaustive study and compile in
this work all our results for exclusive semileptonic and nonleptonic
$B_c^-$ decays, paying an special attention to the theoretical
uncertainties affecting  our predictions and providing reliable
estimates for all of them.

In the present calculation we shall use physical masses taken from
Ref.~\cite{pdg04}.  For the $B_c$ meson mass and lifetime we shall use
the central values of the recent experimental determinations by the
CDF Collaboration of
$m_{B_c}=6285.7\pm5.3\pm1.2$\,MeV/$c^2$~\cite{cdf06-1} and
$\tau_{B_c}=(0.463^{+0.073}_{-0.065}\pm0.036)\times 10^{-12}$\,s
~\cite{cdf06-2}.  This new mass value is very close to the one we
obtain with the different quark-quark potentials that we use in this
work (see below) from where we get $m_{B_c}=6291.6^{+12}_{-33}$\,MeV.

We shall also need  Cabibbo-Kobayashi-Maskawa (CKM) matrix elements and
different meson decay constants.
For the former we shall use the ones quoted in Ref.~\cite{ivanov06} that we
reproduce in Table~\ref{tab:ckm}. All of them are within the ranges quoted by
the  Particle Data Group (PDG)~\cite{pdg04}.
\begin{table}[h!]
\begin{tabular}{c c c c c  c }
$|V_{ud}|$&$|V_{us}|$&$|V_{cd}|$&$|V_{cs}|$&$|V_{cb}|$\\
\hline
&&&&&\\
 0.975  &0.224   & 0.224  &  0.974 & 0.0413 \\
 \end{tabular}
\caption{Values for Cabibbo-Kobayashi-Maskawa matrix elements used 
in
this work.}
\label{tab:ckm}
\end{table}

For the meson decay constants the used values in this work are
compiled in Table~\ref{tab:dc}. They correspond to central values of
experimental measurements or lattice determinations. The results for
$f_\rho $ and $f_{K^*}$ have been obtained by the authors in
Ref.~\cite{ivanov06} using  $\tau $ lepton decay data.
  Our own theoretical calculation, obtained with the model described in
Ref.~\cite{albertus05-1}, give $f_{\rho}=0.189\sim 0.227$\,GeV,
$f_{K^*}=0.180\sim 0.220$\,GeV depending on the inter-quark interaction
used, results  which agree with the determinations in Ref.~\cite{ivanov06}. 
We shall nevertheless use
the latter for our calculations.  
For $f_{\eta_c}$ we have been unable to find an experimental
result or a lattice determination. There are at least two theoretical
determinations that predict $f_{\eta_c}=0.484$\,GeV~\cite{ivanov06}
and $f_{\eta_c}=0.420\pm0.052$\,GeV~\cite{hwang97}. Again our own
calculation gives values in the range $f_{\eta_c}=0.485\sim 0.500$\,GeV
depending on the inter-quark interaction used.  Here we will take
$f_{\eta_c}=0.490$\,GeV.
\begin{table}[h!]
\begin{center}
\begin{tabular}{c c c c c  }
$f_{\pi^-}$&$f_{\pi^0}$&$f_{\rho^-,\, \rho^0}$&$f_{K^-,\,K^0}$
&$f_{K^{*-},\,K^{*0}}$\\
\hline\\
 0.1307~\cite{pdg04}  &0.130~\cite{pdg04}&0.210~\cite{ivanov06}  &
 0.1598~\cite{pdg04}  &  0.217~\cite{ivanov06}\\
&&&\\
 \end{tabular}
 
\begin{tabular}{c c    }
$f_{\eta_c}$&$f_{J/\Psi}$\\
\hline\\
 0.490&0.405~\cite{pdg98}\\
 &\\
 \end{tabular}

\begin{tabular}{c c c c   }
$f_{D^-}$&$f_{D^{*-}}$&$f_{D_s^-}$&$f_{D_s^{*-}}$\\
\hline\\
 0.2226~\cite{artuso05}  &0.245~\cite{becirevic99}   & 0.294~\cite{yao06}  
 &  0.272~\cite{becirevic99}\\
 \end{tabular}

\end{center}
 \caption{Meson decay constants in GeV used in this work.}
\label{tab:dc}
\end{table}%

The rest of the paper is organized as follows. In Sect.~\ref{sect:wf}
we introduce our meson states and the potential models used to obtain
the spatial part of their wave functions. In
Sect.~\ref{sect:semileptonic} we study the $B_c$ meson semileptonic
decays into various $c\bar c $ channels, both for a final light
charged lepton ($e,\,\mu$) and for a heavy one ($\tau$). In
Sect.~\ref{sect:nonlepcc} we study different exclusive nonleptonic
two--meson decay channels of the $B_c^-$ meson with one of the final
mesons being a $c\bar c$ one. In Sect.~\ref{sect:semilepb} we study
semileptonic $B_c^-\to \overline B$ decays and in
Sect.~\ref{sect:nonlepb} nonleptonic two--meson decays with one of
the mesons having a $b$ quark. We briefly summarize our results
in Sect.~\ref{sect:summary}.  The paper also contains four appendices:
in appendix~\ref{app:epsilon} we give different sets of polarization
vectors used in this paper, 
appendices~\ref{app:va} and
\ref{app:sign} collect the expressions for all the matrix elements
needed to evaluate the different observables analyzed, finally in
appendix~\ref{app:hcht} we give the expressions for the helicity components of
the hadron tensor to be defined below.

%
%
%
%
%
\section{Meson states and inter-quark interactions}
\label{sect:wf}
Within a nonrelativistic constituent quark
model, the state of a meson $M$ is given by
\begin{eqnarray}
\label{wf}
&&\hspace{-1cm}\left|{M,\lambda\,\vec{P}}\,\right\rangle_{NR}
=\int d^3p \sum_{\alpha_1,\alpha_2}\hat{\phi}^{(M,\lambda)}_{\alpha_1,\alpha_2}(\,\vec{p}\,)
\nonumber\\ &&\times\frac{(-1)^{(1/2)-s_2}}{(2\pi)^{3/2}
\sqrt{2E_{f_1}(\vec{p}_1)2E_{f_2}(\vec{p}_2)}}\ 
\left|\ q,\ \alpha_1\
\vec{p}_1=\frac{m_{f_1}}{m_{f_1}+m_{f_2}}\vec{P}-\vec{p}\ \right\rangle
\left|\ \bar{q},\ \alpha_2\ \vec{p}_2=\frac{m_{f_2}}{m_{f_1}+m_{f_2}}\vec{P}+\vec{p}\
 \right\rangle
\end{eqnarray}
where $\vec{P}$ stands for the meson three momentum and $\lambda$
represents the spin projection in the meson center of mass. $\alpha_1$
and $\alpha_2$ represent the quantum numbers of spin s, flavor f and
color c ($\alpha\equiv (s,f,c))$, of the quark and the antiquark,
while $(E_{f_1}(\vec{p}_1),\,\vec{p}_1),\ m_{f_1}$ and
$(E_{f_2}(\vec{p}_2),\,\vec{p}_2),\ m_{f_2}$ are their respective
four--momenta and masses. The factor $(-1)^{(1/2)-s_2}$ is
included in order that the antiquark spin states have the correct
relative phase\footnote{Note that under charge conjugation $({\cal
C})$ quark and antiquark creation operators are related via
\hbox{${\cal C}\,c^{\dagger}_\alpha(\,\vec{p}\,)\,{\cal C}^{\dagger}=
(-1)^{(1/2)-s}\,d^{\dagger}_\alpha(\,\vec{p}\,) $}. This implies
that the antiquark states with the correct spin relative phase are not
$d^{\dagger}_\alpha(\,\vec{p}\,)\,|0\rangle= |\,\bar{q},\ \alpha\
\vec{p}\ \rangle$ but are given instead by
$(-1)^{(1/2)-s}\,d^{\dagger}_\alpha(\,\vec{p}\,)\,|0\rangle=
(-1)^{(1/2)-s}\,|\,\bar{q},\ \alpha\ \vec{p}\ \rangle$.}.  The
normalization of the quark and antiquark states is
\begin{eqnarray}
\left\langle\ \alpha^{\prime}\ \vec{p}^{\ \prime}\,|\,\alpha\ \vec{p}\,
\right\rangle=\delta_{\alpha^{\prime}, \alpha}\, (2\pi)^3\, 2E_f(\vec{p}\,)\,\delta(
\vec{p}^{\ \prime}-\vec{p}\,)
\end{eqnarray}
Furthermore,
$\hat{\phi}^{\,(M,\lambda)}_{\alpha_1,\alpha_2}(\,\vec{p}\,)$ is the
momentum space wave function for the relative motion of the
quark-antiquark system.  Its normalization is given by
\begin{equation}
\int\, d^3p\ \sum_{\alpha_1\,\alpha_2} \left(
\hat{\phi}^{\,(M,\lambda')}_{\alpha_1,\,\alpha_2}(\,\vec{p}\,)
\right)^* \hat{\phi}^{\,(M,\lambda)}_{\alpha_1,\,\alpha_2}(\,\vec{p}\,)
=\delta_{\lambda',\,\lambda}
\end{equation}
and, thus,  the normalization of our meson states is 
\begin{equation}
{}_{\stackrel{}{\stackrel{}{NR}}}
\left\langle\, {M,\lambda'\,\vec{P}^{\,\prime}}\,|\,{M,\lambda
\,\vec{P}}\,\right\rangle_{NR}
=\delta_{\lambda',\,\lambda}\,(2\pi)^3\,\delta(\vec{P}^{\,\prime}-\vec{P}\,)
\label{eq:4}
\end{equation}
In this calculation we will need the ground state wave function for
scalar ($0^+$), pseudoscalar($0^-$), vector ($1^-$), axial vector
($1^+$), tensor ($2^+$) and pseudotensor ($2^-$) mesons. Assuming
always the lowest possible value for the orbital angular momentum we
will have for a meson $M$ with scalar, pseudoscalar and vector quantum
numbers:
\begin{eqnarray}
\label{eq:0pm}
\hat{\phi}^{\,(M(0^+))}_{\alpha_1,\,\alpha_2}(\,\vec{p}\,)
&=&\frac{1}{\sqrt{3}}\,\delta_{c_1,\,c_2}\
\hat{\phi}^{\,(M(0^+))}_{(s_1,\,f_1),\,(s_2,\,f_2)}(\,\vec{p}\,)\nonumber\\
&=&\frac{1}{\sqrt{3}}\,\delta_{c_1,\,c_2}\ i
\ \hat{\phi}^{\,(M(0^+))}_{f_1,\,f_2}(\,|\vec{p}\,|)\sum_m\ (1/2,1/2,1\,;\,s_1,s_2,-m)
\ (1,1,0\,;\,m,-m,0)\
Y_{1m}(\,{\vec{p}}\ )\nonumber\\
\hat{\phi}^{\,(M(0^-))}_{\alpha_1,\,\alpha_2}(\,\vec{p}\,)
&=&\frac{1}{\sqrt{3}}\,\delta_{c_1,\,c_2}\
\hat{\phi}^{\,(M(0^-))}_{(s_1,\,f_1),\,(s_2,\,f_2)}(\,\vec{p}\,)\nonumber\\
&=&\frac{1}{\sqrt{3}}\,\delta_{c_1,\,c_2}\ (-i)
\ \hat{\phi}^{\,(M(0^-))}_{f_1,\,f_2}(\,|\vec{p}\,|) 
\ (1/2,1/2,0\,;\,s_1,s_2,0)\ Y_{00}(\,{\vec{p}}\ )\nonumber\\
\hat{\phi}^{\,(M(1^-),\,\lambda)}_{\alpha_1,\,\alpha_2}(\,\vec{p}\,)
&=&\frac{1}{\sqrt{3}}\,\delta_{c_1,\,c_2}\
\hat{\phi}^{\,(M(1^-),\,\lambda)}_{(s_1,\,f_1),\,(s_2,\,f_2)}(\,\vec{p}\,)\nonumber\\
&=&\frac{1}{\sqrt{3}}\,\delta_{c_1,\,c_2}\ (-1)\ 
\hat{\phi}^{\,(M(1^-))}_{f_1,\,f_2}(\,|\vec{p}\,|)\
(1/2,1/2,1\,;\,s_1,s_2,\lambda)\ Y_{00}(\,{\vec{p}}\ )
\end{eqnarray}
where  $(j_1,j_2,j_3\,;\,m_1,m_2,m_3)$ are Clebsch-Gordan coefficients, 
$Y_{lm}(\vec p\,)$ are spherical harmonics, 
 and 
$\hat{\phi}^{\,(M)}_{f_1,\,f_2}(|\,\vec{p}\,|)$ is the
Fourier transform of the radial coordinate space wave function.\\ %
For axial mesons we need orbital angular momentum $L=1$. In this case two values
of the total quark--antiquark spin  $S_{q\bar{q}}=0,1$ are possible, giving
rise to the two states:
\begin{eqnarray}
\label{eq:1pm}
\hat{\phi}^{\,(M(1^+,S_{q\bar{q}}=0),\,\lambda)}_{\alpha_1,\,\alpha_2}(\,\vec{p}\,)
&=&\frac{1}{\sqrt{3}}\,\delta_{c_1,\,c_2}\
\hat{\phi}^{\,(M(1^+,S_{q\bar{q}}=0),\,\lambda)}_{(s_1,\,f_1),\,(s_2,\,f_2)}(\,\vec{p}\,)\nonumber\\
&=&\frac{1}{\sqrt{3}}\,\delta_{c_1,\,c_2}\ (-1)\ 
\hat{\phi}^{\,(M(1^+,S_{q\bar{q}}=0))}_{f_1,\,f_2}(\,|\vec{p}\,|)\
(1/2,1/2,0;s_1,s_2,0)\
Y_{1\lambda}(\,{\vec{p}}\ )\nonumber\\
\hat{\phi}^{\,(M(1^+,S_{q\bar{q}}=1),\,\lambda)}_{\alpha_1,\,\alpha_2}(\,\vec{p}\,)
&=&\frac{1}{\sqrt{3}}\,\delta_{c_1,\,c_2}\
\hat{\phi}^{\,(M(1^+,S_{q\bar{q}}=1),\,\lambda)}_{(s_1,\,f_1),\,(s_2,\,f_2)}(\,\vec{p}\,)\nonumber\\
&=&\frac{1}{\sqrt{3}}\,\delta_{c_1,\,c_2}\ (-1)\ 
\hat{\phi}^{\,(M(1^+,S_{q\bar{q}}=1))}_{f_1,\,f_2}(\,|\vec{p}\,|)\nonumber\\
&&\times \sum_{m}\ (1/2,1/2,1;s_1,s_2,\lambda-m)\ (1,1,1;m,\lambda-m,\lambda)
Y_{1m}(\,{\vec{p}}\ )
\end{eqnarray}
Finally for  tensor and pseudotensor mesons we have the wave functions:
\begin{eqnarray}
\label{eq:2pm}
\hat{\phi}^{\,(M(2^+),\,\lambda)}_{\alpha_1,\,\alpha_2}(\,\vec{p}\,)
&=&\frac{1}{\sqrt{3}}\,\delta_{c_1,\,c_2}\
\hat{\phi}^{\,(M(2^+),\,\lambda)}_{(s_1,\,f_1),\,(s_2,\,f_2)}(\,\vec{p}\,)\nonumber\\
&=&\frac{1}{\sqrt{3}}\,\delta_{c_1,\,c_2}\ \ 
\hat{\phi}^{\,(M(2^+))}_{f_1,\,f_2}(\,|\vec{p}\,|)\
\sum_m\ (1/2,1/2,1;s_1,s_2,\lambda-m)\ (1,1,2;m,\lambda-m,\lambda)\
Y_{1m}(\,{\vec{p}}\ )\nonumber\\
\hat{\phi}^{(\,M(2^-),\,\lambda)}_{\alpha_1,\,\alpha_2}(\,\vec{p}\,)
&=&\frac{1}{\sqrt{3}}\,\delta_{c_1,\,c_2}\
\hat{\phi}^{\,(M(2^-),\,\lambda)}_{(s_1,\,f_1),\,(s_2,\,f_2)}(\,\vec{p}\,)\nonumber\\
&=&\frac{1}{\sqrt{3}}\,\delta_{c_1,\,c_2}\ (-1)\ 
\hat{\phi}^{\,(M(2^-))}_{f_1,\,f_2}(\,|\vec{p}\,|)\
\sum_{m}\ (1/2,1/2,1;s_1,s_2,\lambda-m)\ (2,1,2;m,\lambda-m,\lambda)
Y_{2m}(\,{\vec{p}}\ )
\end{eqnarray}
All phases have been introduced for later convenience. 

To evaluate the coordinate space wave function we use five different
inter-quark interactions, one suggested by Bhaduri and
collaborators~\cite{bhaduri81}, and four others suggested by
Silvestre-Brac and Semay ~\cite{silvestre96,sbs93}. All of them
contain a confinement term, plus Coulomb and hyperfine terms coming
from one-gluon exchange, and differ from one another in the form
factors used for the hyperfine terms, the power of the confining term
or the use of a form factor in the one gluon exchange Coulomb
potential.  All free parameters in the potentials had been adjusted to
reproduce the light ($\pi$, $\rho$, $K$, $K^*$, etc.) and heavy-light
($D$, $D^*$, $B$, $B^*$, etc.) meson spectra. These potentials also
lead to good results for the charmed and bottom baryon
($\Lambda_{c,b}$, $\Sigma_{c,b}$, $\Sigma^*_{c,b}$, $\Xi_{c,b}$,
$\Xi'_{c,b}$, $\Xi^*_{c,b}$, $\Omega_{c,b}$ and $\Omega_{c,b}^*$)
masses~\cite{silvestre96,albertus04}, for the semileptonic
$\Lambda_b^0 \to \Lambda_c^+ l^- {\bar \nu}_l$ and $\Xi_b^0 \to
\Xi_c^+ l^- {\bar \nu}_l$ decays~\cite{albertus05}, for the decay
constants of pseudoscalar $B,D$ and vector $B^*,D^*$ mesons and the
semileptonic $B\to D$ and $B\to D^*$ decays~\cite{albertus05-1}, for
the $B\to\pi$ semileptonic decay~\cite{albertus05-2} , and for the
strong $\Sigma_c\to \Lambda_c\, \pi$, $\Sigma_c^{*}\to \Lambda_c\,
\pi$ and $\Xi_c^{*}\to \Xi_c\, \pi$
decays~\cite{albertus05-3}. Preliminary results for the spectrum of
doubly heavy baryons~\cite{albertus-pre} also show excellent agreement
with previous Faddeev calculations and lattice results. For more details on the
inter-quark interactions see  Ref.~\cite{albertus04} or the original 
works~\cite{bhaduri81,silvestre96,sbs93}.

The use of different inter-quark interactions will provide us with a
spread in the results that we will consider, and quote, as a
theoretical error added to the value obtained with the AL1 potential
or Refs.~\cite{silvestre96,sbs93} that we will use to get our central
results. Another source of theoretical uncertainty, that we can not account
for, is the use of 
nonrelativistic kinematics in the evaluation of the orbital wave
functions and the construction of our states in Eq.(\ref{wf}) above.
While this is a very good approximation for the $B_c$ itself it is not that
good for mesons with a light quark. That notwithstanding note that any
nonrelativistic quark model has free parameters in the inter-quark interaction
that are fitted to experimental data. In that sense we think that at least 
part of the ignored relativistic effects are included in an effective way 
in their fitted values.
 
%
%
%
%
%
\section{ Semileptonic $B_c^-\to c\bar{c}$ decays}
\label{sect:semileptonic}
In this section we will consider the semileptonic decay of the $B_c^-$ meson
into different $c\bar{c}$ states with $0^+$, $0^-$, $1^+$, $1^-$, $2^+$ and $2^-$ 
spin--parity quantum numbers.
  Those decays  correspond to a $b\to c $ transition at the quark level which is
  governed by the current
\begin{equation}
J^{c\,b}_\mu(0)=J^{c\,b}_{V\,\mu}(0)-J^{c\,b}_{A\,\mu}(0)
=
\overline{\Psi}_{c}(0)\gamma_\mu
(I-\gamma_5)\Psi_{b}(0)
\end{equation}
with $\Psi_{f}$  a quark field of a definite  flavor $f$.
%
%
%
\subsection{Form factor decomposition of hadronic matrix elements}
 The hadronic matrix elements involved in these processes can be
 parametrized in terms of a few form factors as
\begin{eqnarray}
\label{eq:ff}
\left\langle\, c\bar{c}\, (0^-),\,\vec{P}_{c\bar{c}}\,\left|\, J^{c\,b}_\mu(0)\,
\right| \, B_c,\,\vec{P}_{B_c}\right\rangle&=&
\left\langle\,  {c\bar{c}}\, (0^-),\,\vec{P}_{c\bar{c}}\,\left|\hspace{.3cm}J^{c\,b}_{V\,\mu}(0)\,
\right| \, B_c,\,\vec{P}_{B_c}\right\rangle=P_\mu\,F_+(q^2)+q_\mu\,F_-(q^2)\nonumber\\
%
%
%
%
\left\langle\, {c\bar{c}}\, (1^-),\,\lambda\,\vec{P}_{c\bar{c}}\,\left|\, J^{c\,b}_\mu(0)\,
\right| \, B_c,\,\vec{P}_{B_c}\right\rangle&=&
\left\langle\, {c\bar{c}}\,(1^-),\,\lambda\,\vec{P}_{c\bar{c}}\,\left|\, J^{c\,b}_{V\mu}(0)-
J^{c\,b}_{A\mu}(0)\,
\right| \, B_c,\,\vec{P}_{B_c}\right\rangle\nonumber\\
&=&\frac{-1}{m_{B_c}+m_{c\bar{c}}} \varepsilon_{\mu\nu\alpha\beta}\ \varepsilon^{\,
\nu\,*}_{(\lambda)}(\,\vec{P
}_{c\bar{c}}\,)\,P^\alpha\,q^{\,
\beta}\,V(q^2)\nonumber\\
&&-\, i\, \bigg\{\ (m_{B_c}-m_{c\bar{c}})\ \varepsilon_{(\lambda)\,\mu}^*
(\,\vec{P}_{c\bar{c}}\,)\,A_0(q^2)\nonumber\\
&&\hspace{+.8cm}
-\frac{P\cdot\varepsilon^*_{(\lambda)}(\,\vec{P}_{c\bar{c}}\,)}
{m_{B_c}+m_{c\bar{c}}}
\left(P_\mu\,A_+(q^2)+q_\mu\,A_-(q^2)\right)\bigg\}\nonumber\\
\left\langle\, {c\bar{c}}\, (2^+),\,\lambda\,\vec{P}_{c\bar{c}}\,\left|\, J^{c\,b}_\mu(0)\,
\right| \, B_c,\,\vec{P}_{B_c}\right\rangle&=&
\left\langle\, {c\bar{c}}\, (2^+),\,\lambda\,\vec{P}_{c\bar{c}}\,\left|\, J^{c\,b}_{V\mu}(0)-
J^{c\,b}_{A\mu}(0)\,
\right| \, B_c,\,\vec{P}_{B_c}\right\rangle\nonumber\\
&=&\ \varepsilon_{\mu\nu\alpha\beta}\ \varepsilon^{\,
\nu\delta\,*}_{(\lambda)}(\,\vec{P
}_ {c\bar{c}}\,)P_\delta\,P^\alpha\,q^{\,
\beta}\,T_4(q^2)\nonumber\\
&&-\, i\, \bigg\{\ \varepsilon_{(\lambda)\,\mu\delta}^*(\,\vec{P}_{c\bar{c}}\,)
P^\delta\,T_1(q^2)\nonumber\\
&&\hspace{+.8cm}
+P^\nu P^\delta\varepsilon_{(\lambda)\nu\delta}^*(\,\vec{P}_{c\bar{c}},)
\left(P_\mu\, T_2(q^2)+q_\mu\,T_3(q^2)\right)\bigg\} \label{eq:9}
\end{eqnarray}
In the above expressions $P=P_{B_c}+P_{c\bar{c}}$,
$q=P_{B_c}-P_{c\bar{c}}$, being $P_{B_c}$ and $P_{c\bar{c}}$ the meson
four--momenta, $m_{B_c}$ and $m_{c\bar{c}}$ are the meson masses,
$\varepsilon^{\mu\nu\alpha\beta}$ is the fully antisymmetric tensor
for which we take the convention $\varepsilon^{0123}=+1$, and
$\varepsilon_{(\lambda)\mu}(\,\vec{P}\,)$ and
$\varepsilon_{(\lambda)\mu\nu}(\,\vec{P}\,)$ are the polarization
vector and tensor of vector and tensor mesons
respectively\footnote{Note we have taken $\lambda$ to be the
third component of the meson spin measured in the meson center of
mass.}. The latter can be evaluated in terms of the former as
\begin{eqnarray}
\varepsilon_{(\lambda)}^{\mu\nu}(\,\vec{P}\,)
=\sum_m (1,1,2\,;\,m,\lambda-m,\lambda)\,
\varepsilon_{(m)}^{\mu}(\,\vec{P}\,)
\varepsilon_{(\lambda-m)}^{\nu}(\,\vec{P}\,)
\end{eqnarray}
Different sets of $\varepsilon_{(\lambda)}(\,\vec{P}\,)$ used in this work
appear in appendix~\ref{app:epsilon}.

Besides the meson states in the Lorenz decompositions of
Eq.~(\ref{eq:9}) are normalized such that
\begin{equation}
\left\langle\, {M,\lambda'\,\vec{P}^{\,\prime}}\,|\,{M,\lambda
\,\vec{P}}\,\right\rangle
=\delta_{\lambda',\,\lambda}\,(2\pi)^3\,2\,E_M(\vec{P})\delta(\vec{P}^{\,\prime}-\vec{P}\,)
\end{equation}
being $E_M(\vec{P})$ the energy of the $M$ meson with three--momentum
$\vec{P}$. Note the factor $2E_M$ of difference with Eq.~(\ref{eq:4})

For the $0^+$, $1^+$ and $2^-$ cases the form factor decomposition is
the same as for the $0^-$, $1^-$ and $2^+$ cases respectively, but
with $-J^{c\,b}_{A\mu}(0)$ contributing where $ J^{c\,b}_{V\mu}(0)$
contributed before and vice versa.

The different form factors in Eq.(\ref{eq:ff}) are all relatively real
thanks to time--reversal invariance. $F_+$ , $F_-$, $V$, $A_0$, $A_+$,
$A_-$ and $T_1$ are dimensionless, whereas $T_2$, $T_3$ and $T_4$ have
dimension of $E^{-2}$.  They can be easily evaluated working in the
center of mass of the $B_c$ meson and taking $\vec{q}$ in the $z$
direction, so that $\vec{P}_{c\bar{c}}=-\vec{q}=-|\vec{q}\,| \vec{k}$,
with $\vec{k}$ representing the unit vector in the $z$ direction.
%
%
%
%
%
%
\subsubsection{$B_c^-\to \eta_c\, l^-\, \bar{\nu}_l,\ \chi_{c0}\, l^-\, 
\bar{\nu}_l $ decays}
Let us start with the $B_c^-$ decays into pseudoscalar $\eta_c$ and scalar 
$\chi_{c0}$ $c\bar{c}$ mesons. For  $B_c^-\to\eta_c$ transitions the form 
factors are given by:
\begin{eqnarray}
\label{eq:fpm0-}
F_+(q^2)=\frac{1}{2m_{B_c}}\,\left(
V^0(|\vec{q}\,|)+\frac{V^3(|\vec{q}\,|)}{|\vec{q}\,|}\,\left(E_{\eta_c}
(-\vec{q}\,)-m_{B_c}\right)\right)\nonumber\\
F_-(q^2)=\frac{1}{2m_{B_c}}\,\left(
V^0(|\vec{q}\,|)+\frac{V^3(|\vec{q}\,|)}{|\vec{q}\,|}\,\left(
E_{\eta_c}
(-\vec{q}\,)+m_{B_c}
\right)\right)
\end{eqnarray}
whereas for $B_c\to\chi_{c0}$ transitions we have:
\begin{eqnarray}
\label{eq:fpm0+}F_+(q^2)=\frac{-1}{2m_{B_c}}\,\left(
A^0(|\vec{q}\,|)+\frac{A^3(|\vec{q}\,|)}{|\vec{q}\,|}\,
\left(E_{\chi_{c0}}(-\vec{q}\,)-m_{B_{c}}\right)\right)\nonumber\\
F_-(q^2)=\frac{-1}{2m_{B_c}}\,\left(
A^0(|\vec{q}\,|)+\frac{A^3(|\vec{q}\,|)}{|\vec{q}\,|}\,\left(
E_{\chi_{c0}}(-\vec{q}\,)+m_{B_{c}}
\right)\right)
\end{eqnarray}%
with $V^\mu(|\vec{q}\,|)$  and $A^\mu(|\vec{q}\,|)$ ($\mu=0,3$)
calculated in our model as
\begin{eqnarray}
&&V^{\mu}(|\vec{q}\,|)=\left\langle\, \eta_c,\hspace{.2cm}-|\vec{q}\,|\,\vec{k}\,\left|\,
 J^{c\,b\ \mu}_V(0)\,
\right| \, B_c^-,\,\vec{0}\right\rangle=\sqrt{2m_{B_c}2E_{\eta_c}(-\vec{q}\,)}\ \ 
\hspace{.2cm}
{}_{\stackrel{}{\stackrel{}{NR}}}
\left\langle\, \eta_c,
\hspace{.2cm}
-|\vec{q}\,|\,{\vec{k}}\,\left|\, J^{c\,b\ \mu}_V(0)\,
\right| \, B_c^-,\,\vec{0}\right\rangle_{NR}\nonumber\\
&&A^{\mu}(|\vec{q}\,|)=\left\langle\, \chi_{c0},\,-|\vec{q}\,|\,\vec{k}\,\left|\,
 J^{c\,b\ \mu}_A(0)\,
\right| \, B_c^-,\,\vec{0}\right\rangle=\sqrt{2m_{B_c}2E_{\chi_{c0}}(-\vec{q}\,)}\ \ 
{}_{\stackrel{}{\stackrel{}{NR}}}\left\langle\, \chi_{c0},\,
-|\vec{q}\,|\,{\vec{k}}\,\left|\, J^{c\,b\ \mu}_A(0)\,
\right| \, B_c^-,\,\vec{0}\right\rangle_{NR}\end{eqnarray}
which expressions are given in appendix~\ref{app:va}.

\begin{figure}[t]
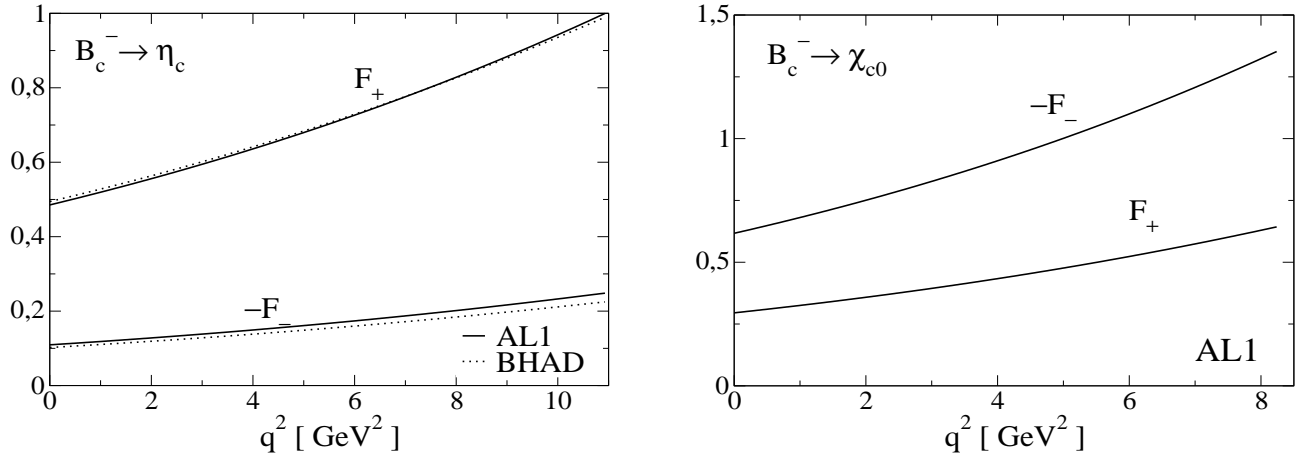

\vspace{1cm}
\centering
\resizebox{8.cm}{6.cm}{\includegraphics{etacelectron.eps}}\hspace{1cm}
\resizebox{8.cm}{6.cm}{\includegraphics{chic0electron.eps}}\hspace{1cm}
\caption{$F_+$ and $F_-$ form factors for  $B_c^-\to\eta_c
$ and $B_c^-\to\chi_{c0}$ semileptonic decay evaluated with the AL1 potential of
Refs.~\cite{silvestre96,sbs93}.  In the first case, and for
comparison, we also show with dotted lines the results obtained with
the Bhaduri (BHAD) potential of Ref.~\cite{bhaduri81}.} 
\label{fig:fpm}
\end{figure}
In Fig.~\ref{fig:fpm} we show our results for the $F_+$ and $F_-$ form
factors for the semileptonic $B_c^-\to \eta_c,\,\chi_{c0}$ transitions.
The minimum $q^2$ value depends on the actual final lepton and it is given, 
neglecting neutrino
masses, by the  lepton mass as
$q^2_{min}=m_l^2$. The form factors have
been evaluated using the AL1 potential of
Refs.~\cite{silvestre96,sbs93}. For decays into $\eta_c$, and for the
sake of comparison, we also show the results obtained with the
potential developed by Bhaduri and collaborators in
Ref.~\cite{bhaduri81} (BHAD).  As seen in the figures the differences
between the form factors evaluated with the two inter-quark
interactions are smaller than 10\%.

In Table~\ref{tab:fpm} we show  $F_+$ and $F_-$ evaluated at
$q^2_{\mathrm{min}}$ and $q^2_{\mathrm{max}}$ for a final light lepton
 ($l=e,\,\mu$)  and  compare them to  the ones
 obtained by Ivanov {\it et al.} in Ref.~\cite{ivanov05}, and, when available, by Ebert {\it et al.} in
 Ref.~\cite{ebert03}.
 For the $B_c\to \eta_c$ transition we also show the
 corresponding values for the $F_0$ form factor defined as
 \begin{equation}
\label{eq:f0}
 F_0(q^2)=F_+(q^2)+\frac{q^2}{m^2_{B_c}-m^2_{\eta_c}}\ F_-(q^2)
 \end{equation}
 Our results for the $\eta_c$ case are in excellent agreement with the ones obtained by  
 Ebert {\it et al.}. Compared to the results by Ivanov {\it et al.} we find
 large discrepancies  for $F_-$. 


\begin{table}[h!]

\begin{tabular}{c| c c c c|  c c}
$B_c^-\to\eta_c\,l^-\bar{\nu}_l\ \ $ & $q^2_{\mathrm{min}}$ &
 $q^2_{\mathrm{max}}$ & $\hspace{1cm}$&
$B_c^-\to\chi_{c0}\,l^-\bar{\nu}_l\ \ $ & $q^2_{\mathrm{min}}$ &
 $q^2_{\mathrm{max}}$  \\
\cline{1-3} \cline{5-7}
&&&&&&\\
 $F_+$ & & & &$F_+$ & &\\
 This work & $0.49^{+0.01}$& $1.00_{-0.01}$ & & This work & $0.30^{+0.01}$& 
 $0.64^{+0.01}_{-0.01}$\\
 \cite{ivanov05} &0.61 &1.14&&\cite{ivanov05} &0.40 &0.65\\
 \cite{ebert03} &0.47 &1.07 & & &  &\\
&&&&&&\\
$F_-$ & & & &$F_-$ & &\\
 This work &$-0.11^{+0.01}_{-0.01}$ &$-0.25^{+0.03}$ & &  This work
 &$-0.62^{+0.01}_{-0.03}$ &$-1.35^{+0.05}$\\
 \cite{ivanov05} & $-0.32$&$-0.61$& & \cite{ivanov05} & $-1.00$&$-1.63$\\
&&\\
$F_0$ & & \\
 This work & $0.49^{+0.01}$ &$ 0.91^{+0.01}$ \\
 \cite{ebert03} & 0.47& 0.92 
 \end{tabular}\hspace{1cm}
\caption{$F_+$ and $F_-$ evaluated at
$q^2_{\mathrm{min}}$ and $q^2_{\mathrm{max}}$ compared to  the ones obtained by 
Ivanov {\it et al.}~\cite{ivanov05} and  Ebert {\it et al.}
Ref.~\cite{ebert03}. Our central values have been obtained with the AL1
potential. For the $\eta_c$ channel we also show $F_0$ (see text for
definition).
Here $l$ stands for $l=e,\,\mu$.}
\label{tab:fpm}
\end{table}

%
%
%
\subsubsection{$B_c^-\to J/\Psi\, l\, \bar{\nu}_l,\  h_c\, l\, \bar{\nu}_l,
\ \chi_{c1}\, l\, \bar{\nu}_l$ decays}
Let us now see the form factors for the
semileptonic $B_c^-$ decays into vector $J/\Psi$ and axial vector
$h_c$ ($S_{q\bar{q}}=0$) and $\chi_{c1}$ ($S_{q\bar{q}}=1$) $c\bar{c}$
mesons. For the decay into $J/\Psi$ the form factors can be evaluated
in terms of matrix elements as:
\begin{eqnarray}
\label{eq:ffv}
V(q^2)&=&\frac{i}{\sqrt2}\,\frac{m_{B_c}+m_{J/\Psi}}{m_{B_c}\,|\vec{q}\,|}\
V^1_{\lambda=-1}(|\vec{q}\,|)\nonumber\\
A_+(q^2)&=&i\,\frac{m_{B_c}+m_{J/\Psi}}{2m_{B_c}}\,
\frac{m_{J/\Psi}}{|\vec{q}\,|\,m_{B_c}}\,
\bigg\{
-A^0_{\lambda=0}(|\vec{q}\,|)+\frac{m_{B_c}-E_{J/\Psi}(-\vec{q}\,)}{|\vec{q}\,|}\,
 A^3_{\lambda=0}(|\vec{q}\,|)\nonumber\\
&&\hspace{4cm}-\sqrt2\,\frac{m_{B_c}E_{J/\Psi}(-\vec{q}\,)-m^2_{J/\Psi}}{|\vec{q}\,|\,
m_{J/\Psi}}\,A^1_{\lambda=-1}(|\vec{q}\,|)\bigg\}\nonumber\\
A_-(q^2)&=&-i\,\frac{m_{B_c}+m_{J/\Psi}}{2m_{B_c}}\,
\frac{m_{J/\Psi}}{|\vec{q}\,|\,m_{B_c}}\,
\bigg\{
\ A^0_{\lambda=0}(|\vec{q}\,|)+\frac{m_{B_c}+E_{J/\Psi}(-\vec{q}\,)}{|\vec{q}\,|}\,
 A^3_{\lambda=0}(|\vec{q}\,|)\nonumber\\
&&\hspace{4cm}-\sqrt2\,\frac{m_{B_c}E_{J/\Psi}(-\vec{q}\,)+m^2_{J/\Psi}}{|\vec{q}\,|\,
m_{J/\Psi}}\,A^1_{\lambda=-1}(|\vec{q}\,|)\bigg\}\nonumber\\
A_0(q^2)&=&-i\sqrt2\,\frac{1}{m_{B_c}-m_{J/\Psi}}\, A^1_{\lambda=-1}(|\vec{q}\,|)
\end{eqnarray}
with $V^\mu_\lambda(|\vec{q}\,|)$  and $A^\mu_\lambda(|\vec{q}\,|)$ 
calculated in our model as
\begin{eqnarray}
\label{eq:vavector}
V^{\mu}_\lambda(|\vec{q}\,|)&=&\left\langle\, J/\Psi,\hspace{.2cm}\lambda\ -|\vec{q}\,|\,\vec{k}\,\left|\,
 J^{c\,b\ \mu}_V(0)\,
\right| \, B_c^-,\,\vec{0}\right\rangle\nonumber\\
&=&\sqrt{2m_{B_c}2E_{J/\Psi}(-\vec{q}\,)}\ 
\ 
{}_{\stackrel{}{\stackrel{}{NR}}}
\left\langle\, J/\Psi,
\hspace{.2cm}\lambda\,
-|\vec{q}\,|\,{\vec{k}}\,\left|\, J^{c\,b\ \mu}_V(0)\,
\right| \, B_c^-,\,\vec{0}\right\rangle_{NR}\nonumber\\
A^{\mu}_\lambda(|\vec{q}\,|)&=&\left\langle\, J/\Psi,\hspace{.2cm}\lambda\ -|\vec{q}\,|\,\vec{k}\,\left|\,
 J^{c\,b\ \mu}_A(0)\,
\right| \, B_c^-,\,\vec{0}\right\rangle\nonumber\\
&=&\sqrt{2m_{B_c}2E_{J/\Psi}(-\vec{q}\,)}\ 
\ 
{}_{\stackrel{}{\stackrel{}{NR}}}
\left\langle\, J/\Psi,
\hspace{.2cm}\lambda\,
-|\vec{q}\,|\,{\vec{k}}\,\left|\, J^{c\,b\ \mu}_A(0)\,
\right| \, B_c^-,\,\vec{0}\right\rangle_{NR}
\end{eqnarray}
which expressions are given in appendix~\ref{app:va}.

The  form factors corresponding to  transitions to the $\chi_{c1}$ 
and $h_c$ axial vector mesons are obtained from the expressions in 
Eq.(\ref{eq:ffv}) by just  changing
\begin{eqnarray}
V^\mu_\lambda(|\vec{q}\,|)\longleftrightarrow -A^\mu_\lambda(|\vec{q}\,|)
\end{eqnarray} 
and using the appropriate mass for the final meson. Obviously in 
Eq.~(\ref{eq:vavector}) $J/\Psi$ has to be replaced by $\chi_{c_1}$ or $h_c$.

In Table~\ref{tab:fva0pm} we show the result for the different form
factors evaluated at $q^2_{\mathrm{min}}$ and $q^2_{\mathrm{max}}$ for
the case where the final lepton is light ($l=e,\,\mu$). For the decay
into $J/\Psi$ we also show the combination of form
factors\footnote{This combination is called $A_0$ by the authors of
Ref.~\cite{ebert03}}:
\begin{eqnarray}
\label{eq:a0tilde}
\widetilde{A}_0(q^2)=\frac{m_{B_c}-m_{J/\Psi}}{2m_{J/\Psi}}\left(
A_0(q^2)-A_+(q^2)\right)-\frac{q^2}{2m_{J/\Psi}\,(m_{B_c}+m_{J/\Psi})}\ A_-(q^2)
\end{eqnarray}
   Our results for the $B_c\to J/\Psi$ decay channel are in  agreement with the ones obtained by  
 Ebert {\it et al.}. They also agree reasonably well, with the exception of
 $A_-$, with the ones obtained by Ivanov 
 {\it et al.}. For the other two cases the discrepancies are in general large.
%
%
%
\begin{table}[h!]
\hspace*{-1cm}
\begin{tabular}{c| c c c c|  c c cc|  c c}
$B_c^-\to J/\Psi\,l^-\bar{\nu}_l\ \ $ & $q^2_{\mathrm{min}}$ &
 $q^2_{\mathrm{max}}$ & $\hspace{.5cm}$&
$B_c^-\to h_{c}\,l^-\bar{\nu}_l\ \ $ & $q^2_{\mathrm{min}}$ &
 $q^2_{\mathrm{max}}$ & $\hspace{.5cm}$&
$B_c^-\to\chi_{c1}\,l^-\bar{\nu}_l\ \ $ & $q^2_{\mathrm{min}}$ &
 $q^2_{\mathrm{max}}$ \\
\cline{1-3} \cline{5-7} \cline{9-11}
&&&&&&&&&\\
 $V$ & & & &$V$ & & & &$V$ & &\\
 This work & $-0.61_{-0.03}$ & $-1.26^{+0.01}$ & & This work & $-0.040_{-0.003}$& $-0.078_{-0.003}$& & This work & 
 $0.92^{+0.04}_{-0.02}$&
 $1.86_{-0.12}$\\
 \cite{ivanov05} &$-0.83$$^*$ &$-1.53$$^*$&&\cite{ivanov05} &$-0.25$
 $^*$ &$-0.365$$^*$&&\cite{ivanov05} 
 &1.18$^*$ &1.81$^*$\\
 \cite{ebert03} &$-0.49$ &$-1.34$ & & &  &&&&&\\
&&&&&&&&&\\
 $A_+$ & & & &$A_+$ & & & &$A_+$ & &\\
 This work & $0.56^{+0.03}$& $1.13^{+0.01}$ & & This work & $-0.85^{+0.01}_{-0.05}$& $-1.90^{+0.06}$& & This work & 
 $-0.44_{-0.03}$& $-0.78^{+0.04}$\\
 \cite{ivanov05} &0.54 &0.97&&\cite{ivanov05} &$-1.08$ &$-1.80$&&\cite{ivanov05} 
 &$-0.39$ &$-0.50$\\
 \cite{ebert03} &0.73 &1.33 & & &  &&&&&\\ 
&&&&&&&&&\\
 $A_-$ & & & &$A_-$ & & & &$A_-$ & &\\
 This work & $-0.60_{-0.03}$& $-1.24_{-0.01}$ & & This work & $0.12^{+0.01}_{-0.02}$& $0.36_{-0.06}$& & This work &
 $ 0.96^{+0.04}_{-0.01}$& $1.97_{-0.13}$\\
 \cite{ivanov05} &$-0.95$ &$-1.76$&&\cite{ivanov05} &0.52 &0.89&&\cite{ivanov05} 
 &1.52 &2.36\\
 &&&&&&&&&\\
$A_0$ & & & &$A_0$ & & & &$A_0$ & &\\
 This work & $1.44^{+0.08}$& $2.58^{+0.01}_{-0.02}$ & & This work & $0.28^{+0.02}$& $0.52^{+0.02}$& & This work &
 $ -0.50_{-0.02}$&$-0.32_{-0.02}$ \\
 \cite{ivanov05} &1.64 &2.50&&\cite{ivanov05} &0.44 &0.54&&\cite{ivanov05} 
 &$-0.064$ &0.46\\
 \cite{ebert03} &1.47 &2.59 & & &  &&&&&\\  
&&&&&&&&&\\
 $\widetilde{A}_0$ & & & & & & & & & &\\
 This work & $0.45^{+0.03}$& $0.96_{-0.01}$ & &  & & & & & & \\
 \cite{ebert03} &0.40 &1.06 & & &  &&&&&\\  
 \end{tabular}
\caption{$V$, $A_+$, $A_-$ and $A_0$ form factors evaluated at
$q^2_{\mathrm{min}}$ and $q^2_{\mathrm{max}}$ compared to  the ones obtained by 
Ivanov {\it et al.}~\cite{ivanov05} and  Ebert {\it et al.} 
Ref.~\cite{ebert03}. Our central values have been evaluated with the AL1
potential. For the $J/\Psi$ channel we also show $\widetilde{A}_0$ (see text
for definition). Here
$l$ stands for $l=e,\,\mu$. The asterisk to the right of a number means we 
have changed its sign to account
 for the different choice of $\varepsilon^{\,0\,1\,2\,3}$
in Ref.~\protect{\cite{ivanov05}}.}
\label{tab:fva0pm}
\end{table}

\begin{figure}[h!!!]
\vspace{1cm}
\centering
\resizebox{8.cm}{6.cm}{\includegraphics{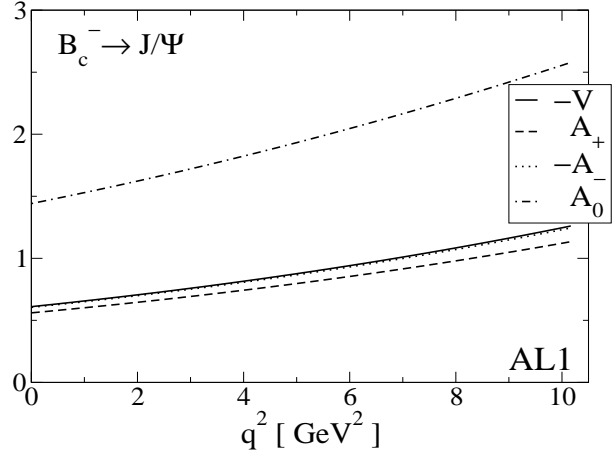}}
\caption{$V$ (solid line), $A_+$ (dashed line), $A_-$ (dotted line)
and $A_0$ (dashed-dotted line) form factors for the $B_c^-\to J/\Psi$
semileptonic decay evaluated with the AL1 potential. }
\label{fig:fva0pmjpsi}
\end{figure}

\begin{figure}[h!!!]
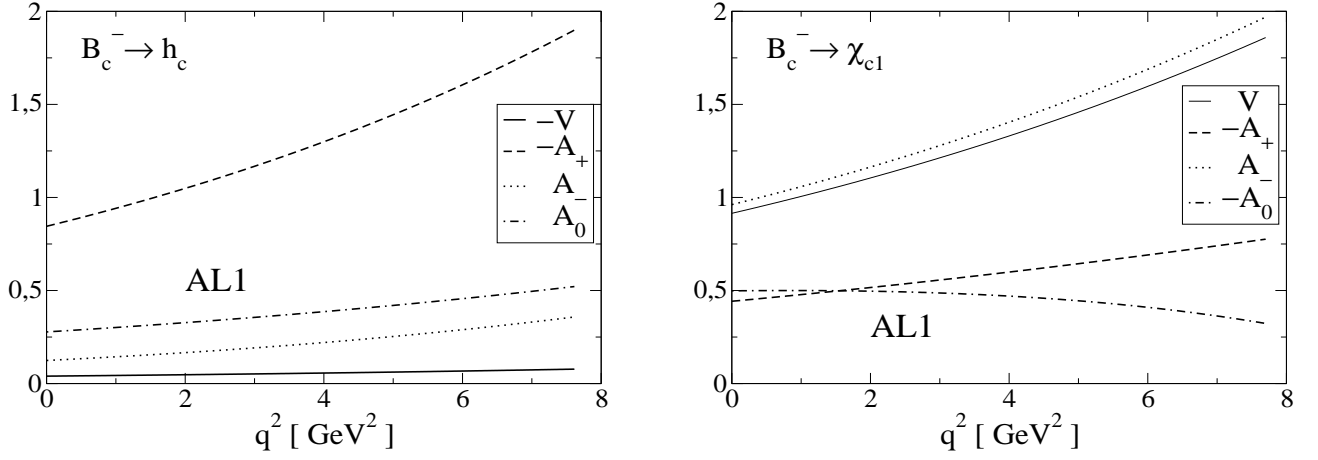

\vspace{1cm}
\centering
\resizebox{8.cm}{6.cm}{\includegraphics{hcelectron.eps}}\hspace{1cm}
\resizebox{8.cm}{6.cm}{\includegraphics{chic1electron.eps}}\hspace{1cm}
\caption{$V$ (solid line), $A_+$ (dashed line), $A_-$ (dotted line) 
and $A_0$ (dashed-dotted line) form factors for  
 $B_c^-\to h_c$ and $B_c^-\to\chi_{c1}$
semileptonic decay evaluated with the AL1 potential.}
\label{fig:fva0pm}
\end{figure}

All the  form factors are depicted in Figs.~\ref{fig:fva0pmjpsi} and \ref{fig:fva0pm}

%
%
%
%
%
%
%
\newpage
\subsubsection{$B_c^-\to \Psi(3836)\, l\, \bar{\nu}_l,\  \chi_{c2}\, l\,
 \bar{\nu}_l$, decays}
Finally let us  see the form factors for the $B_c^-$ decays into tensor
$\chi_{c2}$ and pseudotensor $\Psi(3836)$~\footnote{Note that while 
the $\Psi(3836)$ was still quoted in the particle listings of the former
Review of
Particle Physics~\cite{pdg04}, it has been excluded from the  more recent 
one~\cite{yao06}. We shall nevertheles keep it in our study to illustrate
the results to be expected for a ground state pseudotensor particle.}
 mesons. For the decay into $\chi_{c2}$ the form factors can be evaluated in 
 terms of matrix elements as:
\begin{eqnarray}
\label{eq:fft}
T_1(q^2)&=&-i\,\frac{2m_{\chi_{c2}}}{m_{B_c}|\vec{q}\,|}\
A^1_{T\lambda=+1}(|\vec{q}\,|)\nonumber\\
T_2(q^2)&=&i\,\frac{1}{2m^3_{B_c}}\,
\bigg\{
-\sqrt{\frac{3}{2}}\,\frac{m^2_{\chi_{c2}}}{|\vec{q}\,|^2}\,
A^0_{T\lambda=0}(|\vec{q}\,|)
-\sqrt{\frac{3}{2}}\,\frac{m^2_{\chi_{c2}}}
{|\vec{q}\,|^3}
\, \left(E_{\chi_{c2}}(-\vec{q}\,)-m_{B_c}\right)
 A^3_{T\lambda=0}(|\vec{q}\,|)\nonumber\\
&&\hspace{1.55cm}+\frac{2m_{\chi_{c2}}}{|\vec{q}\,|}
\left(1-\frac{E_{\chi_{c2}}(-\vec{q}\,)}{|\vec{q}\,|^2}
\left(E_{\chi_{c2}}(-\vec{q}\,)-m_{B_c}
\right)
\right)
\,A^1_{T\lambda=+1}(|\vec{q}\,|)\bigg\}\nonumber\\
T_3(q^2)&=&i\,\frac{1}{2m^3_{B_c}}\,
\bigg\{
-\sqrt{\frac{3}{2}}\,\frac{m^2_{\chi_{c2}}}{|\vec{q}\,|^2}\,
A^0_{T\lambda=0}(|\vec{q}\,|)
-\sqrt{\frac{3}{2}}\,\frac{m^2_{\chi_{c2}}}
{|\vec{q}\,|^3}
\, \left(E_{\chi_{c2}}(-\vec{q}\,)+m_{B_c}\right)
 A^3_{T\lambda=0}(|\vec{q}\,|)\nonumber\\
&&\hspace{1.55cm}+\frac{2m_{\chi_{c2}}}{|\vec{q}\,|}
\left(1-\frac{E_{\chi_{c2}}(-\vec{q}\,)}{|\vec{q}\,|^2}
\left(E_{\chi_{c2}}(-\vec{q}\,)+m_{B_c}
\right)
\right)
\,A^1_{T\lambda=+1}(|\vec{q}\,|)\bigg\}\nonumber\\
T_4(q^2)&=&i\,\frac{m_{\chi_{c2}}}{m^2_{B_c}|\vec{q}\,|^2}\, 
V^1_{T\lambda=+1}(|\vec{q}\,|)
\end{eqnarray}

with $V^\mu_{T\lambda}(|\vec{q}\,|)$  and $A^\mu_{T \lambda}(|\vec{q}\,|)$ 
calculated in our model as
\begin{eqnarray}
\label{eq:vatensor}
V^{\mu}_{T \lambda}(|\vec{q}\,|)&=&\left\langle\, \chi_{c2},\hspace{.2cm}\lambda\ -|\vec{q}\,|\,\vec{k}\,\left|\,
 J^{c\,b\ \mu}_V(0)\,
\right| \, B_c^-,\,\vec{0}\right\rangle\nonumber\\
&=&\sqrt{2m_{B_c}2E_{\chi_{c2}}(-\vec{q}\,)}\ 
\ 
{}_{\stackrel{}{\stackrel{}{NR}}}
\left\langle\, \chi_{c2},
\hspace{.2cm}\lambda\,
-|\vec{q}\,|\,{\vec{k}}\,\left|\, J^{c\,b\ \mu}_V(0)\,
\right| \, B_c^-,\,\vec{0}\right\rangle_{NR}\nonumber\\
A^{\mu}_{T\lambda}(|\vec{q}\,|)&=&\left\langle\, \chi_{c2},\hspace{.2cm}\lambda\ -|\vec{q}\,|\,\vec{k}\,\left|\,
 J^{c\,b\ \mu}_A(0)\,
\right| \, B_c^-,\,\vec{0}\right\rangle\nonumber\\
&=&\sqrt{2m_{B_c}2E_{\chi_{c2}}(-\vec{q}\,)}\ 
\ 
{}_{\stackrel{}{\stackrel{}{NR}}}
\left\langle\, \chi_{c2},
\hspace{.2cm}\lambda\,
-|\vec{q}\,|\,{\vec{k}}\,\left|\, J^{c\,b\ \mu}_A(0)\,
\right| \, B_c^-,\,\vec{0}\right\rangle_{NR}
\end{eqnarray}
which expressions are given in appendix~\ref{app:va}.

The  form factors corresponding to  transitions to  $\Psi(3836)$ 
 are obtained from the expressions in 
Eq.(\ref{eq:fft}) by just  changing
\begin{eqnarray}
V^\mu_{T\lambda}(|\vec{q}\,|)\longleftrightarrow -A^\mu_{T\lambda}(|\vec{q}\,|)
\end{eqnarray} 
and using the appropriate mass for the final meson. Besides in 
Eq.~(\ref{eq:vatensor}) $\chi_{c_2}$ has to be replaced by $\Psi(3836)$.

\begin{figure}[h!!!]
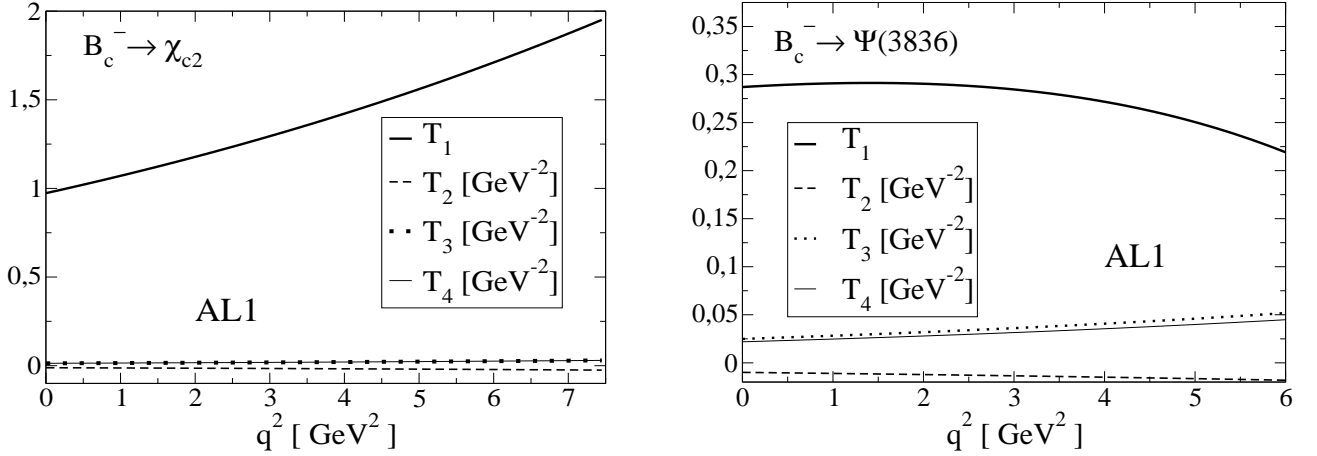

\vspace{1cm}
\centering
\resizebox{8.cm}{6.cm}{\includegraphics{chic2electron.eps}}\hspace{1cm}
\resizebox{8.cm}{6.cm}{\includegraphics{psi3836electron.eps}}\hspace{1cm}
\caption{$T_1$ (bold solid line), $T_2$ (dashed line), $T_3$ (dotted line) 
and $T_4$ (thin solid line) form factors for  
 $B_c^-\to \chi_{c2}$ and
$B_c^-\to \Psi(3836)$
semileptonic decay evaluated with the AL1 potential.}
\label{fig:t1234}
\end{figure}

The results for the different form factors appear in  Fig.~\ref{fig:t1234}.
%
%
%
%
In Table~\ref{tab:t1234} we show $T_1$, $T_2$, $T_3$ and $T_4$ evaluated at
$q^2_{min}$ and $q^2_{max}$ for the case of a final light lepton, and compare
them to the values obtained by Ivanov {\it et al.}~\cite{ivanov05}. For the 
$B_c^-\to \chi_{c2}$ transition
we find a reasonable agreement between the two calculations. For 
$B_c^-\to \Psi (3836)$ there is also a reasonable agreement for the absolute
values of the form factors but we disagree  for some of the signs.

\begin{table}[h!]

\begin{tabular}{c| c c c c|  c c}
$B_c^-\to\chi_{c2}\,l^-\bar{\nu}_l\ \ $ & $q^2_{\mathrm{min}}$ &
 $q^2_{\mathrm{max}}$ & $\hspace{1cm}$&
$B_c^-\to\Psi(3836)\,l^-\bar{\nu}_l\ \ $ & $q^2_{\mathrm{min}}$ &
 $q^2_{\mathrm{max}}$  \\
\cline{1-3} \cline{5-7}
&&&&&&\\
 $T_1$ & & & &$T_1$ & &\\
 This work & $
 0.97^{+0.08}_{-0.01}$&
 $1.95_{-0.06}$ & & This work &$0.29_{-0.02}$& $0.22_{-0.01}$\\
 \cite{ivanov05} &1.22 &1.69&&\cite{ivanov05} &0.052 &0.35\\
&&&&&&\\
$T_2\ [GeV^{-2}]$ & & & &$T_2\ [GeV^{-2}]$ & &\\
 This work & $-0.012_{-0.001}$&$-0.025^{+0.001}$& &  This work &
 $-0.010^{+0.006}$ &$-0.018^{+0.001}$\\
 \cite{ivanov05} & $-0.011$&$-0.018$& & \cite{ivanov05} & 0.0071&0.0090\\
&&&&&&\\
$T_3\ [GeV^{-2}]$ & & & &$T_3\ [GeV^{-2}]$ & &\\
 This work &$0.013^{+0.001}$& $0.030_{-0.001}$ & & 
 This work & $ 0.025_{-0.002}$ & $0.052_{-0.004}$ \\
 \cite{ivanov05} & 0.025&0.040& & \cite{ivanov05} & $-0.036$&$ -0.052$ \\
&&&&&&\\
$T_4\ [GeV^{-2}]$ & & & &$T_4\ [GeV^{-2}]$ & &\\
 This work & $0.013^{+0.001}$& $0.030_{-0.001}$& & 
 This work &  $0.022_{-0.001}$ & $0.045_{-0.003}$ \\
 \cite{ivanov05} & 0.021$^*$&0.033$^*$ & &\cite{ivanov05} 
 &  $-0.026$$^*$&$-0.038$$^*$
 \end{tabular}
 \caption{Values for $T_1$, $T_2$, $T_3$  and $T_4$ evaluated at
$q^2_{\mathrm{min}}$ and $q^2_{\mathrm{max}}$ compared to  the ones obtained by 
Ivanov {\it et al.}~\cite{ivanov05}. Here
$l$ stands for $l=e,\,\mu$. Asterisk as in Table~\ref{tab:fva0pm}}
\label{tab:t1234}
\end{table}
%
%
%
%
%
\subsection{Decay width}
For a $B_c$ at rest the double differential decay width with respect
to $q^2$ and $x_l$, being $x_l$ the cosine of the angle between the
final meson momentum and the momentum of the final charged lepton
measured in the lepton--neutrino center of mass frame (CMF), is given
by\footnote{We shall neglect neutrino masses in the calculation.}
\begin{eqnarray}
\frac{d^2\Gamma}{dq^2dx_l}=\frac{G_F^2}{64m^2_{B_c}}\,
\frac{|V_{bc}|^2}{8\pi^3}\,\frac{\lambda^{1/2}(q^2,m^2_{B_c},m^2_{c\bar{c}})}
{2m_{B_c}}\, \frac{q^2-m^2_{l}}{q^2}
 \ {\cal H}_{\alpha\,\beta} ({P}_{B_c},{P}_{c\bar{c}})
\ {\cal L}^{\alpha\,\beta}({p}_l,{p}_{{\nu}})
\end{eqnarray}
where $G_F=1.16637 (1) \times 10 ^{-5} \mathrm{GeV}^{-2}$~\cite{pdg04}
is the Fermi constant, $\lambda(a,b,c)=(a+b-c)^2-4ab$, $m_l$ is the
mass of the charged lepton, ${\cal H}_{\alpha\,\beta}$ and ${\cal
L}^{\alpha\,\beta}$ are the hadron and lepton tensors, and
${P}_{B_c},\ {P}_{c\bar{c}},\ {p}_l,\ {p}_{{\nu}}$ are the meson and
lepton four--momenta.  The lepton tensor is\footnote{The $\mp$ signs
correspond respectively to decays into $l^-\bar{\nu}_l$ (for $B_c^-$
decays) and $l^+{\nu}_l$ (for $B_c^+$ decays)}
 \begin{eqnarray}
 {\cal L}^{\alpha\,\beta}({p}_l,{p}_{{\nu}})=
 8\left(
 {p}_l^\alpha{p}_{{\nu}}^\beta+
 {p}_l^\beta{p}_{{\nu}}^\alpha
 -g^{\alpha\beta}{p}_l\cdot{p}_{{\nu}}
 \mp i\,\varepsilon^{\alpha\,\beta\,\sigma\,\rho} p_{l\,\sigma}
 p_{\nu\,\rho}\right)
 \end{eqnarray}
As for the hadron tensor it is given by
 \begin{eqnarray}
 {\cal H}_{\alpha\,\beta}({P}_{B_c},{P}_{c\bar{c}})&=&
 \sum_{\lambda}\ h_{(\lambda)\,\alpha}({P}_{B_c},{P}_{c\bar{c}})\,
  h^*_{(\lambda)\,\beta}({P}_{B_c},{P}_{c\bar{c}})
 \end{eqnarray}
 with
 
 \begin{eqnarray}
 h_{(\lambda)\,\alpha}({P}_{B_c},{P}_{c\bar{c}})=
  \left\langle c\bar{c}, \, \lambda\,
 \vec P_{c\bar{c}}|J_{\alpha}^{cb}(0)|
 B_c,\,\vec P_{B_c}
 \right\rangle
 \end{eqnarray}

 The quantity 
\begin{eqnarray}
 {\cal H}_{\alpha\,\beta}({P}_{B_c},{P}_{c\bar{c}})
\ {\cal L}^{\alpha\,\beta}({p}_{l},{p}_{\nu})%
\end{eqnarray}
is a scalar and to evaluate it we can choose $\vec{P}_{c\bar c}$ along
the negative $z$-axis. This implies also that the CMF of the final
leptons moves in the positive $z$-direction.  Furthermore we shall
follow Ref.~\cite{ivanov05} and introduce helicity components for the
hadron and lepton tensors. For that purpose we rewrite
\begin{eqnarray}
 {\cal H}_{\alpha\,\beta}({P}_{B_c},{P}_{c\bar{c}})
\ {\cal L}^{\alpha\,\beta}({p}_{l},{p}_{\nu})
={\cal H}^{\sigma\,\rho}({P}_{B_c},{P}_{c\bar{c}})
\ g_{\sigma\alpha}\,g_{\rho\beta}
\ {\cal L}^{\alpha\,\beta}({p}_{l},{p}_{\nu})
\end{eqnarray}
and use~\cite{korner90}
\begin{eqnarray}
g_{\mu\nu}=\sum_{r=t,\pm 1,0} \ g_{rr}\ \varepsilon_{(r)\,\mu}({q})
\,\varepsilon_{(r)\,\nu}^*({q})
\hspace{0.5cm};\hspace{0.5cm} g_{tt}=1,g_{\pm 1,0}=-1
\end{eqnarray}
with $\varepsilon^\mu_{(t)}({q})=q^\mu/q^2$ and where the
  $\varepsilon_{(r)}({q}),\, r=\pm1,\,0$ are the polarization vectors
  for an on--shell vector particle with four--momentum ${q}$ and
  helicity $r$.  Defining helicity components for the hadron and
  lepton tensors as
\begin{eqnarray}
 {\cal H}_{r\,s}({P}_{B_c},{P}_{c\bar{c}})&=&
\varepsilon_{(r)\,\sigma}^*({q})\,
{\cal H}^{\sigma\,\rho}({P}_{B_c},{P}_{c\bar{c}})
\,\varepsilon_{(s)\,\rho}({q})\nonumber\\
 {\cal L}_{r\,s}({p}_{l},{p}_{\nu})&=&
\varepsilon_{(r)\,\alpha}({q})\,
{\cal L}^{\alpha\,\beta}({p}_{l},{p}_{\nu})
\,\varepsilon_{(s)\,\beta}^*({q})
\end{eqnarray}
we have that
\begin{eqnarray}
 {\cal H}_{\alpha\,\beta}({P}_{B_c},{P}_{c\bar{c}})
\ {\cal L}^{\alpha\,\beta}({p}_{l},{p}_{\nu})
=\sum_{r=t,\pm 1,0}\sum_{s=t,\pm 1,0}\ g_{rr}\,g_{ss}\
 {\cal H}_{r\,s}({P}_{B_c},{P}_{c\bar{c}})\
 {\cal L}_{r\,s}({p}_l,{p}_{\nu})
 \end{eqnarray}

Let us start with the lepton tensor. We can take advantage of the fact
that the Wigner rotation relating the original frame and the CMF of
the final leptons is the identity to evaluate the lepton tensor
helicity components in this latter reference system
\begin{eqnarray}
 {\cal L}_{r\,s}({p}_l,{p}_{\nu})&=&
\varepsilon_{(r)\,\alpha}({q})\,
  {\cal L}^{\alpha\,\beta}({p}_{l},{p}_{\nu})
\,\varepsilon_{(s)\,\beta}^*({q})=
\varepsilon_{(r)\,\alpha}(\widetilde{q})\,
{\cal L}^{\alpha,\beta}(\widetilde{p}_{l},\widetilde{p}_{\nu})
\,\varepsilon_{(s)\,\beta}^*(\widetilde{q})
\end{eqnarray}
were the tilde stands for momenta measured in the final leptons CMF. 
For the purpose of evaluation we can use\footnote{Note this is in accordance
with the definition of $x_l$ and the fact that we have taken $\vec{P}_{c\bar c}$
in the negative $z$ direction. Furthermore there can be no dependence on the
$\varphi_l$ azimuthal angle so that we  can take $\widetilde{\vec{p}}_{l}$, and
then $\widetilde{\vec{p}}_{\nu}$,
 in the $OXZ$ plane.}
\begin{eqnarray}
\label{eq:lm}
\widetilde{p}_l^\alpha&=&(E_l(|\widetilde p_l|),-|\widetilde p_l|
\,\sqrt{1-x_l^2},0,-|\widetilde p_l|\,x_l)\nonumber\\
\widetilde{p}_\nu^\alpha&=&(|\widetilde p_l|,|\widetilde p_l|
\,\sqrt{1-x_l^2},0,|\widetilde p_l|\,x_l)
\end{eqnarray}
with $|\widetilde p_l|$ the modulus of the lepton three-momentum measured in the leptons CMF.

The only  helicity components that we shall need
are the following.
\begin{eqnarray}
{\cal L}_{t\,t}({p}_l,{p}_{\nu})&=&
4 \frac{m_l^2 (q^2-m_l^2)}{q^2}\nonumber\\
{\cal L}_{t\,0}({p}_l,{p}_{\nu})&=&
{\cal L}_{0\,t}({p}_l,{p}_{\nu})=
-4x_l\frac{m_l^2 (q^2-m_l^2)}{q^2}\nonumber\\
{\cal L}_{+1\,+1}({p}_l,{p}_{\nu})&=&
(q^2-m_l^2)\,\left(
4(1\pm x_l)-2(1-x_l^2)\frac{q^2-m_l^2}{q^2}
\right)
\nonumber\\
{\cal L}_{-1\,-1}({p}_l,{p}_{\nu})&=&
(q^2-m_l^2)\,\left(
4(1\mp x_l)-2(1-x_l^2)\frac{q^2-m_l^2}{q^2}
\right)
\nonumber\\
{\cal L}_{0\,0}({p}_l,{p}_{\nu})&=&
4(q^2-m_l^2)\,\left(
1-x_l^2\frac{q^2-m_l^2}{q^2}
\right)
\nonumber\\
\end{eqnarray}
As for the hadron tensor  it is convenient to introduce 
helicity amplitudes defined as
 \begin{eqnarray}
 h_{(\lambda)\,r}({P}_{B_c},{P}_{c\bar{c}})=
\varepsilon_{(r)\,\alpha}^*({q})\ h_{(\lambda)}^\alpha({P}_{B_c},{P}_{c\bar{c}}) 
 \hspace{0.5cm},\hspace{0.5cm} r=t,\,\pm 1,\,0 \end{eqnarray}
  in terms of which
 \begin{eqnarray}
 {\cal H}_{r\,s}({P}_{B_c},{P}_{c\bar{c}})=\sum_\lambda
h_{(\lambda)\,r}({P}_{B_c},{P}_{c\bar{c}})\,
h_{(\lambda)\,s}^*({P}_{B_c},{P}_{c\bar{c}})
\label{eq:hcht}
 \end{eqnarray} 
We now give the expressions for the helicity amplitudes 
evaluated in the original frame. 

\begin{itemize}
\item[-] Case $0^-\to 0^-,0^+$.
\begin{eqnarray}
h_t({P}_{B_c},{P}_{c\bar{c}})&=&\frac{m^2_{B_c}-m^2_{c\bar c}}{\sqrt{q^2}}\,F_+(q^2)+\sqrt{q^2}\,
F_-(q^2)\nonumber\\
h_0({P}_{B_c},{P}_{c\bar{c}})&=&\frac{\lambda^{1/2}(q^2,m^2_{B_c},
m^2_{c\bar c})}{\sqrt{q^2}}\, F_+(q^2)\nonumber\\
h_{+1}({P}_{B_c},{P}_{c\bar{c}})&=&h_{-1}({P}_{B_c},{P}_{c\bar{c}})=0
\end{eqnarray}
\item[-] Case $0^-\to 1^-,1^+$.
\begin{eqnarray}
h_{(\lambda)\,t}({P}_{B_c},{P}_{c\bar{c}})&=&i\delta_{\lambda\,0}\,
\frac{\lambda^{1/2}(q^2,m^2_{B_c},m^2_{c\bar c})}{2m_{c\bar c}\sqrt{q^2}}\,
\left((m_{B_c}-m_{c\bar c})\left(A_0(q^2)-A_+(q^2)\right)
-\frac{q^2}{m_{B_c}+m_{c\bar c}}A_-(q^2)\right)\nonumber\\
h_{(\lambda)\,+1}({P}_{B_c},{P}_{c\bar{c}})&=&-i\delta_{\lambda\,-1}\,
\left(\ \frac{\lambda^{1/2}(q^2,m^2_{B_c},m^2_{c\bar c})}{m_{B_c}+m_{c\bar c}}\,
V(q^2)+(m_{B_c}-m_{c\bar c})\,A_0(q^2)\right)
\nonumber\\
h_{(\lambda)\,-1}({P}_{B_c},{P}_{c\bar{c}})&=&-i\delta_{\lambda\,+1}\,
\left(-\frac{\lambda^{1/2}(q^2,m^2_{B_c},m^2_{c\bar c})}{m_{B_c}+m_{c\bar c}}\,
V(q^2)+(m_{B_c}-m_{c\bar c})\,A_0(q^2)\right)\nonumber\\
h_{(\lambda)\,0}({P}_{B_c},{P}_{c\bar{c}})&=&i\delta_{\lambda\,0}\,
\left( (m_{B_c}-m_{c\bar c})\frac{m^2_{B_c}-q^2-m^2_{c\bar c}}{2m_{c\bar c}\sqrt{q^2}}\,
A_0(q^2)-\frac{\lambda(q^2,m^2_{B_c},m^2_{c\bar c})}{2m_{c\bar
c}\sqrt{q^2}}\,\frac{A_+(q^2)}{m_{B_c}+m_{c\bar c}}
\right)
\end{eqnarray}
\item[-] Case $0^-\to 2^-,2^+$.
\begin{eqnarray}
 h_{(\lambda)\,t}({P}_{B_c},{P}_{c\bar{c}})&=&-i\delta_{\lambda\,0}\,\sqrt{\frac{2}{3}}
\ \frac{\lambda(q^2,m^2_{B_c},m^2_{c\bar c})}{4\,m^2_{c\bar c}\,\sqrt{q^2}}\,
\left(\, T_1(q^2)+(m^2_{B_c}-m^2_{c\bar c})\,T_2(q^2)+q^2\,T_3(q^2)\,
\right)\nonumber\\
h_{(\lambda)\,+1}({P}_{B_c},{P}_{c\bar{c}})&=&i\delta_{\lambda\,-1}\,
\frac{1}{\sqrt2}\,
\frac{\lambda^{1/2}(q^2,m^2_{B_c},m^2_{c\bar c})}{2\,m_{c\bar c}}\,
\left(\,T_1(q^2)-\lambda^{1/2}(q^2,m^2_{B_c},m^2_{c\bar c})\,T_4(q^2)\,
\right)
\nonumber\\
h_{(\lambda)\,-1}({P}_{B_c},{P}_{c\bar{c}})&=&i\delta_{\lambda\,+1}\,
\frac{1}{\sqrt2}\,
\frac{\lambda^{1/2}(q^2,m^2_{B_c},m^2_{c\bar c})}{2\,m_{c\bar c}}\,
\left(\,T_1(q^2)+\lambda^{1/2}(q^2,m^2_{B_c},m^2_{c\bar c})\,T_4(q^2)\,
\right)
\nonumber\\
h_{(\lambda)\,0}({P}_{B_c},{P}_{c\bar{c}})&=&
-i\delta_{\lambda\,0}\,\sqrt{\frac{2}{3}}
\ \frac{\lambda^{1/2}(q^2,m^2_{B_c},m^2_{c\bar c})}{4\,m^2_{c\bar c}\,\sqrt{q^2}}\,
\left(\,(m^2_{B_c}-q^2-m^2_{c\bar c})\,T_1(q^2)+\lambda(q^2,m^2_{B_c},m^2_{c\bar c})\,T_2(q^2)\,
\right)\nonumber\\
\end{eqnarray}

\end{itemize}
We see that the helicity amplitudes, and thus the  helicity components of the
hadron tensor,  only depend on $q^2$. The expressions for the latter are
collected in appendix~\ref{app:hcht}.

We can now define the combinations~\cite{ivanov05}
\begin{eqnarray}
H_U&=&{\cal H}_{+1\,+1}+{\cal H}_{-1\,-1}\nonumber\\
H_L&=&{\cal H}_{0\,0}\nonumber\\
H_P&=&{\cal H}_{+1\,+1}-{\cal H}_{-1\,-1}\nonumber\\
H_S&=&3\,{\cal H}_{t\,t}\nonumber\\
H_{SL}&=&{\cal H}_{t\,0}\nonumber\\
\widetilde{H}_J&=&\frac{m_l^2}{2\,q^2}\,H_J\hspace{0.5cm};\hspace{0.5cm}
J=U,L,S,SL
\label{eq:combinaciones}
\end{eqnarray}
with $U,\,L,\,P,\,S,\,SL$ representing respectively unpolarized--transverse,
longitudinal, parity--odd, scalar and scalar--longitudinal
interference.

Finally the double differential decay width is written in terms of the above
defined combinations 
as
\begin{eqnarray}
\frac{d^2\Gamma}{dq^2dx_l}&=&\frac{G_F^2}{8\pi^3}\,|V_{bc}|^2\,
\frac{(q^2-m^2_{l})^2}{12m^2_{B_c}q^2}\,
\frac{\lambda^{1/2}(q^2,m^2_{B_c},m^2_{c\bar{c}})}
{2m_{B_c}}\, 
\bigg\{\
\frac{3}{8}(1+x_l^2)\,H_U+\frac{3}{4}(1-x_l^2)\,H_L\pm\frac{3}{4}x_lH_P
\nonumber\\
&&\hspace{6.5cm}+\frac{3}{4}(1-x_l^2)\widetilde{H}_U+\frac{3}{2}x^2_l
\widetilde{H}_L
+\frac{1}{2}\widetilde{H}_S+3x_l\widetilde{H}_{SL}
\bigg\}
\end{eqnarray}
Note that  for antiparticle decay $H_P$  has the opposite sign to the case of
particle decay while all other hadron tensor helicity components combinations defined in  Eq.(\ref{eq:combinaciones})
do not change (See
appendix~\ref{app:sign} for details). The sign change  of $H_P$ compensates
 the extra sign  coming from the lepton tensor.  This means that in fact the double
differential decay width is the same for $B_c^-$ or $B_c^+$ decay.

Integrating over $x_l$ we obtain the differential decay width
\begin{eqnarray}
\frac{d\Gamma}{dq^2}&=&\frac{G_F^2}{8\pi^3}\,|V_{bc}|^2\,
\frac{(q^2-m^2_{l})^2}{12m^2_{B_c}q^2}\,
\frac{\lambda^{1/2}(q^2,m^2_{B_c},m^2_{c\bar{c}})}
{2m_{B_c}}\, 
\bigg\{\
H_U+H_L+\widetilde{H}_U+\widetilde{H}_L
+\widetilde{H}_S
\bigg\}
\end{eqnarray}
from where, integrating over $q^2$, we obtain the total decay width that we
write, following Ref.~\cite{ivanov05}, as
\begin{eqnarray}
\Gamma&=&\Gamma_U+\Gamma_L+\widetilde{\Gamma}_U+\widetilde{\Gamma}_L
+\widetilde{\Gamma}_S
\end{eqnarray}
with $\Gamma_J$ and $\widetilde{\Gamma}_J$ partial helicity widths defined as
\begin{eqnarray}
\Gamma_J=\int dq^2 \frac{G_F^2}{8\pi^3}\,|V_{bc}|^2\,
\frac{(q^2-m^2_{l})^2}{12m^2_{B_c}q^2}\,
\frac{\lambda^{1/2}(q^2,m^2_{B_c},m^2_{c\bar{c}})}
{2m_{B_c}}\, H_J
\end{eqnarray}
and similarly for $\widetilde{\Gamma}_J$ in terms of
$\widetilde{H}_J$. 

\begin{table}[t]
\begin{center}
\begin{tabular}{l|c c c c c c c}
$B^-_c\to$ & \hspace{.2cm}$\Gamma_U$\hspace{.2cm} & \hspace{.2cm}$\widetilde{\Gamma}_U$\hspace{.2cm}
 &\hspace{.2cm} $\Gamma_L$\hspace{.2cm} & $\widetilde{\Gamma}_L$ & \hspace{.2cm}$\Gamma_P$
 \hspace{.2cm}& $\widetilde{\Gamma}_S $ & $\widetilde{\Gamma}_{SL}$\\ 
\hline
&&&&&&&\\
$\eta_c\,e^-\bar{\nu}_e$& 0 & 0 &$ 6.95^{+0.31}$&
$0.13^{+0.01}\ 10^{-5}$ & 0 &
 $ 0.44^{+0.03}\ 10^{-5}$  & $ 0.14^{+0.01}\ 10^{-5}$\\
$\eta_c\,\mu^-\bar{\nu}_\mu$& 0 & 0 &$ 6.80^{+0.31}$&
$0.28^{+0.02}\ 10^{-1}$ & 0 &
 $ 0.10^{+0.01}$  & $ 0.31^{+0.01}\ 10^{-1}$\\
$\eta_c\,\tau^-\bar{\nu}_\tau$
 & 0& 0&$0.71^{+0.02}$ &$0.17^{+0.01}$ &0 &$1.58^{+0.04}$
 &$0.29^{+0.01}$ \\
&&&&&&&\\

$\chi_{c0}\,e^-\bar{\nu}_e$& 0 & 0 &$ 1.55^{+0.14}_{-0.02}$&
$0.37^{+0.05}\ 10^{-6}$ & 0 &
 $ 0.11^{+0.01}\ 10^{-5}$  & $ 0.37^{+0.05}\ 10^{-6}$\\
$\chi_{c0}\,\mu^-\bar{\nu}_\mu$& 0 & 0 &$ 1.51^{+0.13}_{-0.02}$&
$0.75^{+0.09}\ 10^{-2}$ & 0 &
 $ 0.23^{+0.02}\ 10^{-1}$  & $ 0.75^{+0.09}\ 10^{-2}$\\
$ \chi_{c0}\,\tau^-\bar{\nu}_\tau$
 &0&0&$0.80^{+0.04}_{-0.02}\ 10^{-1}$
&$0.23^{+0.01}_{-0.01}\ 10^{-1}$
 &0 &
$0.84^{+0.07}\ 10^{-1} $ &$0.25^{+0.02}\ 10^{-1}$ \\
&&&&&&&\\

$J/\Psi\,e^-\bar{\nu}_e$
 & $11.5^{+0.6}$&$0.32^{+0.02}\ 10^{-6}$&$10.4^{+0.6}$ &$0.12^{+0.01}\ 10^{-5}$
 &$-5.48_{-0.24}$ &
$0.32^{+0.03}\ 10^{-5}$ &$0.11^{+0.01}\ 10^{-5}$ \\
$J/\Psi\,\mu^-\bar{\nu}_\mu$
 & $11.4^{+0.6}$&$0.13^{+0.01}\ 10^{-1}$&$10.2^{+0.7}$ &$0.28^{+0.03}\ 10^{-1}$
 &$-5.45_{-0.24}$ &
$0.68^{+0.07}\ 10^{-1}$ &$0.25^{+0.02}\ 10^{-1}$ \\
$J/\Psi\,\tau^-\bar{\nu}_\tau$
 & $2.78^{+0.10}_{-0.01}$&$0.59^{+0.02}$&$1.74^{+0.07}_{-0.01}$ &
 $0.39^{+0.02}$
 &$-1.10_{-0.03}$ &
$0.36^{+0.02}$ &$0.21^{+0.01}$ \\
&&&&&&&\\

$\chi_{c1}\,e^-\bar{\nu}_e$
 &$0.90^{+0.05}_{-0.03}$&$0.43^{+0.03}_{-0.01}\ 10^{-7}$
&$0.35^{+0.03}\ 10^{-1}$&$0.28^{+0.03}\
 10^{-8}$&$-0.75_{-0.04}^{+0.02}$
&$0.57^{+0.07}\ 10^{-8}$&$0.22^{+0.02}\ 10^{-8}$\\
$\chi_{c1}\,\mu^-\bar{\nu}_\mu$
 &$0.89^{+0.05}_{-0.03}$&$0.18^{+0.01}_{-0.01}\ 10^{-2}$
&$0.35^{+0.03}\ 10^{-1}$&$0.77^{+0.08}\
 10^{-4}$&$-0.75_{-0.04}^{+0.02}$
&$0.11^{+0.01}\ 10^{-3}$&$0.49^{+0.05}\ 10^{-4}$\\

$\chi_{c1}\,\tau^-\bar{\nu}_\tau$
 &$0.75^{+0.02}\ 10^{-1}$&$0.21^{+0.01}\ 10^{-1}$&$0.46^{+0.04}\ 10^{-2}$&
 $0.12^{+0.01}\ 10^{-2}$&
$-0.64_{-0.03}\ 10^{-1}$&$0.23^{+0.02}\ 10^{-3}$&$0.28^{+0.02}\ 10^{-3}$\\
&&&&&&&\\

$h_c\,e^-\bar{\nu}_e$ 
&$0.16^{+0.02}$&$0.57^{+0.07}\ 10^{-8}$&$2.23^{+0.12}$&
$0.72^{+0.09}\ 10^{-6}$&$-0.26_{-0.03}\ 10^{-1}$&
$0.23^{+0.03}\ 10^{-5}$&$0.74^{+0.09}\ 10^{-6}$\\
$h_c\,\mu^-\bar{\nu}_\mu$ 
&$0.16^{+0.02}$&$0.24^{+0.03}\ 10^{-3}$&$2.16^{+0.21}_{-0.01}$&
$0.14^{+0.01}\ 10^{-1}$&$-0.26_{-0.03}\ 10^{-1}$&
$0.45^{+0.05}_{-0.01}\ 10^{-1}$&$0.14^{+0.02}\ 10^{-1}$\\

$h_c \,\tau^-\bar{\nu}_\tau$ 
& $0.23^{+0.02} \ 10^{-1}$ & $0.60^{+0.06}\ 10^{-2}$ &
$0.67^{+0.05}\ 10^{-1}$&
$0.20^{+0.01}\ 10^{-1}$&
$-0.27_{-0.03}\ 10^{-2}$ &$0.97^{+0.05}_{-0.03}\ 10^{-1}$&
$0.26_{-0.01}^{+0.02}\ 10^{-1}$\\
& & & & & & &\\

$\chi_{c2}\,e^-\bar{\nu}_e$ 
 &$0.71^{+0.03}_{-0.03}$&$0.37^{+0.02}_{-0.02}\ 10^{-7}$
&$1.17^{+0.08}_{-0.05}$& $0.31^{+0.03}_{-0.01}\ 10^{-6}$
&$-0.35_{-0.02}^{+0.01}$&$0.88^{+0.09}_{-0.03}\ 10^{-6}$
&$0.30^{+0.03}_{-0.01}\ 10^{-6}$\\
$\chi_{c2}\,\mu^-\bar{\nu}_\mu$ 
 &$0.71^{+0.02}_{-0.03}$&$0.15^{+0.01}\ 10^{-2}$
&$1.14^{+0.07}_{-0.05}$& $0.62^{+0.06}_{-0.02}\ 10^{-2}$
&$-0.34_{-0.02}^{+0.01}$&$0.16^{+0.02}\ 10^{-1}$
&$0.57^{+0.06}_{-0.02}\ 10^{-2}$\\
$\chi_{c2}\,\tau^-\bar{\nu}_\tau$ 
 &$0.49^{+0.01}_{-0.03}\ 10^{-1}$ &$0.15_{-0.01}\ 10^{-1}$
 &$0.43^{+0.01}_{-0.02}\ 10^{-1}$ &$0.13_{-0.01}\ 10^{-1}$
 &$-0.18^{+0.01}\ 10^{-1}$ &$0.12_{-0.01}\ 10^{-1}$
 &$0.70^{+0.02}_{-0.04}\ 10^{-2}$\\
& & & & & & &\\

$\Psi(3836)\,e^-\bar{\nu}_e$ 
&$0.58_{-0.07} \ 10^{-1}$&$0.47_{-0.06}\ 10^{-8}$&$0.33^{+0.01}_{-0.02}\ 10^{-2}$
&$0.68^{+0.04}_{-0.04}\ 10^{-9}$&$-0.48^{+0.05}\ 10^{-1}$
&$0.17^{+0.01}_{-0.01}\ 10^{-8}$&$0.60^{+0.03}_{-0.04}\ 10^{-9}$\\
$\Psi(3836)\,\mu^-\bar{\nu}_\mu$ 
&$0.57_{-0.06} \ 10^{-1}$&$0.19_{-0.02}\ 10^{-3}$&$0.32^{+0.01}_{-0.02}\ 10^{-2}$
&$0.15_{-0.01}\ 10^{-4}$&$-0.48^{+0.06}\ 10^{-1}$
&$0.28^{+0.02}_{-0.02}\ 10^{-4}$&$0.11^{+0.01}\ 10^{-4}$\\
$\Psi(3836) \,\tau^-\bar{\nu}_\tau$
&$0.78_{-0.09}\ 10^{-3}$& $0.28_{-0.03}\ 10^{-3}$&$0.74_{-0.06}\ 10^{-4}$
&$0.25_{-0.02}\ 10^{-4}$& $-0.69^{+0.08}\ 10^{-3}$&$0.54_{-0.04}\ 10^{-5}$
&$0.65_{-0.05}\ 10^{-5}$\\
\end{tabular}
\end{center}
\caption{ Partial helicity widths in units of $10^{-15}$\,GeV for $B_c^-$ decays. Central values
have been evaluated with the AL1 potential. } 
\label{tab:ha}
\end{table}

Another quantity of interest is the forward-backward asymmetry of the charged
lepton measured in the leptons CMF. This asymmetry is defined as\footnote{The
forward direction is determined by the momentum of the final meson that we have
chosen in the negative $z$-direction.  }

\begin{eqnarray}
A_{FB}=\frac{\Gamma_{x_l>0}-\Gamma_{x_l<0}}{\Gamma_{x_l>0}+\Gamma_{x_l<0}}
\end{eqnarray}
and it is given in terms or partial helicity widths as
\begin{eqnarray}
=\frac{3}{4}\,\frac{\pm\Gamma_P+4\,\widetilde{\Gamma}_{SL}}
{\Gamma_U+\Gamma_L+\widetilde{\Gamma}_U+\widetilde{\Gamma}_L
+\widetilde{\Gamma}_S}
\end{eqnarray}
being the same for a negative charged lepton $l^-$ ($B_c^-$ decay) than for a
positive charged one $l^+$ ($B_c^+$ decay), as $\Gamma_P$ for antiparticle
decay has the opposite sign as for particle decay.

Finally for the decay channel $B_c\to J/\Psi\ l\bar{\nu}_l$ with the 
$J/\Psi$ decaying into $\mu^-\mu^+$
we can
evaluate the differential cross section
\begin{eqnarray}
\frac{d\Gamma_{B_c\to\mu^-\mu^+ (J/\Psi)\,l\bar{\nu}_l}}{dx_\mu}&=&\left(
1+\frac{\sqrt{1-4m_\mu^2/m_{J/\Psi}^2}\,\bigg(
\Gamma_U+\widetilde{\Gamma}_U-2(\Gamma_L+\widetilde{\Gamma}_L
+\widetilde{\Gamma}_S)\bigg)}{
\left(2-\sqrt{1-4m_\mu^2/m_{J/\Psi}^2}\,
\right)\,\left(\Gamma_U+\widetilde{\Gamma}_U\right)
+2\left(\Gamma_L+\widetilde{\Gamma}_L
+\widetilde{\Gamma}_S\right)
}\ x_\mu^2\right)\nonumber\\
&&\times\frac{\Gamma_{J/\Psi\to\mu^-\mu^+}}{\Gamma_{J/\Psi}}\,
\frac{1}{1+2m_\mu^2/m_{J/\Psi}^2}
\bigg[\ \frac{3}{4}\,\left(\Gamma_L+\widetilde{\Gamma}_L
+\widetilde{\Gamma}_S\right)\nonumber\\
&&\hspace{4.75cm}+\frac{3}{8}\,
\left(2-\sqrt{1-4m_\mu^2/m_{J/\Psi}^2}\ \right)
\left(\Gamma_U+\widetilde{\Gamma}_U\right)
\bigg]
\end{eqnarray}
where $x_\mu$ is the cosine of the polar angle for the final
$\mu^-\mu^+$ pair, relative to the momentum of the decaying $J/\Psi$,
measured in the $\mu^-\mu^+$ CMF, $\Gamma_{J/\Psi\to\mu^-\mu^+}$ is
the $J/\Psi$ decay width into the $\mu^-\mu^+$ channel, and
$\Gamma_{J/\Psi}$ is the total $J/\Psi$ decay width. The asymmetry
parameter
\begin{eqnarray}
\alpha^*=\frac{\sqrt{1-4m_\mu^2/m_{J/\Psi}^2}\,\bigg(
\Gamma_U+\widetilde{\Gamma}_U-2(\Gamma_L+\widetilde{\Gamma}_L
+\widetilde{\Gamma}_S)\bigg)}{
\left(2-\sqrt{1-4m_\mu^2/m_{J/\Psi}^2}\ 
\right)\,\left(\Gamma_U+\widetilde{\Gamma}_U\right)
+2\left(\Gamma_L+\widetilde{\Gamma}_L
+\widetilde{\Gamma}_S\right)
}\
\end{eqnarray}
governs the muons angular distribution in their CMF.

\subsubsection{Results}

\begin{table}[t]
\begin{tabular}{lc|c c c c c c}
&&$A_{FB} (e)$
&$A_{FB} (\mu)$&$\ \ A_{FB} (\tau)\ \ $& $\  \ \alpha^* (e)$&
 $\  \ \alpha^* (\mu)$&$\  \
\alpha^* (\tau)$\\
\hline
&&&\\
$B_c\to \eta_c$&\\
 & This work &$0.60^{+0.01}\ 10^{-6}$&$0.13^{+0.01}\ 10^{-1}$ &$0.35$&&&\\
&\cite{ivanov05}& $0.953\ 10^{-6}$&&0.36\\
$B_c\to \chi_{c0}$ &\\
& This work&$0.72^{+0.02}\ 10^{-6}$&$0.15\ 10^{-1}$&$0.40$&&&\\
&\cite{ivanov05}& $1.31\ 10^{-6}$&&0.39\\
$B_c\to J/\Psi$ &\\
& This work&$-0.19$&$-0.18_{-0.01}$ &$-0.35^{+0.02}\ 10^{-1}$&$-0.29_{-0.01}$
&
$-0.29$&
$-0.19$\\

&\cite{ivanov05}&$-0.21$&&$-0.48\ 10^{-1}$&$-0.34$&&$-0.24$\\
$B_c\to \chi_{c1}$ &\\
&This work &$-0.60_{-0.01}$&$-0.60_{-0.01}$&$-0.46$&&&\\
&\cite{ivanov05}&0.19&&0.34\\
$B_c\to h_c$ &\\
&This work&$-0.83_{-0.05} \ 10^{-2}$&$0.97_{-0.05}^{+0.01}\ 10^{-2}$&
$0.35$&&&\\
&\cite{ivanov05}&$-3.6\ 10^{-2}$&&0.31\\

$B_c\to \chi_{c2}$ &\\
& This work&$-0.14$&$-0.13$&$0.55^{+0.02}\ 10^{-1}$&&&\\
&\cite{ivanov05}&$-0.16$&&$0.44\ 10 ^{-1}$&\\
$B_c\to \Psi (3836)$&\\
&This work &$-0.59$&$-0.59$&$-0.42$&&&\\
&\cite{ivanov05}&0.21&&0.41\\

\end{tabular}
\caption{Asymmetry parameters in semileptonic $B_c\to c\bar c$ decays. Our central values have been evaluated with the
AL1 potential. We also show the results obtained by Ivanov {\it et
al.}~\cite{ivanov05}. } 
\label{tab:asy}
\end{table}

\begin{figure}
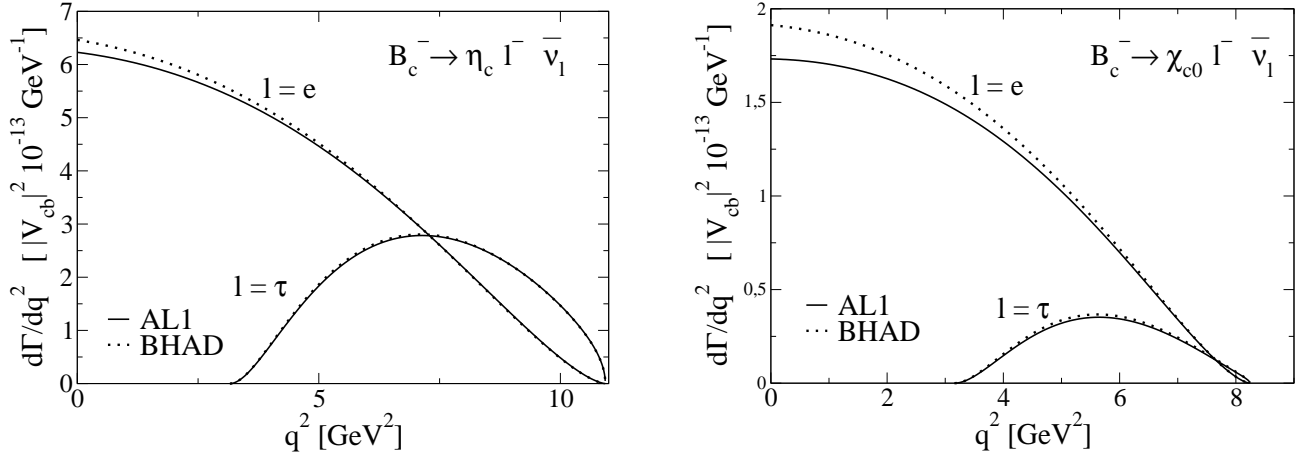

\vspace{1cm}
\centering
\resizebox{8.cm}{6.cm}{\includegraphics{etacdgdq2.eps}}\hspace{1cm}
\resizebox{8.cm}{6.cm}{\includegraphics{chic0dgdq2.eps}}
\caption{Differential decay width for the $B_c^-\to \eta_c
l^-\bar{\nu}_l$ and $B_c^-\to \chi_{c0} l^-\bar{\nu}_l$ processes
obtained with the AL1 potential.  For comparison, we also show with
dotted lines the results obtained with the Bhaduri (BHAD) potential.
 The distribution for a final muon is not explicitly shown. It
differs appreciably from the corresponding to a final electron only
for $q^2$ around $m_\mu^2$.}
\label{fig:0pmdgdq2}
\end{figure}

In Table~\ref{tab:ha} we give our results for the partial helicity
widths corresponding to $B_c^-$ decays. For $B_c^+$ decay the ``$P$''
column changes sign while all others remain the same. The central values have been evaluated with the AL1
potential and the theoretical errors quoted reflect the dependence of the
results on the inter-quark potential.

In Table~\ref{tab:asy} we show the asymmetry parameters. Our values
for $\alpha^*$ compare well with the results obtained in
Ref.~\cite{ivanov05}. The same is true for the forward--backward asymmetry
with some exceptions: most notably we get opposite signs for $B_c\to\chi_{c1}$ and
$B_c\to\Psi(3836)$.

In Fig.~\ref{fig:0pmdgdq2} we show the differential decay width
$d\Gamma/dq^2$ for the decay channels $\eta_c l^-\bar{\nu}_l$ and
$\chi_{c0} l^-\bar{\nu}_l$ for the case where the final lepton is a
light one $l=e,\mu$\footnote{We show the distribution corresponding to
a final electron. The distribution for a final muon differs from the
former only for $q^2$ around $m_\mu^2$} or a heavy one $l=\tau$. We show
the results obtained with the AL1 and BHAD potentials, finding no
significant difference for the $\tau$ case, while for the light final
lepton case the differences are around 10\% at low $q^2$.

\begin{figure}[h!!!]
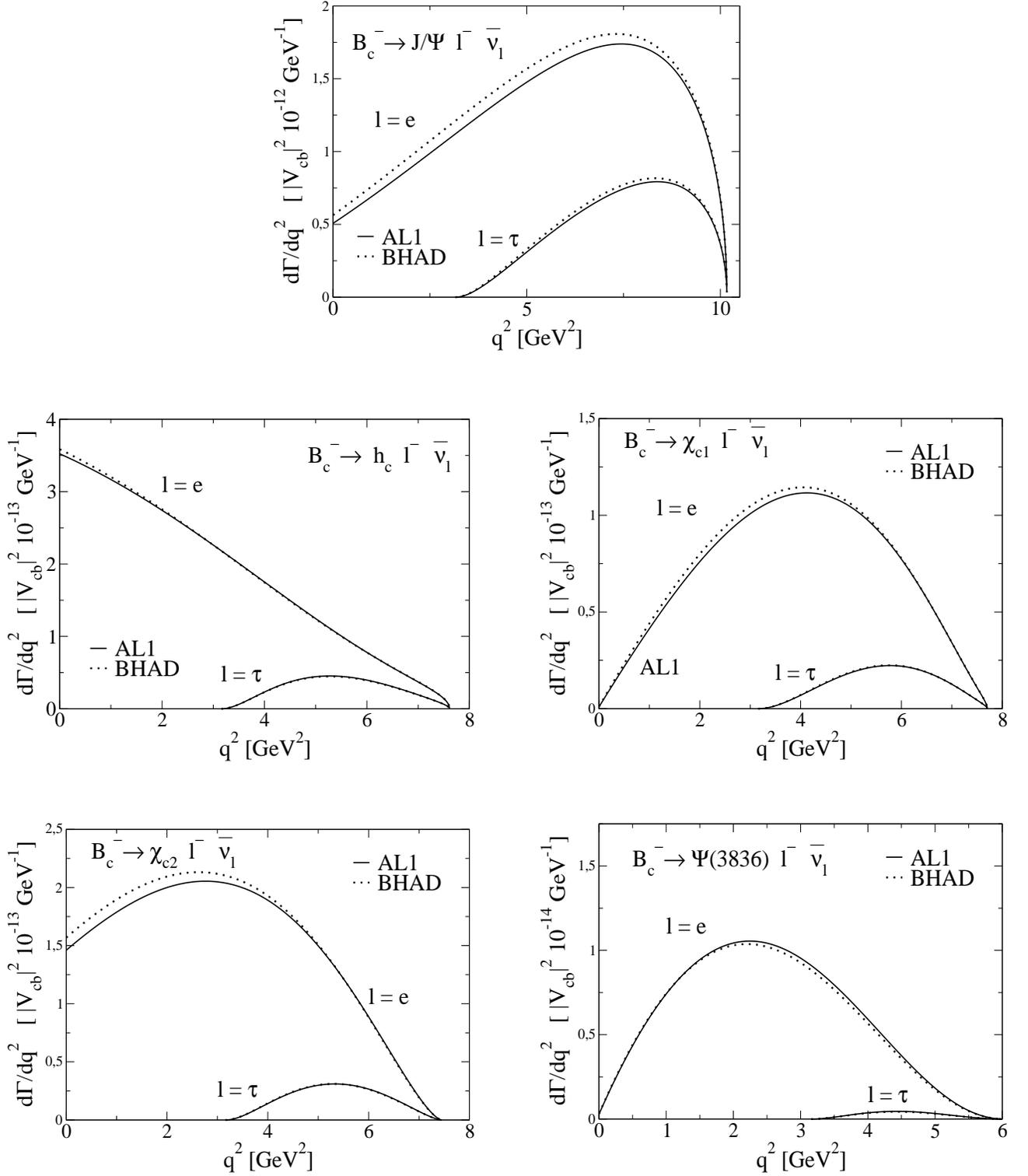

\centering
\resizebox{8.cm}{6.cm}{\includegraphics{jpsidgdq2.eps}}\vspace{1cm}\\
\resizebox{8.cm}{6.cm}{\includegraphics{hcdgdq2.eps}}\hspace{1cm}
\resizebox{8.cm}{6.cm}{\includegraphics{chic1dgdq2.eps}}\vspace{1cm}\\
\resizebox{8.cm}{6.cm}{\includegraphics{chic2dgdq2.eps}}\hspace{1cm}
\resizebox{8.cm}{6.cm}{\includegraphics{psi3836dgdq2.eps}}
\caption{Differential decay width for the $B_c^-\to J/\Psi
  l^-\bar{\nu}_l$,
 $B_c^-\to h_{c} l^-\bar{\nu}_l$,
$B_c^-\to \chi_{c1} l^-\bar{\nu}_l$,
$B_c^-\to \chi_{c2} l^-\bar{\nu}_l$  and
$B_c^-\to \Psi (3836) l^-\bar{\nu}_l$
 decay channels obtained with
 the AL1 potential (solid lines) and  the Bhaduri (BHAD)
  potential (dotted lines)  . The distribution
for a final muon is not explicitly shown. It differs appreciably from the corresponding to a
final electron only for $q^2$ around $m_\mu^2$.}
\label{fig:12pmdgdq2}
\end{figure}

In Fig.~\ref{fig:12pmdgdq2} we show now the
results for vector and tensor mesons. As before only for the case
where the final lepton is light we see  up to 10\% differences 
  between the calculation with the AL1 and the BHAD potentials.

Finally in Tables~\ref{tab:dw},~\ref{tab:br} we give the total decay
widths and corresponding branching ratios for the different
transitions. The branching ratios evaluated by Ivanov {\it et
al.}~\cite{ivanov06}, where they have used the new $B_c$ mass determination by the CDF 
Collaboration~\cite{cdf06-1}, are in reasonable agreement with our results. Discrepancies are
larger for the decay widths in Table~\ref{tab:dw} where they use the larger 
mass value $m_{B_c}=6400$\,MeV quoted by the PDG~\cite{pdg04}.
\begin{table}[t]
\hspace*{-1.5cm}\begin{tabular}{l| c c c c c c c c c c}
\multicolumn{1}{l}{}&\multicolumn{10}{c}
{$\Gamma\ [10^{-15}\ \mathrm{GeV}]$}\\
&&&&\\
&This work&\cite{ivanov05}&\cite{ebert03}&\cite{chang94,chang02}&
\cite{hady00}
&\cite{liu97}&\cite{kiselev00}&\cite{colangelo00}&
\cite{anisimov99}&\cite{sanchis95}\\
\hline
&&&&\\
$B^-_c\to \eta_c\, l^-\,\bar{\nu}_l$ &$6.95^{+0.29}$ 
& 10.7 &5.9&14.2&11.1&8.31&$11\pm 1$& 2.1 (6.9)&8.6&10\\
$B^-_c\to \eta_c\, \tau^-\,\bar{\nu}_\tau$ &$2.46^{+0.07}$ 
&3.52&&&&&&&$3.3\pm0.9$&\\
&&&&\\
$B^-_c\to \chi_{c0}\, l^-\,\bar{\nu}_l$ &$1.55_{-0.02}^{+0.14}$
&2.52&&1.69&&&&&&\\
$B^-_c\to \chi_{c0}\,\tau^-\,\bar{\nu}_\tau$ &$0.19^{+0.01}$
& 0.26&&0.25&&&&&&\\
&&&&\\

$B_c^-\to J/\Psi\, l^-\,\bar{\nu}_l$ &$21.9^{+1.2}$
&28.2&17.7& 34.4&30.2&20.3&$28\pm 5$& 21.6 (48.3)&17.2&42\\
$B_c^-\to J/\Psi\,\tau^-\,\bar{\nu}_\tau$&$5.86^{+0.23}_{-0.03}$
&7.82&&&&&&&$7\pm 2$&\\
&&&&\\

$B_c^-\to \chi_{c1}\, l^-\,\bar{\nu}_l$ &$0.94^{+0.05}_{-0.03}$
&1.40&&2.21&&&&&&\\
$B_c^-\to \chi_{c1}\, \tau^-\,\bar{\nu}_\tau$ &$0.10$
&0.17&&0.35&&&&&&\\
&&&&\\

$B_c^-\to h_c\, l^-\,\bar{\nu}_l$ &$2.40^{+0.23}_{-0.01}$
&4.42&&2.51&&&&&&\\
$B_c^-\to h_c\, \tau^-\,\bar{\nu}_\tau$ &$0.21^{+0.01}$
&0.38&&0.36&&&&&&\\
&&&&\\

$B_c^-\to \chi_{c2}\, l^-\,\bar{\nu}_l$&$1.89^{+0.11}_{-0.08}$
& 2.92&&2.73&&&&&\\
$B_c^-\to \chi_{c2}\, \tau^-\,\bar{\nu}_\tau$&$0.13_{-0.01}^{+0.01}$
& 0.20&&0.42&&&&&\\
&&&&\\

$B_c^-\to \Psi (3836)\, l^-\,\bar{\nu}_l$&$0.062_{-0.008}$ 
& 0.13&&&&&\\
$B_c^-\to \Psi (3836)\, \tau^-\,\bar{\nu}_\tau$ &$0.0012_{-0.0002}$
&0.0031&&&&&\\

\end{tabular}
\caption{Decay widths in units of $10^{-15}$\,GeV for semileptonic $B_c^-\to c\bar c$ decays. Our central values have been evaluated with the
AL1 potential. Here $l$ stands for $l=e,\,\mu$.} 
\label{tab:dw}
\end{table}

\begin{table}[h!!!!!]
\hspace*{-1.5cm}\begin{tabular}{l| c c c c c c c c c}
\multicolumn{1}{l}{}&\multicolumn{6}{c}
{B.R. (\%)}\\
&&&&\\
&This work&\cite{ivanov06}&\cite{ebert03}&\cite{chang94,chang02,chang01}&
\cite{hady00}&\cite{kiselev00-2,kiselev02}&\cite{colangelo00}&
\cite{anisimov99}&\cite{nobes00}\\
\hline
&&&&\\
$B^-_c\to \eta_c\, l^-\,\bar{\nu}_l$ &$0.48^{+0.02}$ 
& 0.81 &0.42&0.97&0.76&0.75&0.15&0.59&0.51\\
$B^-_c\to \eta_c\, \tau^-\,\bar{\nu}_\tau$ &$0.17^{+0.01}$ &0.22&& &&0.23&&0.20&\\
&&&&\\
$B^-_c\to \chi_{c0}\, l^-\,\bar{\nu}_l$ &$0.11^{+0.01}$&0.17&&0.12&&&
&&\\
$B^-_c\to \chi_{c0}\,\tau^-\,\bar{\nu}_\tau$ &$0.013^{+0.001}$&
 0.013&&0.017&&&&&\\
&&&&\\

$B_c^-\to J/\Psi\, l^-\,\bar{\nu}_l$ &$1.54^{+0.06}$&2.07&1.23& 2.35 &2.01&1.9&1.47
&1.20&1.44\\
$B_c^-\to J/\Psi\,\tau^-\,\bar{\nu}_\tau$&$0.41^{+0.02}$&0.49&&&&0.48&&0.34&
\\
&&&&\\

$B_c^-\to \chi_{c1}\, l^-\,\bar{\nu}_l$ &$0.066^{+0.003}_{-0.002}$&0.092&&0.15&
&&&&\\
$B_c^-\to \chi_{c1}\, \tau^-\,\bar{\nu}_\tau$
&$0.0072^{+0.0002}_{-0.0003}$&0.0089&&0.024&&&&&\\
&&&&\\

$B_c^-\to h_c\, l^-\,\bar{\nu}_l$ &$0.17^{+0.02}$&0.27&&0.17&&&&&\\
$B_c^-\to h_c\, \tau^-\,\bar{\nu}_\tau$ &$0.015^{+0.001}$&0.017&&0.024
&&&&&\\
&&&&\\

$B_c^-\to \chi_{c2}\, l^-\,\bar{\nu}_l$&$0.13^{+0.01}$& 0.17&&0.19&&&&&\\
$B_c^-\to \chi_{c2}\, \tau^-\,\bar{\nu}_\tau$&$0.0093^{+0.0002}_{-0.0005}$&
 0.0082&&0.029&&&&&\\
&&&&\\

$B_c^-\to \Psi (3836)\, l^-\,\bar{\nu}_l$&$0.0043_{-0.0005}$ & 0.0066&&&&&&\\
$B_c^-\to \Psi (3836)\, \tau^-\,\bar{\nu}_\tau$ &$0.000083_{-0.000010}$ 
&0.000099&&&&&&\\

\end{tabular}
\caption{Branching ratios in \% for semileptonic $B_c^-\to c\bar c$ decays. Our central values have been evaluated with the
AL1 potential. Here $l$ stands for $l=e,\,\mu$.} 
\label{tab:br}
\end{table}

\subsection{Heavy quark spin symmetry}
As mentioned in the Introduction one can not apply HQS to systems with
two heavy quarks due to flavor symmetry breaking by the kinetic energy
terms. The symmetry that survives for such systems is HQSS amounting
to the decoupling of the two heavy quark spins. Using HQSS Jenkins
{\it et al.}~\cite{jenkins93} obtained relations between different
form factors for semileptonic $B_c$ decays into ground state vector
and pseudoscalar mesons. Let us check the agreement of our
calculations with their results. For that purpose let us re-write the
general form factor decompositions in Eq.~(\ref{eq:ff}) introducing
the four vectors $v$ and $k$ such that
\begin{eqnarray}
P_{B_c}=m_{B_c}\,v\hspace{.5cm};\hspace{.5cm}P_{c\bar c}=m_{c\bar c}\,v+k
\end{eqnarray}
$v$ is the four-velocity of the initial $B_c$ meson whereas $k$ is a residual
momentum. In terms of those we have
\begin{eqnarray}
P=P_{B_c}+P_{c\bar c}=(m_{B_c}+m_{c\bar c})\,v+k\hspace{.5cm};
\hspace{.5cm}q=P_{B_c}-P_{c\bar c}=(m_{B_c}-m_{c\bar c})\,v-k
\end{eqnarray}
and we can write for the $\eta_c\, (c\bar c\,(0^-))$ final state case

\begin{eqnarray}
\left\langle\, \eta_c,\,\vec{P}_{\eta_c}\,\left|\, J^{c\,b}_\mu(0)\,
\right| \, B_c^-,\,\vec{P}_{B_c}\right\rangle&=&
\left\langle\,  \eta_c,\,\vec{P}_{\eta_c}\,\left|\hspace{.3cm}J^{c\,b}_{V\,\mu}(0)\,
\right| \, B_c^-,\,\vec{P}_{B_c}\right\rangle\nonumber\\
&=&\left((m_{B_c}+m_{\eta_c})\,F_+(q^2)
+(m_{B_c}-m_{\eta_c})\,F_-(q^2)\right)\,v_\mu
+(F_+(q^2)-F_-(q^2))\,k_\mu
\nonumber\\
&=&\sqrt{2m_{B_c}2m_{\eta_c}}\left(\Sigma_1^{(0^-)}(q^2)\, v_\mu
+\overline{\Sigma}_2^{(0^-)}(q^2)\,k_\mu\right)
\end{eqnarray}
where we have introduced the new form factors
\begin{eqnarray}
\Sigma_1^{(0^-)}(q^2)&=&\frac{1}{\sqrt{2m_{B_c}2m_{\eta_c}}}\,
\left((m_{B_c}+m_{\eta_c})\,F_+(q^2)
+(m_{B_c}-m_{\eta_c})\,F_-(q^2)\right)\nonumber\\
\overline{\Sigma}_2^{(0^-)}(q^2)&=&\frac{1}{\sqrt{2m_{B_c}2m_{\eta_c}}}\,
(F_+(q^2)-F_-(q^2))
\end{eqnarray}
Similarly for the  $J/\Psi\,(c\bar c\,(1^-))$ final state case we have
\begin{eqnarray}
\left\langle\,J/\Psi ,\,\lambda\,\vec{P}_{J/\Psi}\,\left|\, J^{c\,b}_\mu(0)\,
\right| \, B_c^-,\,\vec{P}_{B_c}\right\rangle&=&
\left\langle\, J/\Psi,\,\lambda\,\vec{P}_{J/\Psi}\,\left|\, J^{c\,b}_{V\mu}(0)-
J^{c\,b}_{A\mu}(0)\,
\right| \, B_c^-,\,\vec{P}_{B_c}\right\rangle\nonumber\\
&=&\frac{2m_{B_c}}{m_{B_c}+m_{J/\Psi}} \varepsilon_{\mu\nu\alpha\beta}\ \varepsilon^{\,
\nu\,*}_{(\lambda)}(\,\vec{P
}_{J/\Psi}\,)\,v^\alpha\,k^{\,
\beta}\,V(q^2)\nonumber\\
&&-\, i\, \bigg\{\ (m_{B_c}-m_{J/\Psi})\ \varepsilon_{(\lambda)\,\mu}^*
(\,\vec{P}_{J/\Psi}\,)\,A_0(q^2)\nonumber\\
&&\hspace{+.8cm}
+\frac{m_{B_c}}{m_{J/\Psi}}{k\cdot\varepsilon^*_{(\lambda)}
(\,\vec{P}_{J/\Psi}\,)}
\bigg(\ \frac{A_+(q^2)}{m_{B_c}+m_{J/\Psi}}\left(
(m_{B_c}+m_{J/\Psi})\,v_\mu+k_\mu\right)\nonumber\\
&&\hspace{4cm} +\frac{A_-(q^2)}{m_{B_c}+m_{J/\Psi}}\left(
(m_{B_c}-m_{J/\Psi})\,v_\mu-k_\mu\right) \bigg)\bigg\}\nonumber\\
&=&-\sqrt{2m_{B_c}2m_{J/\Psi}}\,
\varepsilon_{\mu\nu\alpha\beta}\ \varepsilon^{\,
\nu\,*}_{(\lambda)}(\,\vec{P
}_{J/\Psi}\,)\,v^\alpha\,k^{\,\beta}\,\overline{\Sigma}'^{(1^-)}_2(q^2)\nonumber\\
&&-i\sqrt{2m_{B_c}2m_{J/\Psi}}\bigg\{\
\varepsilon^*_{(\lambda)\,\mu}
(\,\vec{P}_{J/\Psi}\,)\,\Sigma_1^{(1^-)}(q^2)\nonumber\\
&&\hspace{2.55cm}+\left(k\cdot\varepsilon^*_{(\lambda)}
(\,\vec{P}_{J/\Psi}\,)\right)\,\overline{\Sigma}_2^{(1^-)}(q^2)\,v_\mu\nonumber\\
&&\hspace{2.55cm}+\left(k\cdot\varepsilon^*_{(\lambda)}
(\,\vec{P}_{J/\Psi}\,)\right)\,
\overline{\Sigma}_3^{(1^-)}(q^2)\,k_\mu
\bigg\}
\end{eqnarray}
with
\begin{eqnarray}
\Sigma_1^{(1^-)}(q^2)&=&\frac{1}{\sqrt{2m_{B_c}2m_{J/\Psi}}}\,
\left(m_{B_c}-m_{J/\Psi}\right)\,A_0(q^2)
\nonumber\\
\overline{\Sigma}_2^{(1^-)}(q^2)&=&\frac{1}{\sqrt{2m_{B_c}2m_{J/\Psi}}}\,
\frac{m_{B_c}}{m_{J/\Psi}}\,\left(\,A_+(q^2)
+\frac{m_{B_c}-m_{J/\Psi}}{m_{B_c}+m_{J/\Psi}}\,A_-(q^2)
\right)\nonumber\\
\overline{\Sigma}_2'^{(1^-)}(q^2)&=&\frac{1}{\sqrt{2m_{B_c}2m_{J/\Psi}}}\,
\frac{-2m_{B_c}}{m_{B_c}+m_{J/\Psi}}\,V(q^2)
\nonumber\\
\overline{\Sigma}_3^{(1^-)}(q^2)&=&\frac{1}{\sqrt{2m_{B_c}2m_{J/\Psi}}}\,
\frac{m_{B_c}}{m_{J/\Psi}}\,
\frac{1}{m_{B_c}+m_{J/\Psi}}\left(A_+(q^2)-A_-(q^2)\right)
\end{eqnarray}
$\Sigma_1^{(0^-)}$ and $\Sigma_1^{(1^-)}$ are dimensionless,
 $\overline{\Sigma}_2^{(0^-)}$, $\overline{\Sigma}_2^{(1^-)}$, 
$\overline{\Sigma}_2'^{(1^-)}$ 
have dimensions of $E^{-1}$, and $\overline{\Sigma}_3^{(1^-)}$ 
has dimensions of $E^{-2}$.

\begin{figure}[h!!]
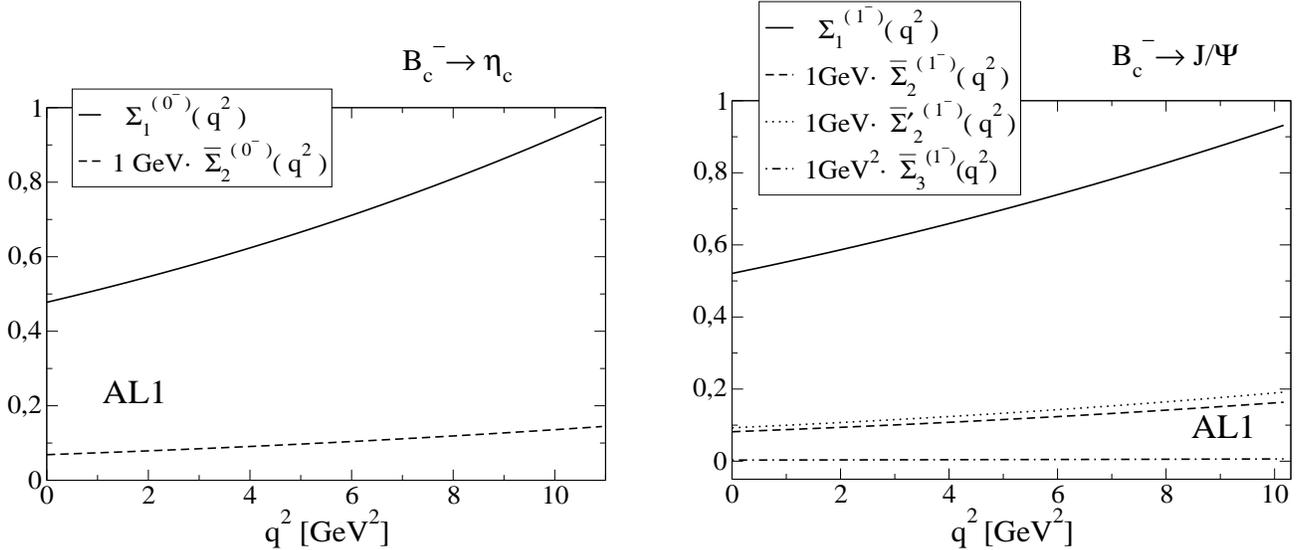

\centering
\resizebox{8.cm}{6.7cm}{\includegraphics{sigescetac.eps}}\hspace{1cm}
\resizebox{8.cm}{7.3cm}{\includegraphics{sigvecjpsi.eps}}\\
\caption{$\Sigma_1^{(0^-)}$ (solid line) and $\overline{\Sigma}_2^{(0^-)}$
(dashed line) of the $B_c^-\to \eta_c$, and
${\Sigma}_1^{(1^-)}$ (solid line), $\overline{\Sigma}_2^{(1^-)}$ (dashed line), 
$\overline{\Sigma}'^{\,(1^-)}_2$ (dotted line) and 
$\overline{\Sigma}_3^{(1^-)}$ (dashed--dotted line) of the
$B_c^-\to J/\Psi$ semileptonic decays evaluated 
 with the AL1 potential.}
\label{fig:sig01m}
\end{figure}
We can take the infinite heavy quark mass limit $m_b\gg m_c\gg \Lambda_{QCD}$
with the result that near zero recoil
\begin{eqnarray}
&&\Sigma_1^{(0^-)}=\Sigma_1^{(1^-)}\nonumber\\
&&\overline{\Sigma}_2^{(0^-)}=\overline{\Sigma}_2^{(1^-)}
=\overline{\Sigma}_2'^{(1^-)}= 0\nonumber\\
&&\overline{\Sigma}_3^{(1^-)}= 0
\end{eqnarray}
This agrees perfectly with the result obtained in
Ref.~\cite{jenkins93} using HQSS\footnote{Note, however, the different
global phases and notation used in Ref.~\cite{jenkins93}.}.

\begin{figure}[h!!!!]
\vspace{.5cmcm}
\centering
\resizebox{8.cm}{6.cm}{\includegraphics{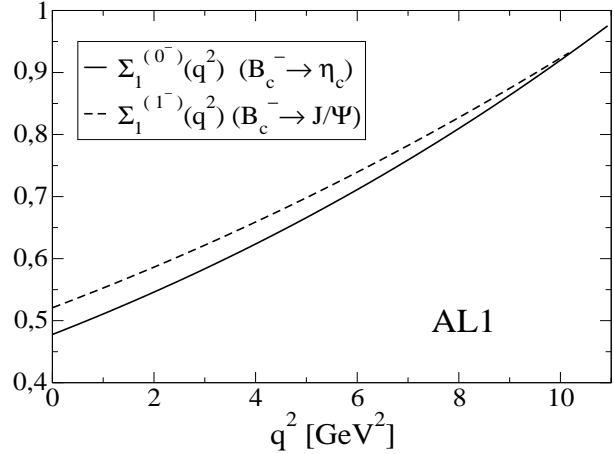}}
\caption{$\Sigma_1^{(0^-)}$ (solid line) of the $B_c^-\to \eta_c$, and
${\Sigma}_1^{(1^-)}$ (dashed line) of the
$B_c^-\to J/\Psi$ semileptonic decays evaluated 
 with the AL1 potential.}
\label{fig:sig101m}
\end{figure}

In Fig.~\ref{fig:sig01m} we give our results for the above quantities
for the semileptonic $B_c^-\to \eta_c$ and $B_c^-\to J/\Psi$ decays
for the actual heavy quark masses. Even though we are not in the
infinite heavy quark mass limit we find that $\Sigma_1^{(0^-)}$ and
$\Sigma_1^{(1^-)}$ dominate over the whole $q^2$ interval.  This
dominant behavior would be more so near the zero--recoil point where
$k\approx 0$ and thus the contributions from the terms in
$\overline{\Sigma}_2^{(0^-)}$, $\overline{\Sigma}_2^{(1^-)}$,
$\overline{\Sigma}_2'^{(1^-)}$ and $\overline{\Sigma}_3^{(1^-)}$ are
even more suppressed. Thus, even for the actual heavy quark masses we
find that near zero recoil
\begin{eqnarray}
\left\langle\, \eta_c,\,\vec{P}_{\eta_c}
\,\left|\, J^{c\,b}_\mu(0)\,
\right| \, B_c^-,\,\vec{P}_{B_c}\right\rangle &\approx&
\sqrt{2m_{B_c}2m_{\eta_c}}\ \Sigma_1^{(0^-)}(q^2)\, v_\mu\nonumber\\
\left\langle\, J/\Psi,\,\lambda\,\vec{P}_{ J/\Psi}\,\left|\, J^{c\,b}_\mu(0)\,
\right| \, B_c^-,\,\vec{P}_{B_c}\right\rangle &\approx&
-i\sqrt{2m_{B_c}2m_{J/\Psi}}\ \varepsilon^*_{(\lambda)\,\mu}
(\,\vec{P}_{J/\Psi}\,)\,\Sigma_1^{(1^-)}(q^2)
\end{eqnarray}
Besides, as seen in In Fig.~\ref{fig:sig101m}, $\Sigma_1^{(0^-)}$ of
the $B_c^-\to \eta_c$, and ${\Sigma}_1^{(1^-)}$ of the $B_c^-\to
J/\Psi$ semileptonic decays are very close to each other over the
whole $q^2$ interval. This implies that the result obtained in
Ref.~\cite{jenkins93} near zero recoil using HQSS seems to be valid,
to a very good approximation, outside the infinite heavy quark mass
limit.
%
%
%
%
%
%
\section{Nonleptonic $B_c^-\to c\bar c\ M_F^-$ two--meson decays.}
\label{sect:nonlepcc}
In this section we will evaluate decay widths for nonleptonic
 $B_c^-\to c\bar c\ M_F^-$ two--meson decays where $M_F^-$ is a
 pseudoscalar or vector meson.  These decay modes involve a $b\to c$
 transition at the quark level and they are governed, neglecting
 penguin operators, by the effective Hamiltonian
 ~\cite{ebert03,colangelo00,ivanov06}
\begin{eqnarray}
H_{eff.}=\frac{G_F}{\sqrt2}\,\left\{ V_{cb}\left[
c_1(\mu)\,Q_1^{cb}+c_2(\mu)\,Q_2^{cb}
\right]
+H.c.\right\}
\end{eqnarray}
where $c_1,\,c_2$ are scale--dependent Wilson coefficients, and
$Q_1^{cb},\,Q_2^{cb}
$ are local four--quark 
operators given by
\begin{eqnarray}
Q_1^{cb}&=&\overline{\Psi}_c(0)\gamma_\mu(I-\gamma_5)\Psi_b(0)\,\bigg[
\ V_{ud}^*\, \overline{\Psi}_d(0)\gamma^\mu(I-\gamma_5)\Psi_u(0)+
V_{us}^*\, \overline{\Psi}_s(0)\gamma^\mu(I-\gamma_5)\Psi_u(0)\nonumber\\
&&\hspace{3.5cm}+
V_{cd}^*\, \overline{\Psi}_d(0)\gamma^\mu(I-\gamma_5)\Psi_c(0)+
V_{cs}^*\, \overline{\Psi}_s(0)\gamma^\mu(I-\gamma_5)\Psi_c(0)\,
\bigg]\nonumber\\
Q_2^{cb}&=&\overline{\Psi}_d(0)\gamma_\mu(I-\gamma_5)\Psi_b(0)\,\bigg[
\ V_{ud}^*\, 
\overline{\Psi}_c(0)\gamma^\mu(I-\gamma_5)\Psi_u(0)
+ V_{cd}^*\, \overline{\Psi}_c(0)\gamma^\mu(I-\gamma_5)\Psi_c(0)\,
\bigg]\nonumber\\
&&\hspace{-.25cm}+\overline{\Psi}_s(0)\gamma_\mu(I-\gamma_5)\Psi_b(0)\, \bigg[
V_{us}^*\, \overline{\Psi}_c(0)\gamma^\mu(I-\gamma_5)\Psi_u(0)\,
+V_{cs}^*\,\overline{\Psi}_c(0)\gamma^\mu(I-\gamma_5)\Psi_c(0)
\bigg]\nonumber\\
\label{eq:q12cb}
\end{eqnarray}
where the different $V_{jk}$ are  CKM matrix elements. 

\begin{figure}[h]
\vspace{1cm}
\centering
\resizebox{9.cm}{4.cm}{\includegraphics{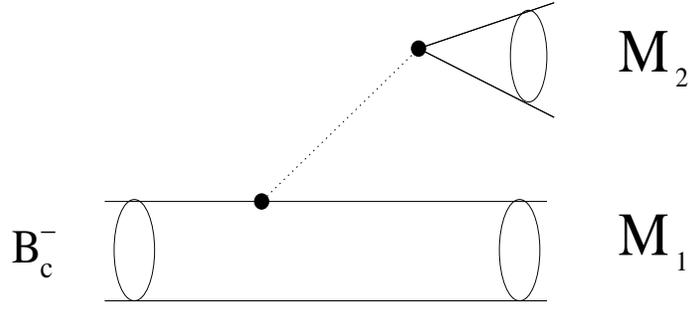}}
\caption{Diagrammatic representation of   $B_c^-$ two--meson decay in the factorization
approximation.}
\label{fig:feynman}
\end{figure}

We shall work in the
factorization approximation which amounts to  evaluate the hadron matrix
elements of the effective Hamiltonian as a product of quark--current matrix
elements: one of these is the matrix element for the $B_c^-$ transition to one
of the final mesons, while the other matrix element corresponds to the transition
from the vacuum to the other final meson. The latter  is given by the corresponding
meson decay constant.
This factorization approximation is schematically represented in Fig.~\ref{fig:feynman}.

When writing the factorization amplitude one has to take into account
the Fierz reordered contribution so that the relevant coefficients are
not $c_1$ and $c_2$ but the combinations
\begin{eqnarray}
a_1(\mu)=c_1(\mu)+\frac{1}{N_C}\,c_2(\mu)\hspace{0.5cm};\hspace{0.5cm}
a_2(\mu)=c_2(\mu)+\frac{1}{N_C}\,c_1(\mu)
\end{eqnarray}
with $N_C=3$ the number of colors. The energy scale $\mu$ appropriate in this
case is $\mu\simeq m_b$ and the values for $a_1$ and $a_2$ that we use
are~\cite{ivanov06}
\begin{eqnarray}
a_1=1.14\hspace{0.5cm};\hspace{0.5cm}
a_2=-0.20
\end{eqnarray}
\subsubsection{$M_F^-=\pi^-,\,\rho^-,\,K^-,\,K^{*-}$ }
\begin{table}[t]
\begin{center}
\begin{tabular}{l|r}
& $\Gamma\ [10^{-15}\ \mathrm{GeV}]$\\
\\
 \hline
$B_c^-\to\eta_{c}\,\pi^-$ & $1.02^{+0.07}\, a_1^2$\\
$B_c^-\to\eta_{c}\,\rho^-$ & $2.60^{+0.16}\, a_1^2$\\
$B_c^-\to\eta_{c}\,K^-$ & $0.082^{+0.004}\, a_1^2$\\
$B_c^-\to\eta_{c}\,K^{*-}$ & $0.15^{+0.01}\,  a_1^2$\\
\hline
$B_c^-\to J/\Psi\,\pi^-$ &$0.83^{+0.09}\, a_1^2$ \\
$B_c^-\to J/\Psi\,\rho^-$ &$2.61^{+0.27}\,a_1^2$\\
$B_c^-\to J/\Psi\,K^-$ &$0.065^{+0.007}\,a_1^2$\\
$B_c^-\to J/\Psi\,K^{*-}$ &$0.16^{+0.01} \,a_1^2$\\
\hline
$B_c^-\to \chi_{c0}\,\pi^-$  &$0.28^{+0.03}\, a_1^2$\\
$B_c^-\to \chi_{c0}\,\rho^-$ &$0.73^{+0.07}\, a_1^2$\\
$B_c^-\to \chi_{c0}\,K^-$    &$0.022^{+0.003} \,a_1^2$\\
$B_c^-\to \chi_{c0}\,K^{*-}$ &$0.041^{+0.005}\, a_1^2$\\
\hline
$B_c^-\to \chi_{c1}\,\pi^-$  &$0.0015^{+0.0002}\, a_1^2$\\
$B_c^-\to \chi_{c1}\,\rho^-$ &$0.11^{+0.01}\, a_1^2$\\
$B_c^-\to \chi_{c1}\,K^-$    &$0.00012^{+0.00001}\, a_1^2$\\
$B_c^-\to \chi_{c1}\,K^{*-}$ &$0.0080^{+0.0007}_{-0.0002}\, a_1^2$\\
\hline
$B_c^-\to h_c\,\pi^-$  &$0.58^{+0.07}\, a_1^2$\\
$B_c^-\to h_c\,\rho^-$ &$1.41^{+0.17}\, a_1^2$\\
$B_c^-\to h_c\,K^-$    &$0.045^{+0.006}\, a_1^2$\\
$B_c^-\to h_c\,K^{*-}$ &$0.078^{+0.009}\,a_1^2$\\
\hline
$B_c^-\to \chi_{c2}\,\pi^-$  &$0.24^{+0.02}\, a_1^2$\\
$B_c^-\to \chi_{c2}\,\rho^-$ &$0.71^{+0.07}_{-0.03} \,a_1^2$\\
$B_c^-\to \chi_{c2}\,K^-$    &$0.018^{+0.002} \,a_1^2$\\
$B_c^-\to \chi_{c2}\,K^{*-}$ &$0.041^{+0.004}_{-0.001}\, a_1^2$\\
\hline
$B_c^-\to \Psi(3836)\,\pi^-$  &$0.00045^{+0.00003}_{-0.00003}\, a_1^2$\\
$B_c^-\to \Psi(3836)\,\rho^-$ &$0.021^{+0.001}_{-0.002}\,a_1^2$\\
$B_c^-\to\Psi(3836) \,K^-$    &$0.000034^{+0.000002}_{-0.000002}\,a_1^2$\\
$B_c^-\to \Psi(3836)\,K^{*-}$ &$0.0015_{-0.0002} \,a_1^2$\\

\end{tabular}
\end{center}
\caption{Decay widths in units of
$10^{-15}$\,GeV, and for general values of the Wilson coefficient $a_1$, 
for exclusive nonleptonic decays of the $B_c^-$ meson. Our central values have
been obtained with the AL1 potential. }
\label{tab:dwccm}
\end{table}

This is the simplest case. 
The decay width is given by
\begin{eqnarray}
&&\Gamma=\frac{G_F^2}{16\pi m^2_{B_c}}\, |V_{cb}|^2\,|V_{F}|^2\ a_1^2\,
\frac{\lambda^{1/2}(m^2_{B_c},\,m^2_{c\bar c},\,m^2_F)}{2m_{B_c}}\,
{\cal H}_{\alpha\beta}(P_{B_c},P_{c\bar c})\, \widehat{\cal H}^{{\alpha\beta}}(P_F)
\nonumber\\
\end{eqnarray}
with $m_F$ the mass of the $M_F^-$ final meson, and $V_{F}=V_{ud}$ or
$V_{F}=V_{us}$ depending on whether $M_F^-=\pi^-,\,\rho^-$ or
$M_F^-=K^-,\,K ^{*-}$. ${\cal H}_{\alpha\beta}(P_{B_c},P_{c\bar c})$
is the hadron tensor for the $B_c^-\to c\bar c$ transition and
$\widehat{\cal H}^{{\alpha\beta}}(P_F)$ is the hadron tensor for the
$\mathrm{vacuum}\to M_F^-$ transition. The latter is
\begin{eqnarray}
\widehat{\cal H}^{{\alpha\beta}}(P_F)&=&P_F^\alpha\,P_F^\beta\,
 f_F^2\hspace{4cm}  M_F^-\equiv0^-\
\mathrm{case}\nonumber\\
\widehat{\cal
H}^{{\alpha\beta}}(P_F)&=&( P_F^\alpha\, P_F^\beta\,-m_F^2\,g^{\alpha\,\beta} )
f_F^2\hspace{2.15cm} M_F^-\equiv1^-\
\mathrm{case}
\end{eqnarray}
with $f_F$ being the $M_F^-$  decay constant.

Similarly to the semileptonic case, the product 
${\cal H}_{\alpha\beta}(P_{B_c},P_{c\bar c})\
\widehat{\cal H}^{{\alpha\beta}}(P_F)$ can now be easily written
in terms of helicity amplitudes for the
$B_c^-\to c\bar c$ transition so that the width is given as~\cite{ivanov06}
\begin{eqnarray}
&&\Gamma=\frac{G_F^2}{16\pi m^2_{B_c}}\, |V_{cb}|^2\,|V_{F}|^2\ a_1^2\,
\frac{\lambda^{1/2}(m^2_{B_c},\,m^2_{c\bar c},\,m^2_F)}{2m_{B_c}}\,
m_F^2\,f^2_F\, {\cal H}^{B_c^-\to c\bar c}_{tt}(m_F^2)\hspace{2.05cm}M_F^-\equiv 0^-\ \mathrm{case}\nonumber\\
&&\Gamma=\frac{G_F^2}{16\pi m^2_{B_c}}\, |V_{cb}|^2\,|V_{F}|^2\ a_1^2\
\frac{\lambda^{1/2}(m^2_{B_c},\,m^2_{c\bar c},\,m^2_F)}{2m_{B_c}}\,
m_F^2\,f^2_F\nonumber\\
 &&\hspace{3cm}\times\left({\cal H}^{B_c^-\to c\bar c}_{+1+1}(m_F^2)
 +{\cal H}^{B_c^-\to c\bar c}_{-1-1}(m_F^2)
 +{\cal H}_{00}^{B_c^-\to c\bar c}(m_F^2)
\right)\hspace{2cm}M_F^-\equiv1^-\ \mathrm{case}
\label{eq:dwccm}
\end{eqnarray}
with the  different ${\cal H}_{rr}$  evaluated at $q^2=m_F^2$.
In Table~\ref{tab:dwccm} we show the decay widths for a general value of the
Wilson coefficient $a_1$, whereas in
Table~\ref{tab:brccm} we give the corresponding branching ratios evaluated
with $a_1=1.14$. Our results for a final $\eta_c$ or $J/\psi$ are in  good
agreement with the ones obtained by Ebert {\it et al.}~\cite{ebert03}, El-Hady
{\it et al.}~\cite{hady00} and Anisimov {\it et al.}~\cite{anisimov99},
 but they are a
factor 2 smaller than the results by Ivanov {\it et al.}~\cite{ivanov06}
and Kiselev~\cite{kiselev02}. Large
discrepancies with Ivanov's results show up for the other transitions.\\

\begin{table}[t]
\begin{center}
\begin{tabular}{l|r c c c c c c c c}
\multicolumn{1}{l}{}&\multicolumn{9}{c}{B.R. (\%)}\\
\multicolumn{1}{l}{}&\multicolumn{9}{c}{}\\
 & This work&\cite{ivanov06}&\cite{ebert03}&\cite{chang94,chang02,chang01}
&\cite{hady00}&\cite{kiselev02,kiselev02-2}
 &\cite{colangelo00}&\cite{anisimov99}&\cite{lopezcastro02}\\
\hline
$B_c^-\to\eta_{c}\,\pi^-$ &$0.094^{+0.006}$
&0.19&0.083&0.18&0.14&0.20&0.025&0.13&\\
$B_c^-\to\eta_{c}\,\rho^-$ & $0.24^{+0.01}$
&0.45&0.20&0.49&0.33&0.42&0.067&0.30&\\
$B_c^-\to\eta_{c}\,K^-$ &$0.0075^{+0.0005}$
&0.015&0.006&0.014&0.011&0.013&0.002&0.013&\\
$B_c^-\to\eta_{c}\,K^{*-}$ &$0.013^{+0.001}$
&0.025&0.011&0.025&0.018&0.020&0.004&0.021&\\
\hline
$B_c^-\to J/\Psi\,\pi^-$ &$0.076^{+0.008}$
&0.17 &0.060&0.18&0.11&0.13&0.13&0.073&\\
$B_c^-\to J/\Psi\,\rho^-$ &$0.24^{+0.02}$
&0.49&0.16&0.53&0.31&0.40&0.37&0.21&\\
$B_c^-\to J/\Psi\,K^-$&$0.0060^{+0.0006}$
&0.013&0.005&0.014&0.008&0.011&0.007&0.007&\\
$B_c^-\to J/\Psi\,K^{*-}$ &$0.014^{+0.002}$
&0.028&0.010&0.029&0.018&0.022&0.020&0.016&\\
\hline
$B_c^-\to \chi_{c0}\,\pi^-$  &$0.026^{+0.003}$
&0.055&&0.028&&0.98&&&\\
$B_c^-\to \chi_{c0}\,\rho^-$ &$0.067^{+0.006}_{-0.001}$
&0.13&&0.072&&3.29&&&\\
$B_c^-\to \chi_{c0}\,K^-$    &$0.0020^{+0.0002} $
&0.0042&&0.00021&&&&&\\
$B_c^-\to \chi_{c0}\,K^{*-}$ &$0.0037^{+0.0005}$
&0.0070&&0.00039&&&&&\\
\hline
$B_c^-\to \chi_{c1}\,\pi^-$  &$0.00014^{+0.00001}$
&0.0068&&0.007&&0.0089&&&\\
$B_c^-\to \chi_{c1}\,\rho^-$ &$0.010^{+0.001}_{-0.001}$
&0.029&&0.029&&0.46&&&\\
$B_c^-\to \chi_{c1}\,K^-$    &$1.1^{+0.1}\,10^{-5}$
&5.1\,$10^{-4}$&&5.2\,$10^{-5}$&&&&&\\
$B_c^-\to \chi_{c1}\,K^{*-}$ &$0.00073^{+0.00007}_{-0.00002}$
&0.0018&&0.00018&&&&&\\
\hline
$B_c^-\to h_c\,\pi^-$  &$0.053^{+0.007}$
&0.11&&0.05&&1.60&&&\\
$B_c^-\to h_c\,\rho^-$ &$0.13^{+0.01}$
&0.25&&0.12&&5.33&&&\\
$B_c^-\to h_c\,K^-$    &$0.0041^{+0.0006}$
&0.0083&&0.00038&&&&&\\
$B_c^-\to h_c\,K^{*-}$ &$0.0071^{+0.0008}$
&0.013&&0.00068&&&&&\\
\hline
$B_c^-\to \chi_{c2}\,\pi^-$  &$0.022^{+0.002}$
&0.046&&0.025&&0.79&&&0.0076\\
$B_c^-\to \chi_{c2}\,\rho^-$ &$0.065^{+0.006}_{-0.002}$
&0.12&&0.051&&3.20&&&0.023\\
$B_c^-\to \chi_{c2}\,K^-$    &$0.0017^{+0.0001} $
&0.0034&&0.00018&&&&&0.00056\\
$B_c^-\to \chi_{c2}\,K^{*-}$
&$0.0038^{+0.0003}_{-0.0002}$
&0.0065&&0.00031&&&&&0.0013\\
\hline
$B_c^-\to \Psi(3836)\,\pi^-$  &$4.1^{+0.03}_{-0.02}\,10^{-5}$
&0.0017&&&&0.030&&&\\
$B_c^-\to \Psi(3836)\,\rho^-$ &$0.0020_{-0.0003}$
&0.0055&&&&0.98&&&\\
$B_c^-\to\Psi(3836) \,K^-$    &$3.1^{+0.2}_{-0.2}\,10^{-6}$
&0.00012&&&&&&&\\
$B_c^-\to \Psi(3836)\,K^{*-}$ &$0.00014_{-0.00002} $
&0.00032&&&&&&&\\

\end{tabular}
\end{center}
\caption{ Branching ratios in \% for exclusive nonleptonic decays 
of the $B_c^-$ meson. Our central values have been obtained with the AL1
potential.  }
\label{tab:brccm}
\end{table}

\subsubsection{$M_F^-=D^-,\,D^{*-},\,D_s^-,\,D_s^{*-}$}
In this subsection we shall evaluate the nonleptonic two--meson  $B_c^-\to \eta_cD^-,\,\eta_c D^{*-},\,
J/\Psi D^-,\,J/\Psi\,D^{*-}$ 
and $B_c^-\to \eta_cD_s^-,\,\eta_c D_s^{*-},\,
J/\Psi D_s^-,\, J/\Psi\,D_s^{*-}$ 
decay widths. 
In these cases there are two different contributions in the factorization
approximation. Following the same steps that lead to Eq.(\ref{eq:dwccm}) we
shall get
\begin{eqnarray}
&&\Gamma=\frac{G_F^2}{16\pi m^2_{B_c}}\, |V_{cb}|^2\,|V_{F}|^2\,
\frac{\lambda^{1/2}(m^2_{B_c},\,m^2_{c\bar c},\,m^2_F)}{2m_{B_c}}\ {\cal HH}
\end{eqnarray}
where now $V_F=V_{cd}$ for $M_F^-=D^-,D^{*-}$ and
$V_F=V_{cs}$ for $M_F^-=D_s^-,D_s^{*-}$. The quantity ${\cal HH}$ incorporates
all information on the hadron matrix elements and depends on the
transition as~\cite{ivanov06}\footnote{Note the different phases used in
Ref.~\cite{ivanov06}.}
\begin{eqnarray}
\label{eq:hh}
{\cal HH}^{B_c^-\to \eta_c D^-}&=&\left|
a_1\,h_t^{B_c^-\to\eta_c}(m^2_{D^-})\,m_{D^-}\,f_{D^-}
+a_2\,h_t^{B_c^-\to D^-}(m_{\eta_c}^2)\,m_{\eta_c}\,f_{\eta_c}
\right|^2\nonumber\\
{\cal HH}^{B_c^-\to \eta_c D^{*-}}&=&\left|
-a_1\,h_0^{B_c^-\to\eta_c}(m^2_{D^{*-}})\,m_{D^{*-}}\,f_{D^{*-}}
+a_2\,i\,h_{(0)\,t}^{B_c^-\to D^{*-}}(m_{\eta_c}^2)\,m_{\eta_c}\,f_{\eta_c}
\right|^2\nonumber\\
{\cal HH}^{B_c^-\to J/\Psi D^{-}}&=&\left|
a_1\, i\,h_{(0)\,t}^{B_c^-\to J/\Psi}(m_{D^{-}}^2)\,m_{D^{-}}\,f_{D^{-}}
-a_2\,\,h_{0}^{B_c^-\to D^{-}}(m_{J/\Psi}^2)\,m_{J/\Psi}\,f_{J/\Psi}
\right|^2\nonumber\\
{\cal HH}^{B_c^-\to J/\Psi D^{*-}}&=&\sum_{r=+1,\,-1,\,0}\,\left|
a_1\, \,h_{(r)\,r}^{B_c^-\to J/\Psi}(m_{D^{*-}}^2)\,m_{D^{*-}}\,f_{D^{*-}}
+a_2\,\,h_{(r)\,r}^{B_c^-\to D^{*-}}(m_{J/\Psi}^2)\,m_{J/\Psi}\,f_{J/\Psi}
\right|^2\nonumber\\
\end{eqnarray}
and similarly for $D^-_s,\,D^{*-}_s$. Note  that the helicity amplitudes
corresponding to $B_c^-\to D^-,\,D^{*-},\,D_s^-,\,D_s^{*-}$ have been evaluated
from the matrix elements for  the  effective current operators
$\overline{\Psi}_{d,s}(0)\gamma^\mu(I-\gamma_5)\Psi_b(0)$  in
Eq.~(\ref{eq:q12cb}). While in practice this is a $b\to d,\,s$ transition, the
momentum transfer ($m_{\eta_c^2}$ or $m_{J/\Psi}^2$) is neither too high,
so that one has to include a $B_c^*$ resonance, nor too low, so as to have too high
three-momentum transfers\footnote{Our experience with the $B\to \pi$ 
decay~\cite{albertus05-2}, where we have a similar $b\to u$ quark transition, shows that the naive nonrelativistic quark model
gives reliable results 
for $q^2\approx 9$\,GeV$^2$.}. Besides the contribution is weighed by the much smaller $a_2$ Wilson
coefficient.
 In Table~\ref{tab:dwcccm} we
give the decay widths for general values of the Wilson coefficients
$a_1$ and $a_2$, and in  Table~\ref{tab:brcccm} we show the branching ratios.
 We are in reasonable agreement with the
results by Ivanov {\it et al.}~\cite{ivanov06}, El-Hady {\it et
al.}~\cite{hady00} and Kiselev~\cite{kiselev02}. For decays with a final
$D_s^-,D_s^{*-}$ the agreement is also reasonable with the results by
Colangelo {\it et
al.}~\cite{colangelo00} and Anisimov {\it et
al.}~\cite{anisimov99}.
\begin{table}[h]
\begin{center}
\begin{tabular}{l|c }

 & $\Gamma\ [10^{-15}\ \mathrm{GeV}]$\\
 &\\
\hline
$B_c^-\to\eta_{c}\,D^-$ & $(0.438^{+0.010} a_1+ 0.236^{+0.030}_{-0.023} a_2)^2$
\\
$B_c^-\to\eta_{c}\,D^{*-}$ & $(-0.390_{-0.009} a_1- 0.136_{-0.022}^{+0.015}a_2)^2$
\\
$B_c^-\to J/\Psi\,D^-$ & $( -0.328_{-0.012}a_1-0.156_{-0.019}^{+0.016} a_2)^2$
\\
$B_c^-\to J/\Psi\,D^{*-}$ & \ $(-0.195_{-0.008} a_1-0.066^{+0.006}_{-0.011} a_2)^2$\\
                          &  +$(-0.390_{-0.018} a_1-0.209^{+0.019}_{-0.032} a_2)^2$\\
                          & +$( 0.447_{-0.016} a_1+0.167_{-0.027}^{+0.016} a_2)^2$
		   \\
\hline
$B_c^-\to\eta_{c}\,D_s^-$ & $(2.54^{+0.05} a_1+ 1.93^{+0.10}a_2)^2$
\\
$B_c^-\to\eta_{c}\,D_s^{*-}$ & $(-1.84_{-0.04} a_1- 1.17_{-0.14}^{+0.02}a_2)^2$
\\
$B_c^-\to J/\Psi\,D_s^-$ & $(-1.85_{-0.06}^{+0.01} a_1-1.23_{-0.06} a_2)^2$
\\
$B_c^-\to J/\Psi\,D_s^{*-}$
& $(-1.01_{-0.04} a_1-0.60^{+0.02}_{-0.07} a_2)^2$\\ 
& $+(-2.00_{-0.06} a_1-1.71^{+0.03}_{-0.18} a_2)^2$\\
& $+( 2.17_{-0.08} a_1+1.42_{-0.16}^{+0.02} a_2)^2$
\\
\end{tabular}
\end{center}
\caption{Decay widths in units of $10^{-15}$\,GeV, and for general
values of the Wilson coefficients $a_1$ and $a_2$, for exclusive
nonleptonic decays of the $B_c^-$ meson. Our central values have been
obtained with the AL1 potential.  For vector--vector final state we
show the three different contributions corresponding to
$r=+1,\,-1,\,0$ (see Eq.(\ref{eq:hh})).}
\label{tab:dwcccm}
\end{table}
\begin{table}[h]
\begin{center}
\begin{tabular}{l|c c c c c c c}
\multicolumn{1}{l}{}&\multicolumn{7}{c}{B.R. (\%)}\\
\multicolumn{1}{l}{}&\multicolumn{7}{c}{}\\
 & This work&\cite{ivanov06}&\cite{chang94}&\cite{hady00}&\cite{kiselev02}&\cite{colangelo00}
& \cite{anisimov99}\\
\hline
$B_c^-\to\eta_{c}\,D^-$ & $0.014^{+0.001}$
&0.019&0.0012&0.014&0.015&0.005&0.010\\
$B_c^-\to\eta_{c}\,D^{*-}$ &
$0.012^{+0.001}$
&0.019&0.0010&0.013&0.010&0.002&0.0055\\
$B_c^-\to J/\Psi\,D^-$ &
$0.0083^{+0.0005}$
&0.015&0.0009&0.009&0.009&0.013&0.0044\\
$B_c^-\to J/\Psi\,D^{*-}$ & $0.031^{+0.001}$
&0.045&&0.028&0.028&0.019&0.010\\
\hline
$B_c^-\to\eta_{c}\,D_s^-$ & $0.44^{+0.02}$
&0.44&0.054&0.26&0.28&0.50&0.35\\
$B_c^-\to\eta_{c}\,D_s^{*-}$ & $0.24^{+0.02}$
&0.37&0.044&0.24&0.27&0.038&0.36\\
$B_c^-\to J/\Psi\,D_s^-$ & $0.24^{+0.02}$
&0.34&0.041&0.15&0.17&0.34&0.12\\
$B_c^-\to J/\Psi\,D_s^{*-}$ & $0.68^{+0.03}$
&0.97&&0.55&0.67&0.59&0.62\\
\end{tabular}
\end{center}
\caption{Branching ratios in \% 
for exclusive nonleptonic decays of the $B_c^-$ meson. Our central values have
been obtained with the AL1 potential. }
\label{tab:brcccm}
\end{table}
%
%
%
%
\section{semileptonic $B_c^-\to \overline{B}^0,\, \overline{B}^{*0},\,
 \overline{B}_s^0,\, \overline{B}_s^{*0}$  decays}
\label{sect:semilepb}
In this section we shall study the semileptonic $B_c^-\to
 \overline{B}^0,\, \overline{B}^{*0},\, \overline{B}_s^0,\,
 \overline{B}_s^{*0}$ decays. With obvious changes the calculations
 are done as before, with the only novel thing that now it is the
 antiquark that suffers the transition (we have $\bar c\to \bar d,\,
 \bar s$ ), and thus we have to take into account the changes in
 the form factors according to the results in appendix~\ref{app:sign}.

\subsection{Form factors}
\begin{figure}[t]
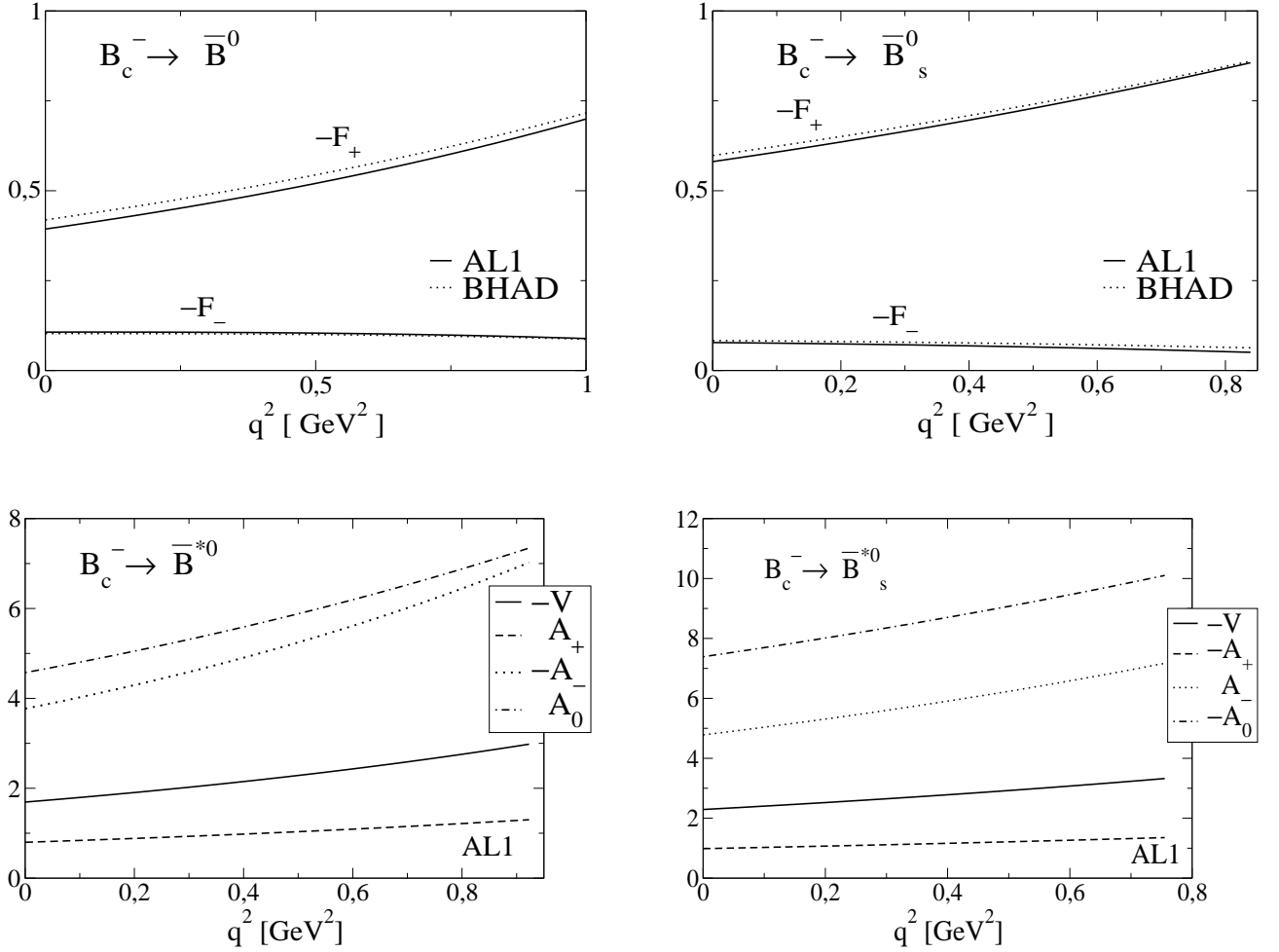

\vspace{1cm}
\centering
\resizebox{8.cm}{6.cm}{\includegraphics{bctobelectron.eps}}\hspace{1cm}
\resizebox{8.cm}{6.cm}{\includegraphics{bctobselectron.eps}}\vspace{.9cm}\\
\resizebox{8.cm}{6.cm}{\includegraphics{bctobstarelectron.eps}}\hspace{1cm}
\resizebox{8.cm}{6.cm}{\includegraphics{bctobsstarelectron.eps}}
\caption{$F_+$ and $F_-$ form factors (solid lines) for 
$B_c^-\to\overline{B}^0$ and
$B_c^-\to\overline{B}^0_s$ semileptonic decay, and $V$ (solid line),
$A_+$ (dashed line), $A_-$ (dotted line) and $A_0$ (dashed--dotted line) form factors for
$B_c^-\to\overline{B}^{*0}$ and
$B_c^-\to\overline{B}^{*0}_s$ semileptonic decay evaluated
with the AL1 potential.  For the first two cases, and for comparison,
we also show with dotted lines the results obtained with the Bhaduri (BHAD)
potential.}
\label{fig:fpmvapm0}
\end{figure}

In Fig.~\ref{fig:fpmvapm0} we show the form factors for the above
transitions evaluated with the AL1 potential.  For the
$\overline{B}^0$ and $\overline{B}^0_s$ cases we also show the results
obtained with the BHAD potential. Although they are less visible in
the figures, the larger differences, of up to 25\%, occur for the
$F_-$ form factor.
    
In Table~\ref{tab:fpmb} we show $F_+$, $F_-$ and $F_0$ (defined as in
Eq.~(\ref{eq:f0}) changing  the mass of the final meson) of the $B_c^-\to
\overline{B}^0,\,\overline{B}^0_s$ transitions evaluated at
$q^2_{min}$ and $q^2_{max}$ and compare them with the results by
Ivanov {\it et al.}~\cite{ivanov01} and Ebert {\it et
al.}~\cite{ebert03-2}.  Notice that, to favor comparison, we have
changed the signs of the form factors by Ebert {\it et al.} (they
evaluate $B_c^+$ decay) in accordance with the results in
appendix~\ref{app:sign}.  The agreement with the results by Ebert {\it
et al.}  is good. We also agree with Ivanov {\it et al.}  for $F_+$,
but get very different results for $F_-$.  As fermion masses are very
small the disagreement in the $F_-$ form factor will have a negligible
effect on the decay width.
 \begin{table}[h!!!!!!!]

\begin{tabular}{c| c c c c|  c c}
$B_c^-\to\overline{B}^0\,l^-\bar{\nu}_l\ \ $ & $q^2_{\mathrm{min}}$ &
 $q^2_{\mathrm{max}}$ & $\hspace{1cm}$&
$B_c^-\to\overline{B}^0_s\,l^-\bar{\nu}_l\ \ $ & $q^2_{\mathrm{min}}$ &
 $q^2_{\mathrm{max}}$  \\
\cline{1-3} \cline{5-7}
&&&&&&\\
 $F_+$ & & & &$F_+$ & &\\
 This work & $-0.39^{+0.03}_{-0.03}$& $-0.70^{+0.004}_{-0.02}$ & & 
 This work & $-0.58^{+0.01}_{-0.02}$& 
 $-0.86^{+0.01}_{-0.01}$\\
 \cite{ivanov01} &$-0.58$ &$-0.96$&&\cite{ivanov01} &$-0.61$ &$-0.92$\\
 \cite{ebert03-2} &$-0.39$&$-0.96$&&\cite{ebert03-2}&$-0.50$&$-0.99$\\
&&&&&\\
$F_-$ & & & &$F_-$ & &\\
 This work &$-0.11^{+0.02}_{-0.03}$ &$-0.09^{+0.01}_{-0.06}$ & &  This work
 &$-0.08^{+0.01}_{-0.02}$ &$-0.05^{+0.01}_{-0.03}$\\
 \cite{ivanov01} & 2.14&2.98& & \cite{ivanov01} & 1.83&2.35\\
 &&&&&&\\
$F_0$ & & & &$F_0$ & &\\
 This work &$-0.39^{+0.03}_{-0.03}$ &$-0.71^{+0.05}_{-0.02}$ & &  This work
 &$-0.58_{-0.02}$ &$-0.86^{+0.01}_{-0.01}$\\
 \cite{ebert03-2} &$-0.39$&$-0.80$&&\cite{ebert03-2}&$-0.50$ &$-0.86$\\
 \end{tabular}\hspace{1cm}
\caption{$F_+$, $F_-$ and $F_0$ evaluated at
$q^2_{\mathrm{min}}$ and $q^2_{\mathrm{max}}$ compared to  the ones obtained by 
Ivanov {\it et al.}~\cite{ivanov01} and Ebert{\it et al.}~\cite{ebert03-2}. 
Our central values have been obtained with the AL1
potential. Here $l$ stands for $l=e,\,\mu$.}
\label{tab:fpmb}
\end{table}

In Table~\ref{tab:fva0pmb} we show $V$, $A_+$, $A_-$, $A_0$ and
$\widetilde{A}_0$ (defined as in
Eq.~(\ref{eq:a0tilde}) changing  the mass of the final meson) 
 of the
$B_c^-\to \overline{B}^{*0},\,\overline{B}^{*0}_s$ evaluated at
$q^2_{min}$ and $q^2_{max}$ and compare them with the results by
Ivanov {\it et al.}~\cite{ivanov01} and Ebert {\it et
al.}~\cite{ebert03-2}.  With some exceptions the agreement with Ebert's results 
is bad in this case. We are also in clear disagreement with Ivanov's results.

\begin{table}[h!]
\begin{tabular}{c| c c c c|  c c}
$B_c^-\to\overline{B}^{*0}\,l^-\bar{\nu}_l\ \ $ & $q^2_{\mathrm{min}}$ &
 $q^2_{\mathrm{max}}$ & $\hspace{1cm}$&
$B_c^-\to\overline{B}^{*0}_s\,l^-\bar{\nu}_l\ \ $ & $q^2_{\mathrm{min}}$ &
 $q^2_{\mathrm{max}}$  \\
\cline{1-3} \cline{5-7}
&&&&&&\\
 $V$ & & & &$V$ & &\\
 This work & $-1.69^{+0.11}_{-0.09}$& $-2.98^{+0.17}_{-0.03}$ & & 
 This work & $-2.29^{+0.02}_{-0.09}$& 
 $-3.32^{+0.04}_{-0.01}$\\
 \cite{ivanov01} & &$-5.32^*$&&\cite{ivanov01} & &$-4.91^*$\\
 \cite{ebert03-2} &$-3.94$&$-8.91$&&\cite{ebert03-2}&$-3.44$&$-6.25$\\
&&&&&&\\
$A_+$ & & & &$A_+$ & &\\
 This work &$-0.80^{+0.02}_{-0.02}$ &$-1.30^{+0.02}$ & &  This work
 &$-0.98^{+0.01}_{-0.03}$ &$-1.35^{+0.03}_{-0.02}$\\
 \cite{ivanov01} & &0.49& & \cite{ivanov01} & &0.21\\
 \cite{ebert03-2} &$-2.89$ &$-2.83$& & \cite{ebert03-2} &$-2.19$ &$-2.62$\\
 
 &&&&&&\\
$A_-$ & & & &$A_-$ & &\\
 This work &$3.77^{+0.15}_{-0.17}$ &$7.02_{-0.35}$ & &  This work
 &$4.78^{+0.23}_{-0.01}$ &$7.17^{+0.06}_{-0.18}$\\
 \cite{ivanov01} & &18.0& & \cite{ivanov01} & &15.9\\
 &&&&&&\\
$A_0$ & & & &$A_0$ & &\\
 This work &$-4.57^{+0.27}_{-0.33}$ &$-7.34^{+0.28}_{-0.49}$ & &  This work
 &$-7.39^{+0.14}_{-0.30}$ &$-10.10^{+0.22}_{-0.16}$\\
 \cite{ivanov01} & &$-5.07$& & \cite{ivanov01} & &$-6.60$\\
 \cite{ebert03-2} & $-5.08$&$-8.70$& & \cite{ebert03-2} & $-6.60$&$-10.23$\\
 
 &&&&&&\\
$\widetilde{A}_0$ & & & &$\widetilde{A}_0$ & &\\
 This work &$-0.34^{+0.03}_{-0.03}$ &$-0.60^{+0.05}_{-0.02}$ & &  This work
 &$-0.51^{+0.01}_{-0.03}$ &$-0.74^{+0.01}_{-0.01}$\\
 \cite{ebert03-2} &$-0.20$&$-1.06$&&\cite{ebert03-2}&$-0.35$ &$-0.91$\\
 \end{tabular}\hspace{1cm}
\caption{$V$, $A_+$, $A_-$, $A_0$ and $\widetilde A_0$ evaluated at
$q^2_{\mathrm{min}}$ and $q^2_{\mathrm{max}}$ compared to the ones
obtained by Ivanov {\it et al.}~\cite{ivanov01} and Ebert {\it et
al.}~\cite{ebert03-2}. Our central  values have been obtained with
the AL1 potential. Here $l$ stands for $l=e,\,\mu$. Asterisk as in
Table~\ref{tab:fva0pm}.}
\label{tab:fva0pmb}
\end{table}

\subsection{Decay width}
In Tables~\ref{tab:habtob}, \ref{tab:asybtob} we give respectively 
our results for the partial helicity widths and forward-backward asymmetries.

\begin{table}[h!!!]
\begin{center}
\begin{tabular}{l|c c c c c c c}
$B^-_c\to$ & \hspace{.2cm}$\Gamma_U$\hspace{.2cm} & \hspace{.2cm}$\widetilde{\Gamma}_U$\hspace{.2cm}
 &\hspace{.2cm} $\Gamma_L$\hspace{.2cm} & $\widetilde{\Gamma}_L$ & \hspace{.2cm}$\Gamma_P$
 \hspace{.2cm}& $\widetilde{\Gamma}_S $ & $\widetilde{\Gamma}_{SL}$\\ 
\hline
&&&&&&&\\
$\overline{B}^0\,e^-\bar{\nu}_e$& 0 & 0 &$ 0.65^{+0.07}_{-0.09}$&
$0.14^{+0.02}_{-0.02}\ 10^{-5}$ & 0 &
 $ 0.47^{+0.05}_{-0.07}\ 10^{-5}$  & $ 0.15^{+0.01}_{-0.02}\ 10^{-5}$\\
$\overline{B}^0\,\mu^-\bar{\nu}_\mu$& 0 & 0 &$ 0.55^{+0.06}_{-0.07}$&
$0.15^{+0.01}_{-0.02}\ 10^{-1}$ & 0 &
 $ 0.64^{+0.16}_{-0.09}\ 10^{-1}$  & $ 0.17^{+0.02}_{-0.02}\ 10^{-1}$\\
&&&&&&&\\
$\overline{B}_s^0\,e^-\bar{\nu}_e$& 0 & 0 &$ 15.1^{+0.7}_{-0.3}$&
$0.43^{+0.02}_{-0.01}\ 10^{-4}$ & 0 &
 $ 0.14^{+0.01}\ 10^{-3}$  & $ 0.45^{+0.03}_{-0.01}\ 10^{-4}$\\
$\overline{B}_s^0\,\mu^-\bar{\nu}_\mu$& 0 & 0 &$ 12.4^{+0.5}_{-0.3}$&
$0.40^{+0.02}_{-0.01}$ & 0 &
 $ 1.69^{+0.08}_{-0.03}$  & $ 0.47^{+0.02}_{-0.01}$\\
&&&&&&&\\

$\overline{B}^{*0}\,e^-\bar{\nu}_e$& $0.83^{+0.08}_{-0.11}$ 
& $0.26^{+0.03}_{-0.04}\ 10^{-6}$ &$ 0.76^{+0.09}_{-0.11}$&
$0.10^{+0.02}_{-0.01}\ 10^{-5}$ & $0.36_{-0.05}^{+0.03}$ &
 $ 0.27^{+0.04}_{-0.05}\ 10^{-5}$  & $ 0.96^{+0.15}_{-0.15}\ 10^{-6}$\\
 $\overline{B}^{*0}\,\mu^-\bar{\nu}_\mu$& $0.79^{+0.10}_{-0.11}$ 
& $0.97^{+0.11}_{-0.13}\ 10^{-2}$ &$ 0.68^{+0.08}_{-0.10}$&
$0.14^{+0.01}_{-0.03}\ 10^{-1}$ & $0.34_{-0.05}^{+0.02}$ &
 $ 0.28^{+0.04}_{-0.05}\ 10^{-1}$  & $ 0.11^{+0.02}_{-0.02}\ 10^{-1}$\\

&&&&&&&\\
$\overline{B}_s^{*0}\,e^-\bar{\nu}_e$& $16.7^{+0.8}_{-0.7}$ 
& $0.66^{+0.04}_{-0.03}\ 10^{-5}$ &$ 16.8^{+1.1}_{-0.8}$&
$0.33^{+0.02}_{-0.02}\ 10^{-4}$ & $6.11_{-0.20}^{+0.21}$ &
 $ 0.88^{+0.08}_{-0.05}\ 10^{-4}$  & $ 0.30^{+0.03}_{-0.01}\ 10^{-4}$\\
 $\overline{B}_s^{*0}\,\mu^-\bar{\nu}_\mu$& $15.6^{+0.8}_{-0.6}$ 
& $0.24^{+0.02}_{-0.01}$ &$ 14.5^{+1.0}_{-0.6}$&
$0.37^{+0.02}_{-0.02}$ & $5.66_{-0.19}^{+0.18}$ &
 $ 0.77^{+0.06}_{-0.04}$  & $ 0.30^{+0.02}_{-0.01}$\\

\end{tabular}
\end{center}
\caption{ Partial helicity widths in units of $10^{-15}$\,GeV for $B_c^-$ decay. Central values
have been evaluated with the AL1 potential. } 
\label{tab:habtob}
\end{table}

\begin{table}[h]
\begin{tabular}{l|c c }
&$A_{FB} (e)$  &$A_{FB} (\mu)$  \\
\hline
&\\
$B_c^-\to \overline{B}^0$ &$0.67^{+0.02}\ 10^{-5}$&$0.82^{+0.01}\ 10^{-1}$\\
$B_c^-\to \overline{B}_s^0$ &$0.89^{+0.01}\ 10^{-5}$  &$0.96^{+0.01}\ 10^{-1}$\\
$B_c^-\to \overline{B}^{*0}$ &$0.17_{-0.01}$ &$0.19_{-0.01}$ \\
$B_c^-\to \overline{B}_s^{*0}$ &$0.14_{-0.01}$&$0.16^{+0.01}$\\

\end{tabular}
\caption{Forward-backward asymmetry. Our central values have been
evaluated with the AL1 potential. We would obtain the same results for
$B_c^+\to B^0$ decays. }
\label{tab:asybtob}
\end{table}

\begin{figure}[h!!]
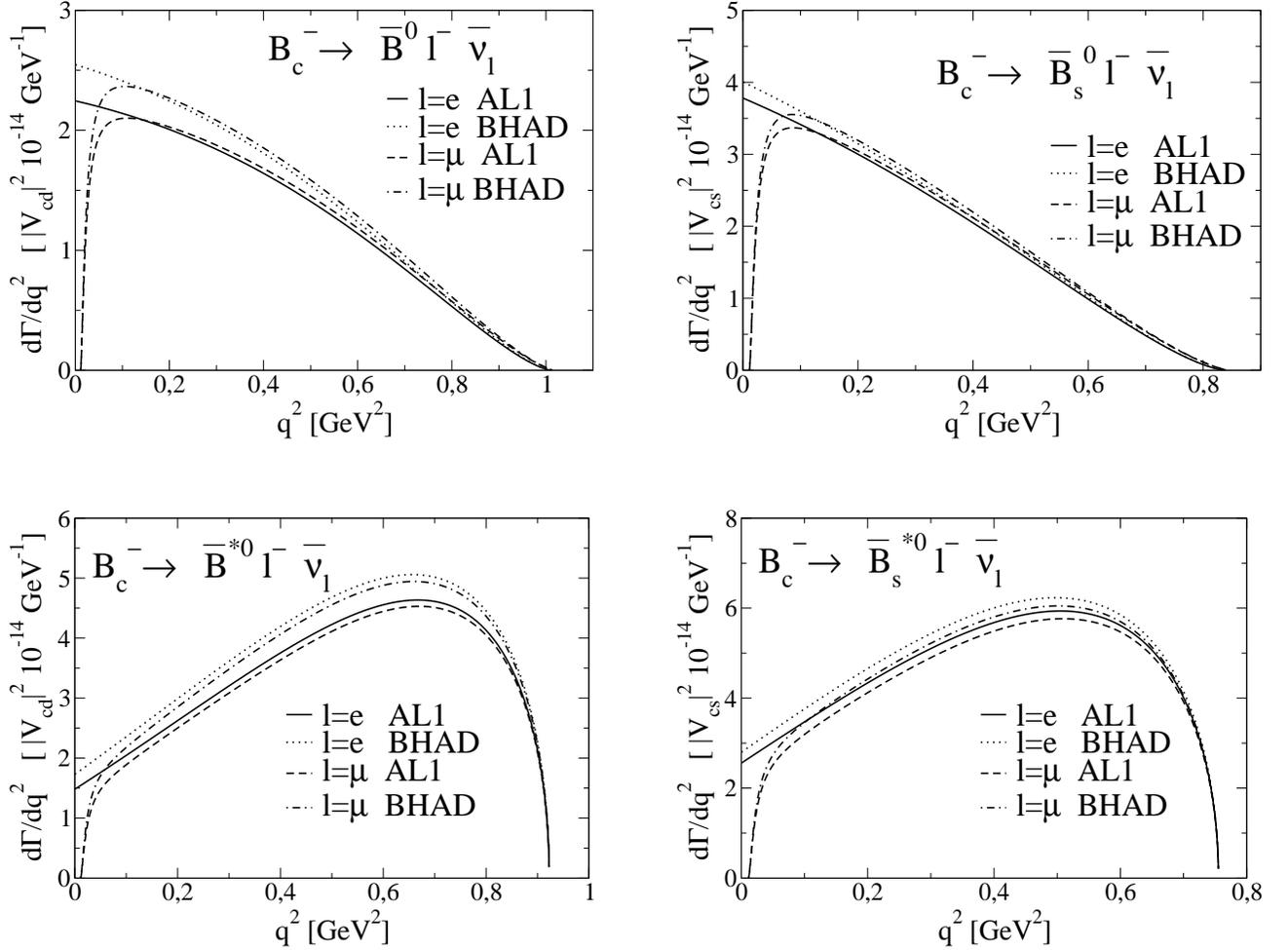

\vspace{1cm}
\centering
\resizebox{8.cm}{6.cm}{\includegraphics{dgdq2bctob.eps}}\hspace{1cm}
\resizebox{8.cm}{6.cm}{\includegraphics{dgdq2bctobs.eps}}\vspace{.9cm}\\
\resizebox{8.cm}{6.cm}{\includegraphics{dgdq2bctobstar.eps}}\hspace{1cm}
\resizebox{8.cm}{6.cm}{\includegraphics{dgdq2bctobsstar.eps}}
\caption{Differential decay width for the for the $B_c^-\to\overline{B}^0\, 
l^-\bar{\nu}_l$ and $B_c^-\to\overline{B}^0_s\, 
l^-\bar{\nu}_l$,
$B_c^-\to\overline{B}^{*0}\, 
l^-\bar{\nu}_l$ and $B_c^-\to\overline{B}^{*0}_s\, 
l^-\bar{\nu}_l$ ($l=e,\,\mu$)
transitions. Solid line: results for a final $e$ evaluated with the
 AL1 potential; dotted line: results for a final $e$ evaluated with the
Bhaduri (BHAD)  potential; dashed line: results for a final 
$\mu$ evaluated with the
 AL1 potential; dashed--dotted line: results for a final $\mu$ evaluated with the
Bhaduri (BHAD)  potential.}
\label{fig:dgdq2btob}
\end{figure}

In Fig.~\ref{fig:dgdq2btob} we show the differential decay width  for the $B_c^-\to\overline{B}^0\, 
l^-\bar{\nu}_l$, $B_c^-\to\overline{B}^0_s\, 
l^-\bar{\nu}_l$,
$B_c^-\to\overline{B}^{*0}\, 
l^-\bar{\nu}_l$ and $B_c^-\to\overline{B}^{*0}_s\, 
l^-\bar{\nu}_l$
transitions ($l=e,\,\mu$). In Tables~\ref{tab:dwbtob}, \ref{tab:brbtob} we give the total decay 
widths and branching ratios and
compare them with determinations by other groups. Our results are in better
agreement with the ones obtained by Ebert {\it et al.}~\cite{ebert03-2}, 
Colangelo {\it et al.}~\cite{colangelo00},
Anisimov {\it et al.}~\cite{anisimov99} 
and Lu {\it et al.}~\cite{lu95}.

\begin{table}[h!]
\begin{center}
\hspace*{-1.5cm}\begin{tabular}{l| c c c c c c c c c c c}
\multicolumn{1}{l}{}&\multicolumn{11}{c}
{$\Gamma\ [10^{-15}\ \mathrm{GeV}]$}\\
\multicolumn{1}{l}{}&\multicolumn{11}{c}{}\\
&This work&\cite{ivanov01}&\cite{ebert03-2}
&\cite{chang94}&\cite{hady00}&\cite{liu97}&\cite{kiselev02}&\cite{colangelo00}&
\cite{anisimov99}&\cite{nobes00}&
\cite{lu95} \\
\hline
&\\
$B^-_c\to \overline{B}^0\, e\,\bar{\nu}_e$&$0.65^{+0.07}_{-0.09}$
&2.1&0.6&2.30&1.14&1.90&4.9&0.9(1.0)&&&0.59\\
$B^-_c\to \overline{B}^0\, \mu\,\bar{\nu}_\mu$&$0.63^{+0.07}_{-0.09}$\\
$B^-_c\to \overline{B}_s^0\, e\,\bar{\nu}_e$&$15.1^{+0.7}_{-0.3}$
&29&12&26.6&14.3&26.8&59&11.1(12.9)&15&12.3&11.75\\
$B^-_c\to \overline{B}_s^0\, \mu\,\bar{\nu}_\mu$&$14.5^{+0.6}_{-0.3}$\\
$B^-_c\to \overline{B}^{*0}\, e\,\bar{\nu}_e$&$1.59^{+0.17}_{-0.22}$
&2.3&1.7&3.32&3.53&2.34&8.5&2.8(3.2)&&&2.44\\
$B^-_c\to \overline{B}^{*0}\, \mu\,\bar{\nu}_\mu$&$1.52^{+0.17}_{-0.21}$\\
$B^-_c\to \overline{B}_s^{*0}\, e\,\bar{\nu}_e$&$33.5^{+1.9}_{-1.5}$
&37&25&44.0&50.4&34.6&65&33.5(37.0)&34&19.0&32.56\\
$B^-_c\to \overline{B}_s^{*0}\, \mu\,\bar{\nu}_\mu$&$31.5^{+1.8}_{-1.4}$
\end{tabular}
\end{center}
\caption{Decay widths in units of $10^{-15}$\,GeV.
 Our central values have been evaluated with the
AL1 potential.} 
\label{tab:dwbtob}
\end{table}

\begin{table}[h!!!!]
\begin{center}

\begin{tabular}{l| c c c c c c c c  c}
\multicolumn{1}{l}{}&\multicolumn{8}{c}
{B.R. (\%)}\\
\multicolumn{1}{l}{}&\multicolumn{8}{c}{}\\
&This work&\cite{ivanov06}&\cite{ebert03-2}&\cite{chang94}&\cite{hady00}
&\cite{kiselev02}&\cite{colangelo00}&
\cite{anisimov99}&
\cite{nobes00}\\
\hline
&\\
$B^-_c\to \overline{B}^0\, e\,\bar{\nu}_e$&$0.046^{+0.004}_{-0.007}$
&0.071&0.042&0.16&0.078&0.34&0.06&&0.048\\
$B^-_c\to \overline{B}^0\, \mu\,\bar{\nu}_\mu$&$0.044^{+0.005}_{-0.006}$\\
$B^-_c\to \overline{B}_s^0\, e\,\bar{\nu}_e$&$1.06^{+0.05}_{-0.02}$
&1.10&0.84&1.82&0.98&4.03&0.8&0.99&0.92\\
$B^-_c\to \overline{B}_s^0\, \mu\,\bar{\nu}_\mu$&$1.02^{+0.04}_{-0.02}$\\
$B^-_c\to \overline{B}^{*0}\, e\,\bar{\nu}_e$&$0.11^{+0.01}_{-0.01}$
&0.063&0.12&0.23&0.24&0.58&0.19&&0.051\\
$B^-_c\to \overline{B}^{*0}\, \mu\,\bar{\nu}_\mu$&$0.11^{+0.01}_{-0.02}$\\
$B^-_c\to \overline{B}_s^{*0}\, e\,\bar{\nu}_e$&$2.35^{+0.14}_{-0.10}$
&2.37&1.75&3.01&3.45&5.06&2.3&2.30&1.41\\
$B^-_c\to \overline{B}_s^{*0}\, \mu\,\bar{\nu}_\mu$&$2.22^{+0.12}_{-0.10}$\\

\end{tabular}

\end{center}
\caption{Branching ratios in \% .
 Our central values have been evaluated with the
AL1 potential.} 
\label{tab:brbtob}
\end{table}

\subsection{Heavy quark spin symmetry}

\begin{figure}[t]
\vspace{1cm}
\centering
\resizebox{8.cm}{6.cm}{\includegraphics{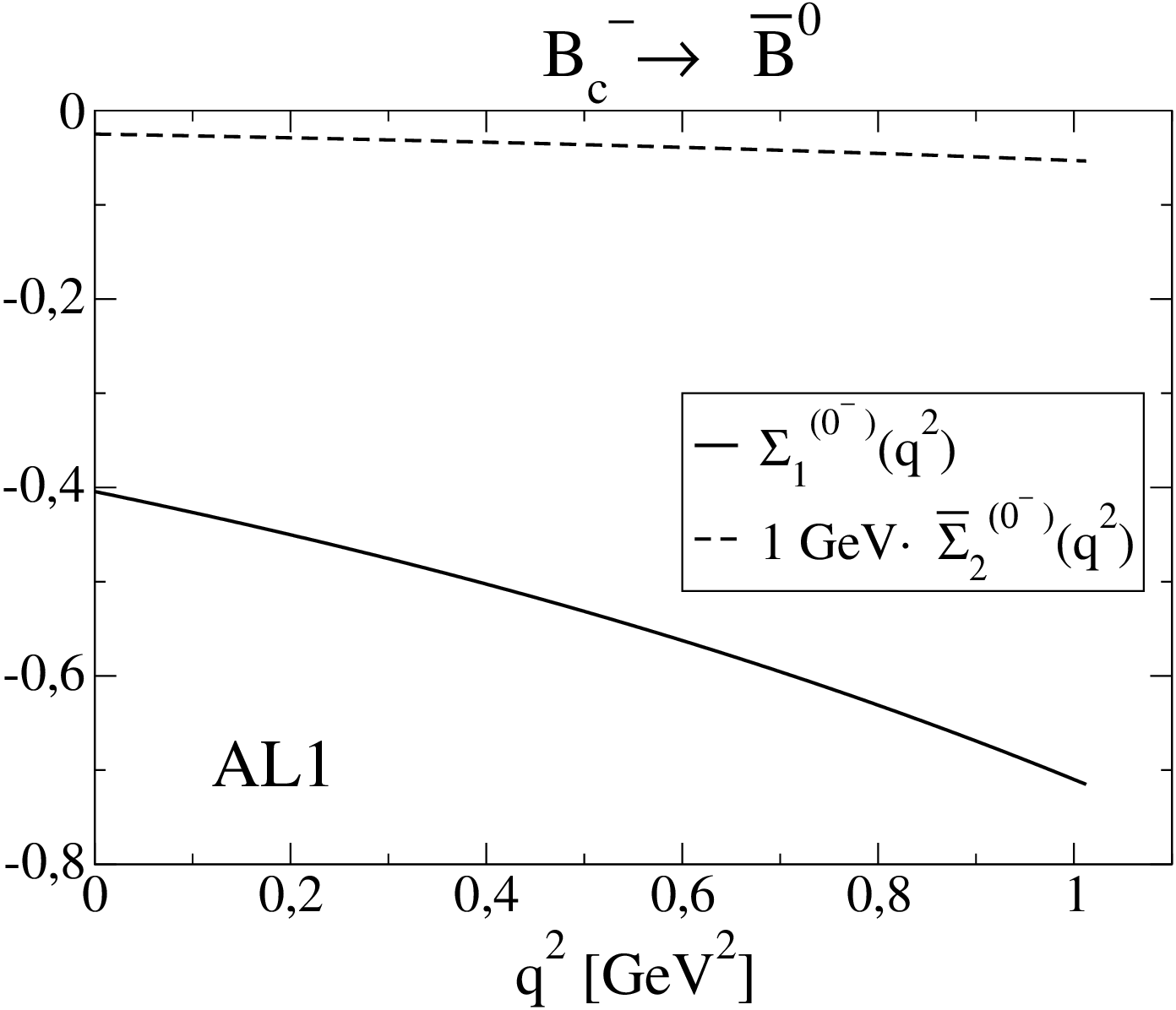}}\hspace{1cm}
\resizebox{8.cm}{6.3cm}{\includegraphics{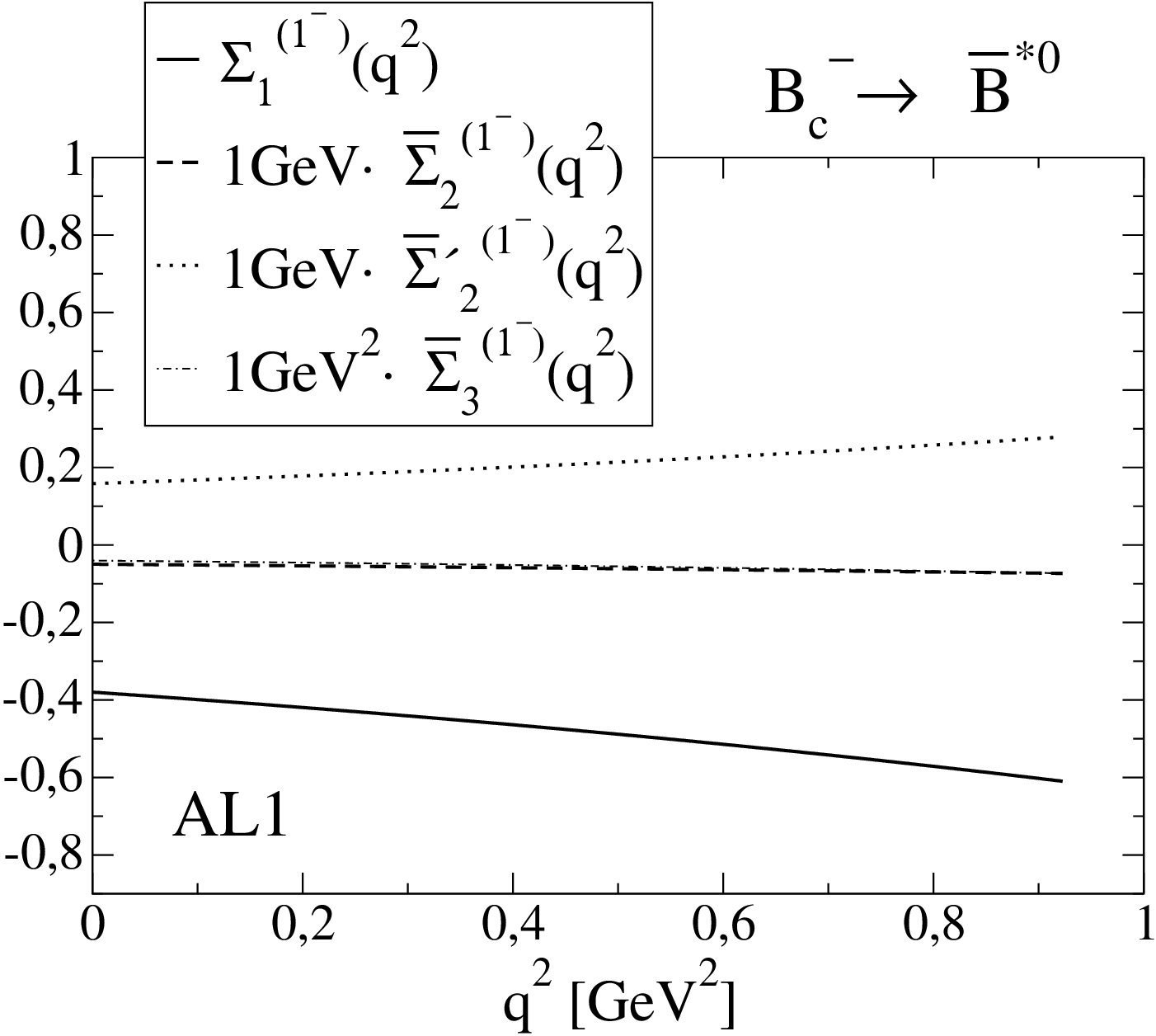}}\vspace{.9cm}\\
\resizebox{8.cm}{6.cm}{\includegraphics{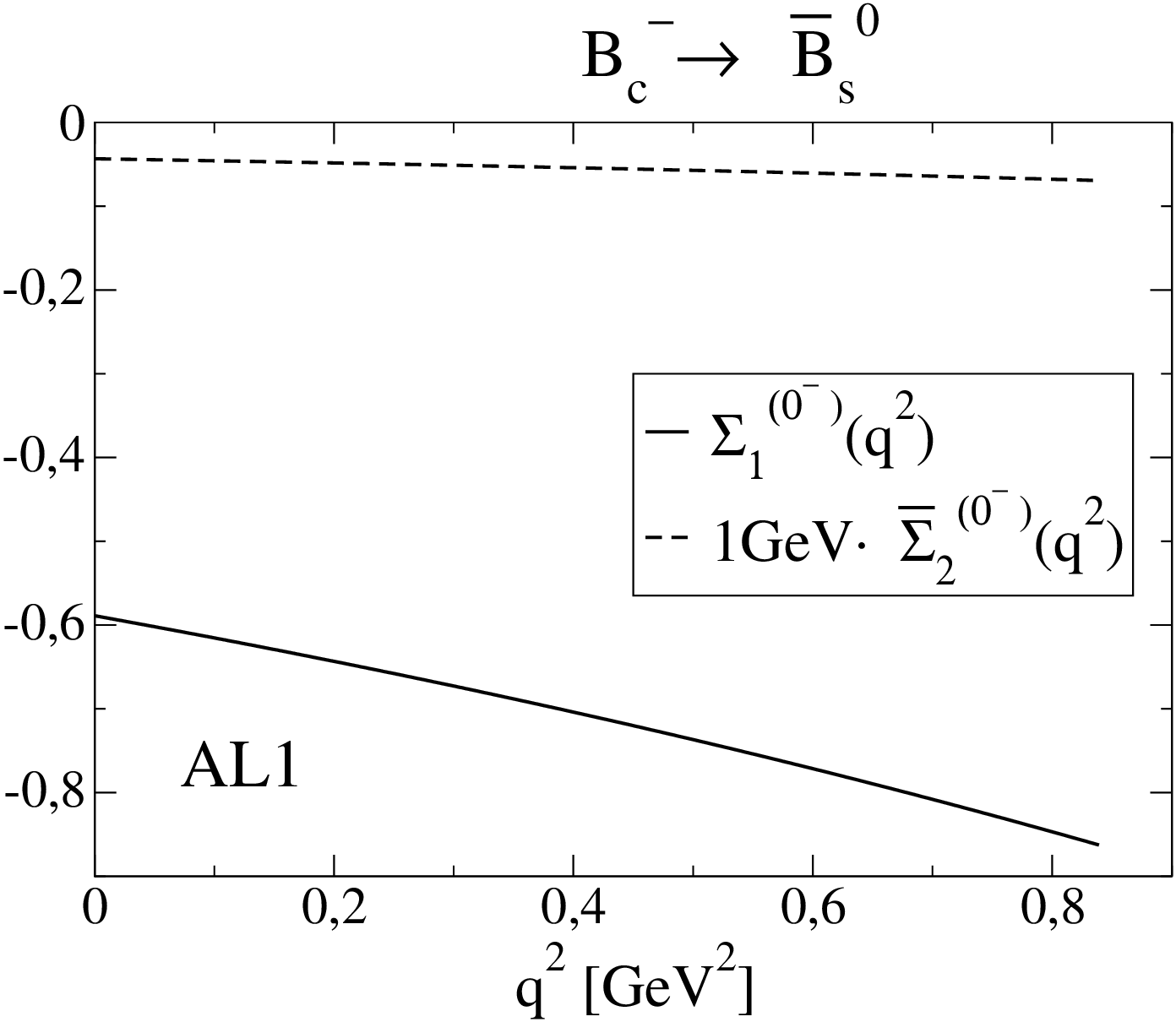}}\hspace{1cm}
\resizebox{8.cm}{6.2cm}{\includegraphics{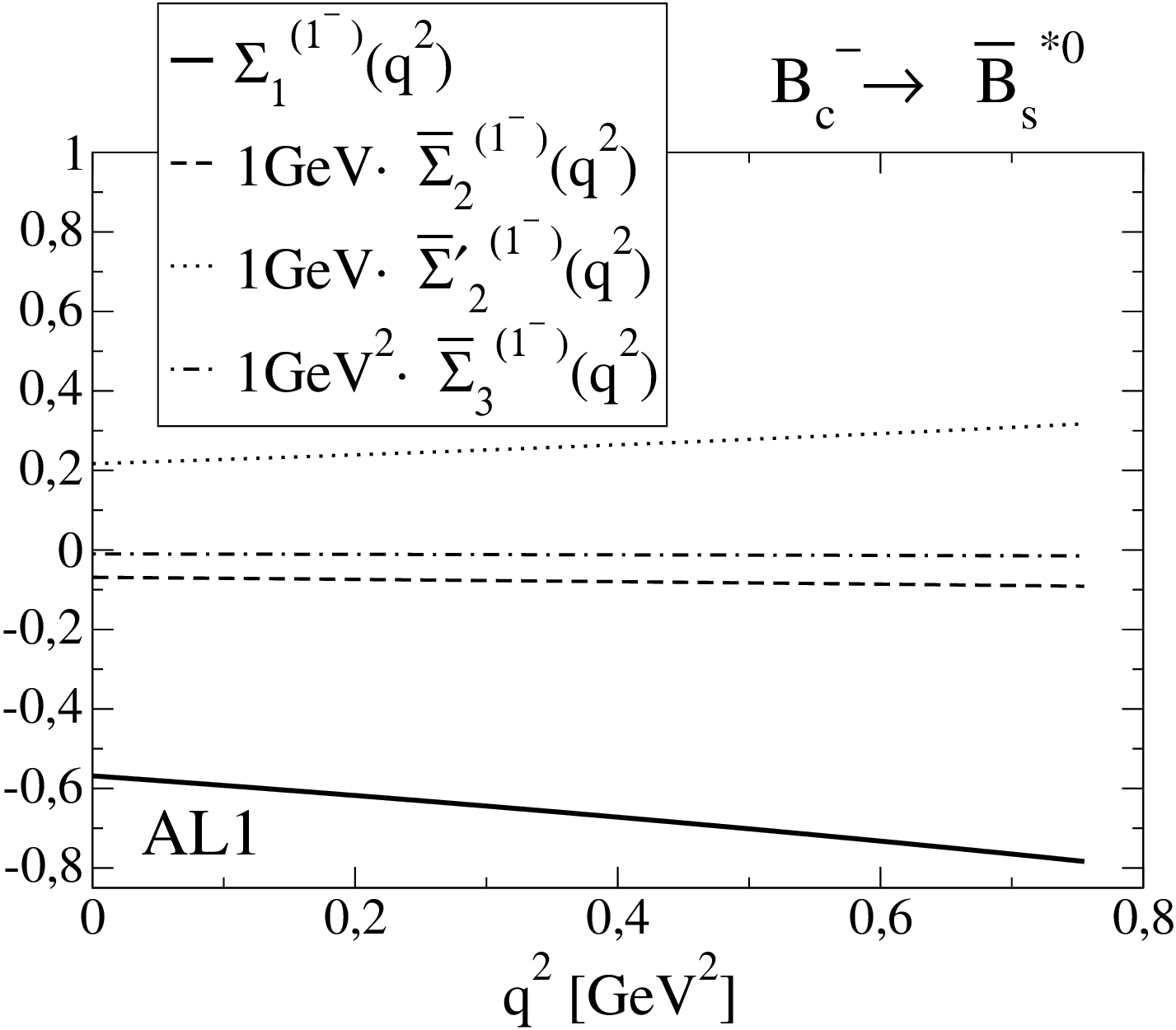}}
\caption{$\Sigma_1^{(0^-)}$ (solid line) and $\Sigma_2^{(0^-)}$ (dashed line) of the $B_c^-\to
\overline{B}^0$ and $B_c^-\to \overline{B}^0_s$ transitions, and
$\Sigma_1^{(1^-)}$ (solid line), $\overline{\Sigma}_2^{(1^-)}$ (dashed line),
$\overline{\Sigma}_2'^{(1^-)}$ (dotted line) and 
$\overline{\Sigma}_3^{(1^-)}$ (dashed dotted line) of
the $B_c^-\to \overline{B}^{*0}$ and $B_c^-\to \overline{B}^{*0}_s$
transitions  evaluated with the AL1 potential.  }
\label{fig:sigbtob}
\end{figure} 
In Fig.~\ref{fig:sigbtob} we give $\Sigma_1^{(0^-)}$ and
$\Sigma_2^{(0^-)}$ of the $B_c^-\to \overline{B}^0$ and $B_c^-\to
\overline{B}^0_s$ transitions, and $\Sigma_1^{(1^-)}$,
$\overline{\Sigma}_2^{(1^-)}$, $\overline{\Sigma}_2'^{(1^-)}$ and
$\overline{\Sigma}_3^{(1^-)}$ of the $B_c^-\to \overline{B}^{*0}$ and
$B_c^-\to \overline{B}^{*0}_s$ transitions.

We can take the infinite heavy quark mass limit on our analytic expressions
with the result that near zero recoil 
\begin{eqnarray}
\label{eq:sigbtob}
\Sigma_1^{(1^-)}&=&\Sigma_1^{(0^-)}\nonumber\\
\overline{\Sigma}_2^{(1^-)}&=&\overline{\Sigma}_2'^{(1^-)}
=-\overline{\Sigma}_2^{(0^-)}\nonumber\\
\overline{\Sigma}_3^{(1^-)}&=&0
\end{eqnarray}
When compared to the results of HQSS by Jenkins {\it et
 al.}~\cite{jenkins93} we see differences. In Ref.~\cite{jenkins93} we
 find\footnote{Note the different notation and global phases used.}
 $\overline{\Sigma}_2^{(1^-)}=\overline{\Sigma}_2^{(0^-)}$
 instead. This is wrong as there is a misprint in
 Ref.~\cite{jenkins93} that has not been noted before: the sign of the
 term in $v^\mu$ in the last expression of Eqs.~(2.9) and~(2.10) in
 Ref.~\cite{jenkins93} should be a minus~\cite{jenkins}. Also from
 Ref.~\cite{jenkins93} one would expect
\footnote{One would have to look at Eq.~(2.10) in
Ref.~\cite{jenkins93}, even though it refers to $B_c^+$ decay into
$D^0,\,D^{*0}$, because that is the reaction where you have antiquark
decay in their case.}  $\overline{\Sigma}_2
'^{(1^-)}=\overline{\Sigma}_2^{(0^-)}$ contradicting our result in
Eq.~(\ref{eq:sigbtob}) were we find $\overline{\Sigma}_2
'^{(1^-)}=-\overline{\Sigma}_2^{(0^-)}$.  Our result is a clear
prediction of the quark model and comes from the extra signs that
appear due to the fact that it is the antiquark that decays (See
appendix~\ref{app:sign}). This difference between quark
and antiquark decay was not properly reflected in their published work~\cite{jenkins}.
\begin{figure}[h!!!!]
\centering
\resizebox{8.cm}{7.cm}{\includegraphics{sig01btob.eps}}
\caption{$\Sigma_1^{(0^-)}$ of the semileptonic $B_c^-\to \overline{B}^0$
(solid line) and
$B_c^-\to \overline{B}_s^0$ (dotted line) transitions, and
$\Sigma_1^{(1^-)}$ of the semileptonic 
$B_c^-\to \overline{B}^{*0}$ (dashed line) and
$B_c^-\to \overline{B}_s^{*0}$ (dashed--dotted line) transitions
evaluated with the AL1 potential.}
\label{fig:sig1btob}
\end{figure} 

\begin{figure}[h!!]
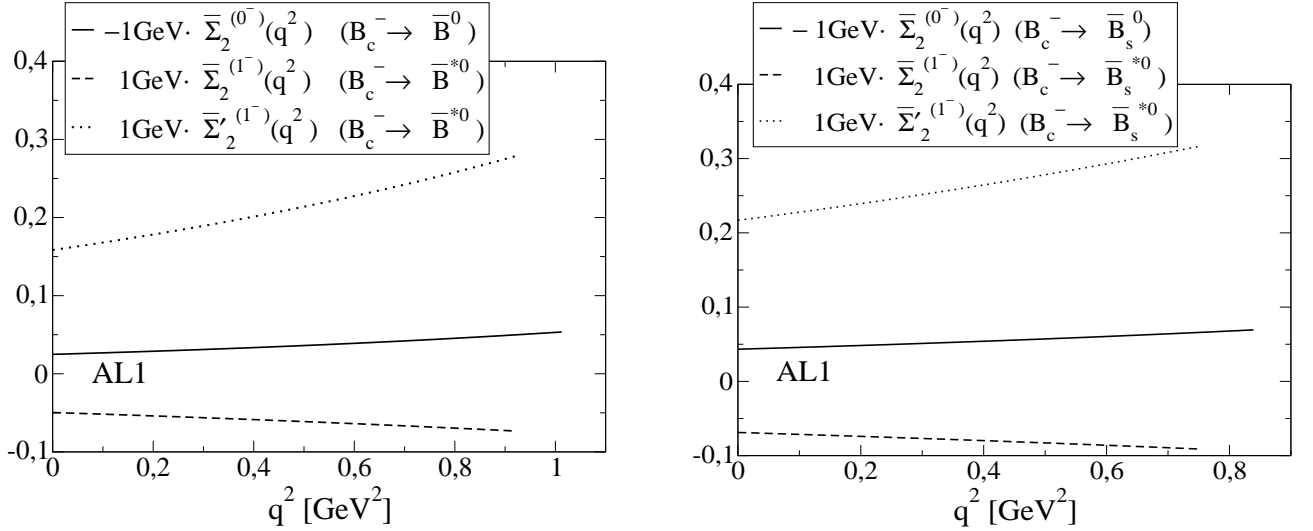

\centering
\resizebox{8.cm}{7.cm}{\includegraphics{sig2btob.eps}}\hspace{1cm}
\resizebox{8.cm}{7.cm}{\includegraphics{sig2btobs.eps}}\hspace{1cm}
\caption{$-\overline{\Sigma}_2^{(0^-)}$ (solid line) of the semileptonic $B_c^-\to \overline{B}^0$ and
$B_c^-\to \overline{B}_s^0$ transitions, and
$\overline{\Sigma}_2^{(1^-)}$ (dashed line), $\overline{\Sigma}_2'^{(1^-)}$
(dotted line) of the semileptonic $B_c^-\to \overline{B}^{*0}$ and
$B_c^-\to \overline{B}_s^{*0}$ transitions
evaluated with the AL1 potential.}
\label{fig:sig2btob}
\end{figure}%
How far are we from the infinite heavy quark mass limit?. In
Fig.~\ref{fig:sig1btob} we show $\Sigma_1^{(0^-)}$ of the semileptonic
$B_c^-\to \overline{B}^0$ and $B_c^-\to \overline{B}_s^0$ transitions,
and $\Sigma_1^{(1^-)}$ of the semileptonic $B_c^-\to
\overline{B}^{*0}$ and $B_c^-\to \overline{B}_s^{*0}$ transitions. The
differences between the corresponding $\Sigma_1^{(0^-)}$ and
$\Sigma_1^{(1^-)}$ are at the level of 10\%. The differences are much
more significant for $\overline{\Sigma}_2^{(0^-)}$,
$\overline{\Sigma}_2^{(1^-)}$ and $\overline{\Sigma}_2'^{(1^-)}$ that
we show in Fig.~\ref{fig:sig2btob}. In each case the three curves
shown would be the same in the infinite heavy quark mass limit.
Clearly in this case corrections on the inverse of the heavy quark
masses seem to be important.

\section{Nonleptonic $B_c^-\to {B}\, M_F$ two-meson decays}
\label{sect:nonlepb}
In this section we will evaluate decay widths for nonleptonic
 $B_c^-\to {B}\ M_F$ two--meson decays where $M_F$ is a pseudoscalar
 or vector meson with no $b$ quark content, and, at this point, ${B}$
 represents a meson with a $b$ quark.  These decay modes involve a
 $\bar c\to \bar d$ or $\bar c \to \bar s$ transition at the quark
 level and they are governed, neglecting penguin operators, by the
 effective Hamiltonian ~\cite{ivanov06,ebert03-2}
\begin{eqnarray}
H_{eff.}=\frac{G_F}{\sqrt2}\,\left\{ V_{cd}\left[
c_1(\mu)\,Q_1^{cd}+c_2(\mu)\,Q_2^{cd}
\right]
+V_{cs}\left[
c_1(\mu)\,Q_1^{cs}+c_2(\mu)\,Q_2^{cs}
\right]
+H.c.\right\}
\end{eqnarray}
where $c_1,\,c_2$ are scale--dependent Wilson coefficients, and
$Q_1^{cd},\,Q_2^{cd},\,Q_1^{cs},\,Q_2^{cs}$ are local four--quark 
operators given by
\begin{eqnarray}
Q_1^{cd}&=&\overline{\Psi}_c(0)\gamma_\mu(I-\gamma_5)\Psi_d(0)\,\bigg[
\ V_{ud}^*\, \overline{\Psi}_d(0)\gamma^\mu(I-\gamma_5)\Psi_u(0)+
V_{us}^*\, \overline{\Psi}_s(0)\gamma^\mu(I-\gamma_5)\Psi_u(0)
\bigg]\nonumber\\
Q_2^{cd}&=&\overline{\Psi}_c(0)\gamma_\mu(I-\gamma_5)\Psi_u(0)\,\bigg[
\ V_{ud}^*\, 
\overline{\Psi}_d(0)\gamma^\mu(I-\gamma_5)\Psi_d(0)
+V_{us}^*\, \overline{\Psi}_s(0)\gamma^\mu(I-\gamma_5)\Psi_d(0)
\bigg]\nonumber\\
Q_1^{cs}&=&\overline{\Psi}_c(0)\gamma_\mu(I-\gamma_5)\Psi_s(0)\,\bigg[
\ V_{ud}^*\, \overline{\Psi}_d(0)\gamma^\mu(I-\gamma_5)\Psi_u(0)+
V_{us}^*\, \overline{\Psi}_s(0)\gamma^\mu(I-\gamma_5)\Psi_u(0)
\bigg]\nonumber\\
Q_2^{cs}&=&\overline{\Psi}_c(0)\gamma_\mu(I-\gamma_5)\Psi_u(0)\,\bigg[
\ V_{ud}^*\, 
\overline{\Psi}_d(0)\gamma^\mu(I-\gamma_5)\Psi_s(0)
+V_{us}^*\, \overline{\Psi}_s(0)\gamma^\mu(I-\gamma_5)\Psi_s(0)\,
\bigg]
\end{eqnarray}

We shall work again in the factorization approximation taking into
account the Fierz reordered contribution so that the relevant
coefficients are not $c_1$ and $c_2$ but the combinations
\begin{eqnarray}
a_1(\mu)=c_1(\mu)+\frac{1}{N_C}\,c_2(\mu)\hspace{0.5cm};\hspace{0.5cm}
a_2(\mu)=c_2(\mu)+\frac{1}{N_C}\,c_1(\mu)
\end{eqnarray}
The energy scale $\mu$ appropriate in this
case is $\mu\simeq m_c$ and the values for $a_1$ and $a_2$ that we use
are~\cite{ivanov06}
\begin{eqnarray}
a_1=1.20\hspace{0.5cm};\hspace{0.5cm}
a_2=-0.317
\end{eqnarray}
\subsubsection{$M_F=\pi^-,\,\rho^-,\,K^-,\,K^{*-}$ }

\begin{table}[h]
\begin{tabular}{l|r}
\multicolumn{1}{l}{}&\multicolumn{1}{c}{$\Gamma$ [$10^{-15}$\,GeV]}\\
\multicolumn{1}{l}{}&\multicolumn{1}{c}{}\\
 & This work\\
\hline
$B_c^-\to\overline{B}^0\,\pi^-$ & $1.10^{+0.14}_{-0.16}\, a_1^2$\\
$B_c^-\to\overline{B}^0\,\rho^-$ & $1.41^{+0.12}_{-0.19}\, a_1^2$\\
$B_c^-\to\overline{B}^0\,K^-$ & $0.098^{+0.012}_{-0.012}\, a_1^2$\\
$B_c^-\to\overline{B}^0\,K^{*-}$ & $0.038^{+0.003}_{-0.005}\,  a_1^2$\\
\hline
$B_c^-\to \overline{B}^{*0}\,\pi^-$ &$0.71^{+0.12}_{-0.11}\, a_1^2$ \\
$B_c^-\to \overline{B}^{*0}\,\rho^-$ &$5.68^{+0.55}_{-0.77}\,1a_1^2$\\
$B_c^-\to \overline{B}^{*0}\,K^-$ &$0.047^{+0.007}_{-0.007}\,a_1^2$\\
$B_c^-\to \overline{B}^{*0}\,K^{*-}$ &$0.29^{+0.03}_{-0.04} \,a_1^2$\\
\hline
$B_c^-\to\overline{B}_s^0\,\pi^-$ & $34.7^{+2.0}_{-0.6}\, a_1^2$\\
$B_c^-\to\overline{B}_s^0\,\rho^-$ & $23.1^{+0.5}_{-0.6}\, a_1^2$\\
$B_c^-\to\overline{B}_s^0\,K^-$ & $2.87^{+0.13}_{-0.06}\, a_1^2$\\
$B_c^-\to\overline{B}_s^0\,K^{*-}$ & $0.13_{-0.01}\,  a_1^2$\\
\hline
$B_c^-\to \overline{B}_s^{*0}\,\pi^-$ &$22.8^{+2.2}_{-1.0}\, a_1^2$ \\
$B_c^-\to \overline{B}_s^{*0}\,\rho^-$ &$132^{+5}_{-6}\,a_1^2$\\
$B_c^-\to \overline{B}_s^{*0}\,K^-$ &$1.29^{+0.10}_{-0.06}\,a_1^2$\\
\end{tabular}\hspace{0.5cm}
\begin{tabular}{l|r c c c c c c c}
\multicolumn{1}{l}{}&\multicolumn{8}{c}{B.R. in \%}\\
\multicolumn{1}{l}{}&\multicolumn{8}{c}{}\\
 & This work&\cite{ivanov06}&\cite{ebert03-2}&\cite{chang94}&\cite{hady00}
&\cite{kiselev02} &\cite{colangelo00}
 &\cite{anisimov99}\\
\hline
$B_c^-\to\overline{B}^0\,\pi^-$ & $0.11^{+0.01}_{-0.01}$
&0.20&0.10&0.32&0.10&1.06&0.19&0.15\\
$B_c^-\to\overline{B}^0\,\rho^-$ & $0.14^{+0.02}_{-0.02}$
&0.20&0.13&0.59&0.28&0.96&0.15&0.19\\
$B_c^-\to\overline{B}^0\,K^-$ & $0.010^{+0.001}_{-0.001}$
&0.015&0.009&0.025&0.010&0.07&0.014&\\
$B_c^-\to\overline{B}^0\,K^{*-}$ &
$0.0039^{+0.0003}_{-0.0005}$
&0.0048&0.004&0.018&0.012&0.015&0.003&\\
\hline
$B_c^-\to \overline{B}^{*0}\,\pi^-$ &$0.072^{+0.012}_{-0.012}$
&0.057&0.026&0.29&0.076&0.95&0.24&0.077 \\
$B_c^-\to \overline{B}^{*0}\,\rho^-$ &$0.58^{+0.05}_{-0.08}$
&0.30&0.67&1.17&0.89&2.57&0.85&0.67\\
$B_c^-\to \overline{B}^{*0}\,K^-$
&$0.0048^{+0.0007}_{-0.0008}$
&0.0036&0.004&0.019&0.006&0.055&0.012&\\
$B_c^-\to \overline{B}^{*0}\,K^{*-}$
&$0.030^{+0.002}_{-0.004}$
&0.013&0.032&0.037&0.065&0.058&0.033&\\
\hline
$B_c^-\to\overline{B}_s^0\,\pi^-$ & $3.51^{+0.19}_{-0.06}$
&3.9&2.46&5.75&1.56&16.4&3.01&3.42\\
$B_c^-\to\overline{B}_s^0\,\rho^-$ & $2.34^{+0.05}_{-0.06}$
&2.3&1.38&4.41&3.86&7.2&1.34&2.33\\
$B_c^-\to\overline{B}_s^0\,K^-$ & $0.29^{+0.01}_{-0.01}$
&0.29&0.21&0.41&0.17&1.06&0.21&\\
$B_c^-\to\overline{B}_s^0\,K^{*-}$ & $0.013_{-0.001}$
&0.011&0.0030&&0.10&&0.0043&\\
\hline
$B_c^-\to \overline{B}_s^{*0}\,\pi^-$ &$2.34^{+0.19}_{-0.14}$
&2.1&1.58&5.08&1.23&6.5&3.50&1.95 \\
$B_c^-\to \overline{B}_s^{*0}\,\rho^-$ &$13.4^{+0.5}_{-0.6}$
&11&10.8&14.8&16.8&20.2&10.8&12.1\\
$B_c^-\to \overline{B}_s^{*0}\,K^-$ &$0.13^{+0.01}_{-0.01}$
&0.13&0.11&0.29&0.13&0.37&0.16&\\

\end{tabular}
\caption{Decay widths in units of
$10^{-15}$\,GeV, and for general values of the Wilson coefficient $a_1$, and
branching ratios in \%
for exclusive nonleptonic decays of the $B_c^-$ meson. Our central values have
been obtained with the AL1 potential. }
\label{tab:dwbrbtobpik}
\end{table}

In this case $B$ denotes one of the 
$\overline{B}^{0},\,\overline{B}^{*0},\,\overline{B}_s^{0},\,\overline{B}_s^{*0}$.
The decay widths  are
\begin{eqnarray}
&&\Gamma=\frac{G_F^2}{16\pi m^2_{B_c}}\, |V_{cd}|^2\,|V_{F}|^2\ a_1^2\,
\frac{\lambda^{1/2}(m^2_{B_c},\,m^2_{\overline{B}^{0},\,\overline{B}^{*0}},
\,m^2_F)}{2m_{B_c}}\,
m_F^2\,f^2_F\, {\cal H}_{tt}^{B_c^-\to\overline{B}^{0},\,\overline{B}^{*0}}
(m_F^2)\hspace{1.cm}M_F\equiv 0^-\ \mathrm{case}\nonumber\\
&&\Gamma=\frac{G_F^2}{16\pi m^2_{B_c}}\, |V_{cd}|^2\,|V_{F}|^2\ a_1^2\
\frac{\lambda^{1/2}(m^2_{B_c},\,m^2_{\overline{B}^{0},\,
\overline{B}^{*0}},\,m^2_F)}{2m_{B_c}}\,
m_F^2\,f^2_F
\nonumber\\
 &&\hspace{2cm}\times\left({\cal H}_{+1+1}^{B_c^-\to\overline{B}^{0},\,\overline{B}^{*0}}(m_F^2)
 +{\cal H}_{-1-1}^{B_c^-\to\overline{B}^{0},\,\overline{B}^{*0}}(m_F^2)
 +{\cal H}_{00}^{B_c^-\to\overline{B}^{0},\,\overline{B}^{*0}}(m_F^2)
\right)\hspace{1.cm}M_F\equiv1^-\ \mathrm{case}
\label{eq:dwb}
\end{eqnarray}
and similarly for $\overline{B}^{0}_s,\,\overline{B}^{*0}_s$ with
$|V_{cd}|\to|V_{cs}|;\,\overline{B}^{0},\,\overline{B}^{*0}\to
\overline{B}^{0}_s,\,\overline{B}^{*0}_s$.  $V_F=V_{ud}$ or
$V_F=V_{us}$ depending on whether $M_F=\pi^-,\,\rho^-$ or
$M_F=K^-,\,K^{*-}$, $f_F$ is the decay constant of the $M_F$ meson, and
 the different ${\cal H}_{rr}$ have been evaluated at $q^2=m_F^2$.  In
Table~\ref{tab:dwbrbtobpik} we show the decay widths for a general
value of the Wilson coefficient $a_1$, and the corresponding branching
ratios evaluated with $a_1=1.20$. The transition
$B_c^-\to\overline{B}^{*0}_s K^{*-}$ is not allowed with the new $B_c$
mass value from Ref.~\cite{cdf06-1}. Our branching ratios
for a final $\overline{B}^0_s$
or $\overline{B}^{*0}_s$ are in very good agreement with the results by
Ivanov {\it et al.}~\cite{ivanov06}, while for a final $\overline{B}^0$ or
$\overline{B}^{*0}$
we are in very good agreement with the results by Ebert {\it et
al.}~\cite{ebert03-2} (with the exception of the $B_c\to
\overline{B}^{*0}\pi^-$ decay).

\subsubsection{$M_F=\pi^0,\,\rho^0,\,K^0,\,K^{*0}$}
Here the generic name ${B}$ stands for a ${B}^-$ or a ${B}^{*-}$
meson. The different decay widths are given by
\begin{eqnarray}
&&\Gamma=\frac{G_F^2}{16\pi m^2_{B_c}}\, |V_{ud}|^2\,|V_{F}|^2\ a_2^2\,
\frac{\lambda^{1/2}(m^2_{B_c},\,m^2_{ B^{-},\, B^{*-}},
\,m^2_F)}{2m_{B_c}}\,
m_F^2\,\widetilde{f}^2_F\, {\cal H}_{tt}^{B_c^-\to {B}^{-},\,{B}^{*-}}
(m_F^2)\hspace{1.cm}M_F\equiv 0^-\ \mathrm{case}\nonumber\\
&&\Gamma=\frac{G_F^2}{16\pi m^2_{B_c}}\, |V_{ud}|^2\,|V_{F}|^2\ a_2^2\
\frac{\lambda^{1/2}(m^2_{B_c},\,m^2_{{B}^{-},\,{B}^{*-}},
\,m^2_F)}{2m_{B_c}}\,
m_F^2\,\widetilde{f}^2_F
\nonumber\\
 &&\hspace{2cm}\times\left({\cal H}_{+1+1}^{B_c^-\to {B}^{-},\,{B}^{*-}}(m_F^2)
 +{\cal H}_{-1-1}^{B_c^-\to {B}^{-},\,{B}^{*-}}(m_F^2)
 +{\cal H}_{00}^{B_c^-\to {B}^{-},\,{B}^{*-}}(m_F^2)
\right)\hspace{1.cm}M_F\equiv1^-\ \mathrm{case}
\label{eq:dwcc2m}
\end{eqnarray}
where $V_F=V_{cd}$ or $V_F=V_{cs}$ depending on whether
$M_F=\pi^0,\,\rho^0$ or $M_F=K^0,\,K^{*0}$, $\widetilde{f}_F=f_F$ for
$M_F=K^0,\,K^{*0}$ whereas $\widetilde{f}_F=f_F/\sqrt2$ for
$M_F=\pi^0,\,\rho^0$, with $f_F$ the $M_F$ meson decay constant, and
the different ${\cal H}_{rr}$  evaluated at $q^2=m_F^2$. The latter
 have been obtained
from the matrix elements for  the  effective current operator
$\overline{\Psi}_{c}(0)\gamma^\mu(I-\gamma_5)\Psi_u(0)$.
The decay
widths, for a general value of the Wilson coefficient $a_2$, and the
corresponding branching ratios are shown in
Table~\ref{tab:dwbrbtobm}. With the exception of the $B_c^-\to B^{*-}\pi^0$
case, our results are in a global  good agreement with
the ones by Ebert {\it et al.}~\cite{ebert03-2}.

\begin{table}[h]
\begin{tabular}{l|r}
\multicolumn{1}{l}{}&\multicolumn{1}{c}{$\Gamma$ [$10^{-15}$\,GeV]}\\
\multicolumn{1}{l}{}&\multicolumn{1}{c}{}\\
 & This work\\
\hline
$B_c^-\to{B}^-\,\pi^0$ & $0.54^{+0.07}_{-0.07}\, a_2^2$\\
$B_c^-\to{B}^-\,\rho^0$ & $0.71^{+0.06}_{-0.10}\, a_2^2$\\
$B_c^-\to{B}^-\,K^0$ & $35.3^{+4.0}_{-4.9}\, a_2^2$\\
$B_c^-\to{B}^-\,K^{*0}$ & $13.1^{+0.9}_{-0.7}\,  a_2^2$\\
\hline
$B_c\to {B}^{*-}\,\pi^0$ &$0.35^{+0.06}_{-0.05}\, a_2^2$ \\
$B_c\to {B}^{*-}\,\rho^0$ &$2.84^{+0.27}_{-0.39}\,a_2^2$\\
$B_c\to {B}^{*-}\,K^0$ &$16.9^{+2.4}_{-2.7}\,a_2^2$\\
$B_c\to {B}^{*-}\,K^{*0}$ &$103^{+8}_{-13} \,a_2^2$\\

\end{tabular}\hspace{1cm}
\begin{tabular}{l|r c c c  c c c}
\multicolumn{1}{l}{}&\multicolumn{7}{c}{B.R. in \%}\\
\multicolumn{1}{l}{}&\multicolumn{7}{c}{}\\

 & This work&\cite{ivanov06}&\cite{ebert03-2}&\cite{chang94}&\cite{hady00}
&\cite{kiselev02} &\cite{anisimov99}\\
 \hline
$B_c^-\to{B}^-\,\pi^0$ & $0.0038^{+0.0005}_{-0.0006}$
&0.007&0.004&0.011&0.004&0.037&0.007\\
$B_c^-\to{B}^-\,\rho^0$ & $0.0050^{+0.0004}_{-0.0007}$
&0.0071&0.005&0.020&0.010&0.034&0.009\\
$B_c^-\to{B}^-\,K^0$ & $0.25^{+0.03}_{-0.04}$
&0.38  &0.24&0.66 &0.27&1.98 &0.17 \\
$B_c^-\to{B}^-\,K^{*0}$ &$0.093^{+0.006}_{-0.013}$
&0.11 &0.09 &0.47&0.32 &0.43 &0.095\\
\hline
$B_c\to {B}^{*-}\,\pi^0$ &$0.0025^{+0.0004}_{-0.0005}$
&0.0020&0.001&0.010&0.003&0.033&0.004 \\
$B_c\to {B}^{*-}\,\rho^0$ &$0.020^{+0.002}_{-0.003}$
&0.011 &0.024&0.041&0.031&0.09&0.031\\
$B_c\to {B}^{*-}\,K^0$&$0.12^{+0.02}_{-0.02}$
&0.088 &0.11&0.50 &0.16 &1.60&0.061\\
$B_c\to {B}^{*-}\,K^{*0}$&$0.73^{+0.06}_{-0.10}$
&0.32 &0.84 &0.97 &1.70 &1.67&0.57\\
\end{tabular}
\caption{Decay widths in units of
$10^{-15}$\,GeV, and for general values of the Wilson coefficient $a_2$, and
branching ratios in \%
for exclusive nonleptonic decays of the $B_c^-$ meson. Our central values have
been obtained with the AL1 potential. }
\label{tab:dwbrbtobm}
\end{table}

%
%
%
%
\section{Summary}
\label{sect:summary}
We have made a comprehensive and exhaustive study of exclusive
semileptonic and nonleptonic two--meson decays of the $B_c$ meson
within a nonrelativistic quark model. We have left out semileptonic 
processes involving  a $b\to u$ transition at the quark level to avoid known
deficiencies both at high and low $q^2$
transfers~\cite{albertus05-2}. For similar reasons we have only considered
two--meson nonleptonic decay channels that include  a $c\bar c$ or $B$
meson.  Our model respects 
HQSS constraints in the infinite heavy quark mass limit but hints at
sizeable corrections away from that limit for some form
factors. Unfortunately such corrections have not been worked out in
perturbative QCD as they have for heavy--light
mesons~\cite{neubert92}.

 To check the sensitivity of our results
to the inter--quark interaction we have used five different
quark--quark potentials.  Most observables change only at the level of
a few per cent when changing the interaction.  There is another source 
of theoretical uncertainty in the  use of 
nonrelativistic kinematics in the evaluation of the orbital wave
functions and the construction of our states in Eq.(\ref{wf}).
While this is a very good approximation for the $B_c$ itself it is not 
necessarily so for mesons with a light quark. We nevertheless think that, 
to a certain extent, the ignored relativistic
effects are contained in an effective way in the free parameters of the
inter-quark interaction, which are fitted to experimental data.

Our  results for the observables analyzed are in a
general good agreement (whenever comparison is possible)
 with the results obtained within the quasi-potential approach to the
relativistic quark model of Ebert {\it et al.}~\cite{ebert03,ebert03-2}. 

The branching ratios for the leptonic $B_c^-\to c\bar c$ and $B_c^-\to\overline{B}$ 
 decays are also in reasonable  agreement with the relativistic constituent 
quark model results of
Ivanov {\it et al.}~\cite{ivanov05,ivanov06}.

For the nonleptonic $B_c^-\to \eta_c\,M_F^-$ and $B_c^-\to
J/\Psi\,M_F^-$ two--meson decay channels with
$M_F^-=\pi^-,\,\rho^-,\,K^-,\,K^{*-}$, we find also reasonable
agreement (better for the $J/\Psi$ channel) with the Bethe--Salpeter calculation
by El-Hady {\it et
al.}~\cite{hady00} and the light front calculation by 
Anisimov {\it et al.}~\cite{anisimov99},
while our results are a factor of two smaller than the ones by Ivanov
{\it et al.}~\cite{ivanov06}, and Chang {\it et al.}
~\cite{chang94,chang02,chang01}, the latter obtained  within the
nonrelativistic approach to the Bethe--Salpeter equation. For the two--meson decay channels
with $\chi_{c0},\,\chi_{c1},\,h_c,\,\chi_{c2}$ or $\Psi(3836)$ as the
final $c\bar c$ meson and $M_F=\pi^-,\,\rho^-,\,K^-,\,K^{*-}$ our
results are generally a factor of two smaller than the ones of Ivanov
{\it et al.}~\cite{ivanov06}, whereas for some channels
($\chi_{c0}\,\pi^-$, $\chi_{c0}\,\rho^-$, $h_c\,\pi^-$, $h_c\,\rho^-$,
$\chi_{c2}\,\pi^-$ and $\chi_{c2}\,\rho^-$,) we find very good
agreement with the results by Chang {\it et
al.}~\cite{chang94,chang02,chang01}. The disagreement with Ivanov {\it
et al.}  extend to the two-meson decay channels with a final $c\bar c$
and $D$ mesons. There we find good agreement with the results by
El-Hady {\it et al.}~\cite{hady00} and the ones obtained within the 
sum rules of QCD and nonrelativistic QCD by Kiselev~\cite{kiselev02}.

As for the two--meson nonleptonic decay channels $B_c^-\to
\overline{B}\, M_F$ with $M_F=\pi^-,\,\rho^-,\,K^-,\,K^{*-}$ we are in a
reasonable good agreement with the results by Anisimov {\it et al.}~\cite{anisimov99}.
For the case of a final $\overline{B}^{0}_s$ or $\overline{B}^{*0}_s$ 
the agreement with the results by Ivanov {\it et al.}~\cite{ivanov06} is 
very good. For the $B_c^-\to B^-\
M_F$ case with $M_F=\pi^0,\,\rho^0,\,K^0,\,K^{*0}$, and apart from the results 
by Ebert {\it et al.}~\cite{ebert03-2}, we find no global agreement with other 
calculations, neither do they agree with each other.

From the above comparison one sees  that there are
different models producing sometimes very different results for the same
observables. Accurate experimental data will shed light into this issue.
\begin{acknowledgments}
We thank J.G. K\"orner for a critical reading  of our
work. We also thank A.V. Manohar for discussions on the form factor
decomposition in the the infinite heavy quark
mass limit.
 This research was supported by DGI and FEDER funds, under contracts
FIS2005-00810, BFM2003-00856 and FPA2004-05616, by Junta de
Andaluc\'\i a and Junta de Castilla y Le\'on under contracts FQM0225
and SA104/04, and it is part of the EU integrated infrastructure
initiative Hadron Physics Project under contract number
RII3-CT-2004-506078.  J. M. V.-V. acknowledges an E.P.I.F. contract with the
University of Salamanca.
\end{acknowledgments}
%
%
\appendix
\section{$\varepsilon_{(\lambda)}^\mu(\vec P)$ Polarization vectors}
\label{app:epsilon}
Different sets of polarization vectors  used in this paper:
\begin{itemize}
\item[] \underline{$\vec P=\vec 0$}\\
\begin{eqnarray}
\mathrm{Spin\ or\ heliciy\ bases}\ \left\{\begin{array}{l}
\varepsilon_{(+1)}^\mu(\vec
P\,)=(0,-\frac{1}{\sqrt2},-\frac{i}{\sqrt2},0)\\
\\
\varepsilon_{(-1)}^\mu(\vec
P\,)=(0,\frac{1}{\sqrt2},-\frac{i}{\sqrt2},0)\\
\\
\varepsilon_{(0)}^\mu(\vec
P\,)\hspace{.25cm}=(0,0,0,1)
\end{array}\right.
\end{eqnarray}
\item[] \underline{$\vec P=|\vec P| \vec k$}
\begin{eqnarray}
\mathrm{Spin\ or\ heliciy\ bases}\ \left\{\begin{array}{l}
\varepsilon_{(+1)}^\mu(\vec
P\,)=(0,-\frac{1}{\sqrt2},-\frac{i}{\sqrt2},0)\\
\\
\varepsilon_{(-1)}^\mu(\vec
P\,)=(0,\frac{1}{\sqrt2},-\frac{i}{\sqrt2},0)\\
\\
\varepsilon_{(0)}^\mu(\vec
P\,)\hspace{.25cm}=(\frac{|\vec P|}{m},0,0,\frac{E(\vec P)}{m})
\end{array}\right.
\end{eqnarray}
\item[] \underline{$\vec P=-|\vec P| \vec k$}
\begin{eqnarray}
\mathrm{Spin \ base}\ \left\{ \begin{array}{l}\varepsilon_{(+1)}^\mu(\vec
P\,)=(0,-\frac{1}{\sqrt2},-\frac{i}{\sqrt2},0)\\
\\
\varepsilon_{(-1)}^\mu(\vec
P\,)=(0,\frac{1}{\sqrt2},-\frac{i}{\sqrt2},0)\\
\\
\varepsilon_{(0)}^\mu(\vec
P\,)\hspace{.25cm}=(-\frac{|\vec P|}{m},0,0,\frac{E(\vec P)}{m})
\end{array}\right.\hspace{1cm}
\mathrm{Helicity \ base}\ \left\{ \begin{array}{l}\varepsilon_{(+1)}^\mu(\vec
P\,)=(0,\frac{1}{\sqrt2},-\frac{i}{\sqrt2},0)\\
\\
\varepsilon_{(-1)}^\mu(\vec
P\,)=(0,-\frac{1}{\sqrt2},-\frac{i}{\sqrt2},0)\\
\\
\varepsilon_{(0)}^\mu(\vec
P\,)\hspace{.25cm}=(\frac{|\vec P|}{m},0,0,-\frac{E(\vec P)}{m})
\end{array}\right.
\end{eqnarray}

\end{itemize}

%
%
%
\section{Expression for the $V^\mu(|\vec{q}\,|),\,V^\mu_{(\lambda)}(|\vec{q}\,|)
,\,V^\mu_{T(\lambda)}(|\vec{q}\,|)$ and 
$A^\mu(|\vec{q}\,|),\,A^\mu_{(\lambda)}(|\vec{q}\,|)
,\,A^\mu_{T(\lambda)}(|\vec{q}\,|)$ 
matrix elements}
\label{app:va}
Here we give general expressions valid for transitions between a pseudoscalar
meson $M_I$ at rest with quark content $q_{f_1}\overline{q}_{f_2}$ and a final 
$M_F$ meson with total angular momentum and parity $J^\pi=0^-,0^+,1^-,1^+,2^-,2^+$,
three-momentum $-|\vec{q}\,|\vec{k}$ and  quark content
$q_{f'_1}\overline{q}_{f_2}$. In the  transition it is the quark that
changes flavor. The phases of the wave functions are the ones chosen in
Eqs.~(\ref{eq:0pm},\ref{eq:1pm},\ref{eq:2pm}).
We generally have
\begin{eqnarray}
{\cal V}^\mu(|\vec{q}\,|)-{\cal A}^\mu(|\vec{q}\,|)&=&\sqrt{2m_{I}2E_{F}(-\vec{q\,})}\
{}_{\stackrel{}{\stackrel{}{NR}}}
\left\langle\, M_F(J^\pi),\,\lambda\,
-|\vec{q}\,|\,\vec{k}\,\bigg|\, J^{f'_1\,f_1\, \mu}(0)\,
\bigg| \, M_I(0^-),\,\vec{0}\right\rangle_{NR}\nonumber\\
&=&\sqrt{2m_{I}2E_{F}(-\vec{q\,})}\ \int\,d^3p\sum_{s'_1}\sum_{s_1,s_2}
\left(\hat{\phi}^{(M_F(J^\pi),\lambda)}_{(s'_1,f'_1),\,(s_2,f_2)}(\vec{p}\,)\right)^*\ 
\hat{\phi}^{(M_I(0^-))}_{(s_1,f_1),\,(s_2,f_2)}(\vec{p}
-\frac{m_{f_2}}{m_{f'_1}+m_{f_2}}\,|\vec{q}\,|\vec{k})
\nonumber\\
&&\hspace{1cm}\frac{1}
{\sqrt{2E_{f'_1}\,2E_{f_1}}}\ \bar{u}_{s'_1,f'_1}(-\frac{m_{f'_1}}{m_{f'_1}+m_{f_2}}
|\vec{q}\,|\vec{k}-\vec{p}\,)\, 
\gamma^\mu(I-\gamma_5)\,u_{s_1,f_1}(\frac{m_{f_2}}{m_{f'_1}+m_{f_2}}
|\vec{q}\,|\vec{k}-\vec{p}\,)\nonumber\\
\end{eqnarray}
where ${\cal V}^\mu ({\cal A}^\mu)$ represent any of the $V^\mu\, (A^\mu)$,
$V^\mu_{(\lambda)}\, (A^\mu_{(\lambda)})$ or
$V^\mu_{T(\lambda)}\, (A^\mu_{T(\lambda)})$,
and where $E_{f'_1}$ and $E_{f_1}$ are shorthand notations for
$E_{f'_1}(-\frac{m_{f'_1}}{m_{f'_1}+m_{f_2}}
|\vec{q}\,|\vec{k}-\vec{p}\,)$ and $E_{f_1}(\frac{m_{f_2}}{m_{f'_1}+m_{f_2}}
|\vec{q}\,|\vec{k}-\vec{p}\,)$ respectively. Defining also
$\widehat{E}_{f'_1}=E_{f'_1}+m_{f'_1}$ and $\widehat{E}_{f_1}=E_{f_1}+m_{f_1}$
 we arrive at the following final expressions:\\
 \begin{itemize}
 \item {\large Case $J^\pi=0^-$}
 \end{itemize}
\begin{eqnarray}
V^0(|\vec{q}\,|)&=&\sqrt{2m_I2E_F(-\vec{q}\,)}\ \int\,d^3p\ \frac{1}{4\pi}
\left(\hat{\phi}^{(M_F(0^-))}_{f'_1,\,f_2}(|\vec{p}\,|)\right)^*\,
\hat{\phi}^{(M_I(0^-))}_{f_1,\,f_2}\left(\bigg|\,\vec{p}-\frac{m_{f_2}}{m_{f'_1}+m_{f_2}}
|\vec{q}\,|
\vec{k} \bigg|\right)\nonumber\\
&&\hspace{3cm}\sqrt{\frac{\widehat{E}_{f'_1}\widehat{E}_{f1}}{4E_{f'_1}
E_{f_1}}}
\left(
1+\frac{(-\frac{m_{f'_1}}{m_{f'_1}+m_{f_2}}\,|\vec{q}\,|\vec{k}-\vec{p}\,)
\cdot(\frac{m_{f_2}}{m_{f'_1}+m_{f_2}}\,|\vec{q}\,|\vec{k}-\vec{p}\,)}{\widehat{E}_{f'_1}\widehat{E}_{f_1}}
\right) \nonumber\\
V^3(|\vec{q}\,|)&=&\sqrt{2m_I2E_F(-\vec{q}\,)}\ \int\,d^3p\ \frac{1}{4\pi}
\left(\hat{\phi}^{(M_F(0^-))}_{f'_1,\,f_2}(|\vec{p}\,|)\right)^*\,
 \hat{\phi}^{(M_I
 (0^-))}_{f_1,\,f_2}\left(\bigg|\,\vec{p}-\frac{m_{f_2}}{m_{f'_1}+m_{f_2}}
|\vec{q}\,|
\vec{k} \bigg|\right)
 \nonumber\\
&&\hspace{3cm}\sqrt{\frac{\widehat{E}_{f'_1}\widehat{E}_{f1}}{4E_{f'_1}
E_{f_1}}}
\left(\frac{\frac{m_{f_2}}{m_{f'_1}+m_{f_2}}\,|\vec{q}\,|-p_z}{\widehat{E}_{f_1}}+
\frac{-\frac{m_{f'_1}}{m_{f'_1}+m_{f_2}}\,|\vec{q}\,|-p_z}{\widehat{E}_{f'_1}}
\right) \nonumber\\
\end{eqnarray}
\begin{itemize}
\item {\large Case $J^\pi= 0^+$}
\end{itemize}
\begin{eqnarray}
A^0(|\vec{q}\,|)&=&\sqrt{2m_I2E_F(-\vec{q}\,)}\ \int\,d^3p\ \frac{1}{4\pi
|\vec{p}\,|}
\left(\hat{\phi}^{(M_F(0^+))}_{f'_1,\,f_2}(|\vec{p}\,|)\right)^*\,
\hat{\phi}^{(M_I
(0^-))}_{f_1,\,f_2}\left(\bigg|\,\vec{p}-\frac{m_{f_2}}{m_{f'_1}+m_{f_2}}
|\vec{q}\,|
\vec{k} \bigg|\right)\nonumber\\
&&\hspace{1cm}\sqrt{\frac{\widehat{E}_{f'_1}\widehat{E}_{f1}}{4E_{f'_1}
E_{f_1}}}
\left(\frac{\vec{p}\cdot(\frac{m_{f_2}}{m_{f'_1}+m_{f_2}}\,|\vec{q}\,|\vec{k}-\vec{p}\,)}{\widehat{E}_{f_1}}
+\frac{\vec{p}\cdot(-\frac{m_{f'_1}}{m_{f'_1}+m_{f_2}}\,|\vec{q}\,|\vec{k}-\vec{p}\,)}
{\widehat{E}_{f'_1}}
\right) \nonumber\\
A^3(|\vec{q}\,|)&=&\sqrt{2m_I2E_F(-\vec{q}\,)}\ \int\,d^3p\
\frac{1}{4\pi|\vec{p}\,|}
\left(\hat{\phi}^{(M_F (0^+))}_{f'_1,\,f_2}(|\vec{p}\,|)\right)^*\,
\hat{\phi}^{(M_I(0^-))}_{f_1,\,f_2}\left(\bigg|\,\vec{p}-\frac{m_{f_2}}{m_{f'_1}+m_{f_2}}
|\vec{q}\,|
\vec{k} \bigg|\right)
 \nonumber\\
&&\hspace{1cm}\sqrt{\frac{\widehat{E}_{f'_1}\widehat{E}_{f1}}{4E_{f'_1}
E_{f_1}}}\bigg\{\
p_z\bigg(  
1-\frac{(-\frac{m_{f'_1}}{m_{f'_1}+m_{f_2}}\,|\vec{q}\,|\vec{k}-\vec{p}\,)
\cdot(\frac{m_{f_2}}{m_{f'_1}+m_{f_2}}\,|\vec{q}\,|\vec{k}-\vec{p}\,)}{\widehat{E}_{f'_1}\widehat{E}_{f_1}}
\bigg)\nonumber\\ 
&&\hspace{3.cm}+\frac{1}{\widehat{E}_{f'_1}\widehat{E}_{f1}}
\bigg[\hspace{.35cm} (-\frac{m_{f'_1}}{m_{f'_1}+m_{f_2}}|\vec{q}\,|-p_z)\ \ \
\vec{p}\cdot\bigg(\hspace{.4cm}\frac{m_{f_2}}{m_{f'_1}+m_{f_2}}|\vec{q}\,|\vec{k}-\vec{p}
\bigg)\nonumber\\
&&\hspace{4.75cm}+(\hspace{.4cm} \frac{m_{f_2}}{m_{f'_1}+m_{f_2}}|\vec{q}\,|-p_z
)\ \ \
\vec{p}\cdot\bigg(-\frac{m_{f'_1}}{m_{f'_1}+m_{f_2}}|\vec{q}\,|\vec{k}-\vec{p}
\bigg)
\bigg]\bigg\}\nonumber\\
\end{eqnarray}

\begin{itemize}
\item {\large Case $J^\pi=1^-$}
\end{itemize}

\begin{eqnarray}
V^{(1^-)\,1}_{\lambda=-1}(|\vec{q}\,|)&=&\frac{-i}{\sqrt2}
\sqrt{2m_I2E_F(-\vec{q}\,)}\ \int\,d^3p\ \frac{1}{4\pi}
\left(\hat{\phi}^{(M_F(1^-))}_{f'_1,\,f_2}(|\vec{p}\,|)\right)^*\,
\hat{\phi}^{(M_I(0^-))}_{f_1,\,f_2}\left(\bigg|\,\vec{p}-\frac{m_{f_2}}{m_{f'_1}+m_{f_2}}
|\vec{q}\,|
\vec{k} \bigg|\right)\nonumber\\
&&\hspace{3cm}\sqrt{\frac{\widehat{E}_{f'_1}\widehat{E}_{f1}}{4E_{f'_1}
E_{f_1}}}
\left(-\frac{\frac{m_{f_2}}{m_{f'_1}+m_{f_2}}\,|\vec{q}\,|-p_z}{\widehat{E}_{f_1}}+
\frac{-\frac{m_{f'_1}}{m_{f'_1}+m_{f_2}}\,|\vec{q}\,|-p_z}{\widehat{E}_{f'_1}}
\right)\nonumber\\
\end{eqnarray}
and similarly
\begin{eqnarray}
A^{(1^-)\,0}_{\lambda=0}(|\vec{q}\,|)&=&i\
\sqrt{2m_I2E_F(-\vec{q}\,)}\ \int\,d^3p\ \frac{1}{4\pi}
\left(\hat{\phi}^{(M_F(1^-))}_{f'_1,\,f_2}(|\vec{p}\,|)\right)^*\,
\hat{\phi}^{(M_I(0^-))}_{f_1,\,f_2}\left(\bigg|\,\vec{p}-\frac{m_{f_2}}{m_{f'_1}+m_{f_2}}
|\vec{q}\,|
\vec{k} \bigg|\right)\nonumber\\
&&\hspace{3cm}\sqrt{\frac{\widehat{E}_{f'_1}\widehat{E}_{f1}}{4E_{f'_1}
E_{f_1}}}
\left(\frac{\frac{m_{f_2}}{m_{f'_1}+m_{f_2}}\,|\vec{q}\,|-p_z}{\widehat{E}_{f_1}}+
\frac{-\frac{m_{f'_1}}{m_{f'_1}+m_{f_2}}\,|\vec{q}\,|-p_z}{\widehat{E}_{f'_1}}
\right)\nonumber\\
A^{(1^-)\,1}_{\lambda=-1}(|\vec{q}\,|)&=&\frac{i}{\sqrt2}
\sqrt{2m_I2E_F(-\vec{q}\,)}\ \int\,d^3p\ \frac{1}{4\pi}
\left(\hat{\phi}^{(M_F(1^-))}_{f'_1,\,f_2}(|\vec{p}\,|)\right)^*\,
\hat{\phi}^{(M_I(0^-))}_{f_1,\,f_2}\left(\bigg|\,\vec{p}-\frac{m_{f_2}}{m_{f'_1}+m_{f_2}}
|\vec{q}\,|
\vec{k} \bigg|\right)\nonumber\\
&&\hspace{3cm}\sqrt{\frac{\widehat{E}_{f'_1}\widehat{E}_{f1}}{4E_{f'_1}
E_{f_1}}}
\left(1+\frac{2p_x^2-(-\frac{m_{f'_1}}{m_{f'_1}+m_{f_2}}\,|\vec{q}\,|\vec{k}-\vec{p}\,)
\cdot(\frac{m_{f_2}}{m_{f'_1}+m_{f_2}}\,|\vec{q}\,|\vec{k}-\vec{p}\,)}
{\widehat{E}_{f'_1}\widehat{E}_{f_1}}
\right)
\nonumber\\
A^{(1^-)\,3}_{\lambda=0}(|\vec{q}\,|)&=&i
\sqrt{2m_I2E_F(-\vec{q}\,)}\ \int\,d^3p\ \frac{1}{4\pi}
\left(\hat{\phi}^{(M_F(1^-))}_{f'_1,\,f_2}(|\vec{p}\,|)\right)^*\,
\hat{\phi}^{(M_I(0^-))}_{f_1,\,f_2}\left(\bigg|\,\vec{p}-\frac{m_{f_2}}{m_{f'_1}+m_{f_2}}
|\vec{q}\,|
\vec{k} \bigg|\right)\nonumber\\
&&\hspace{3cm}\sqrt{\frac{\widehat{E}_{f'_1}\widehat{E}_{f1}}{4E_{f'_1}
E_{f_1}}}
\ \Bigg(
1+\frac{2(-\frac{m_{f'_1}}{m_{f'_1}+m_{f_2}}\,|\vec{q}\,|-p_z\,)
\cdot(\frac{m_{f_2}}{m_{f'_1}+m_{f_2}}\,|\vec{q}\,|-p_z\,)}
{\widehat{E}_{f'_1}\widehat{E}_{f_1}}\nonumber\\
&&\hspace{5.45cm}-\frac{(-\frac{m_{f'_1}}{m_{f'_1}+m_{f_2}}\,|\vec{q}\,|\vec{k}-\vec{p}\,)
\cdot(\frac{m_{f_2}}{m_{f'_1}+m_{f_2}}\,|\vec{q}\,|\vec k-\vec{p}\,)}
{\widehat{E}_{f'_1}\widehat{E}_{f_1}}
\Bigg)\nonumber\\
\end{eqnarray}

\begin{itemize}
\item {\large Case $J^\pi=1^+$}
\end{itemize}

\begin{eqnarray}
V^{(1^+,S_{q\bar{q}}=0)\,0}_{\lambda=0}(|\vec{q}\,|)&=&i\sqrt3
\sqrt{2m_I2E_F(-\vec{q}\,)}\ \int\,d^3p\ \frac{1}{4\pi|\vec p\,|}
\left(\hat{\phi}^{(M_F(1^+,S_{q\bar{q}}=0))}_{f'_1,\,f_2}(|\vec{p}\,|)\right)^*\,
\hat{\phi}^{(M_I(0^-))}_{f_1,\,f_2}\left(\bigg|\,\vec{p}-\frac{m_{f_2}}{m_{f'_1}+m_{f_2}}
|\vec{q}\,|
\vec{k} \bigg|\right)\nonumber\\
&&\hspace{3cm}\sqrt{\frac{\widehat{E}_{f'_1}\widehat{E}_{f1}}{4E_{f'_1}
E_{f_1}}}\,p_z\,
\left(
1+\frac{(-\frac{m_{f'_1}}{m_{f'_1}+m_{f_2}}\,|\vec{q}\,|\vec{k}-\vec{p}\,)
\cdot(\frac{m_{f_2}}{m_{f'_1}+m_{f_2}}\,|\vec{q}\,|\vec{k}-\vec{p}\,)}{\widehat{E}_{f'_1}\widehat{E}_{f_1}}
\right)
\nonumber\\
V^{(1^+,S_{q\bar{q}}=1)\,0}_{\lambda=0}(|\vec{q}\,|)&=&-i\sqrt{\frac{3}{2}}
\sqrt{2m_I2E_F(-\vec{q}\,)}\ \int\,d^3p\ \frac{1}{4\pi|\vec p\,|}
\left(\hat{\phi}^{(M_F(1^+,S_{q\bar{q}}=1))}_{f'_1,\,f_2}(|\vec{p}\,|)\right)^*\,
\hat{\phi}^{(M_I(0^-))}_{f_1,\,f_2}\left(\bigg|\,\vec{p}-\frac{m_{f_2}}{m_{f'_1}+m_{f_2}}
|\vec{q}\,|
\vec{k} \bigg|\right)\nonumber\\
&&\hspace{3cm}\sqrt{\frac{\widehat{E}_{f'_1}\widehat{E}_{f1}}{4E_{f'_1}
E_{f_1}}}
\ \frac{|\vec{q}\,| (p_z^2-\vec{p}^{\,2})}{\widehat{E}_{f'_1}\widehat{E}_{f1}}\nonumber\\
V^{(1^+,S_{q\bar{q}}=0)\,1}_{\lambda=-1}(|\vec{q}\,|)&=&-i\sqrt{\frac{3}{2}}
\sqrt{2m_I2E_F(-\vec{q}\,)}\ \int\,d^3p\ \frac{1}{4\pi|\vec{p}\,|}
\left(\hat{\phi}^{(M_F(1^+,S_{q\bar{q}}=0))}_{f'_1,\,f_2}(|\vec{p}\,|)\right)^*\,
\hat{\phi}^{(M_I(0^-))}_{f_1,\,f_2}\left(\bigg|\,\vec{p}-\frac{m_{f_2}}{m_{f'_1}+m_{f_2}}
|\vec{q}\,|
\vec{k} \bigg|\right)\nonumber\\
&&\hspace{3cm}\sqrt{\frac{\widehat{E}_{f'_1}\widehat{E}_{f1}}{4E_{f'_1}
E_{f_1}}}
\ p_x^2\,\left(\frac{1}{\widehat{E}_{f_1}}+\frac{1}{\widehat{E}_{f'_1}}
\right)\nonumber\\%
V^{(1^+,S_{q\bar{q}}=1)\,1}_{\lambda=-1}(|\vec{q}\,|)&=&i\frac{\sqrt{3}}{2}
\sqrt{2m_I2E_F(-\vec{q}\,)}\ \int\,d^3p\ \frac{1}{4\pi|\vec{p}\,|}
\left(\hat{\phi}^{(M_F(1^+,S_{q\bar{q}}=1))}_{f'_1,\,f_2}(|\vec{p}\,|)\right)^*\,
\hat{\phi}^{(M_I(0^-))}_{f_1,\,f_2}\left(\bigg|\,\vec{p}-\frac{m_{f_2}}
{m_{f'_1}+m_{f_2}}
|\vec{q}\,|
\vec{k} \bigg|\right)\nonumber\\
&&\hspace{3cm}\sqrt{\frac{\widehat{E}_{f'_1}\widehat{E}_{f1}}{4E_{f'_1}
E_{f_1}}}
\ \left(\frac{p_y^2+p_z^2+p_z|\vec{q}\,|
\frac{m_{f'_1}}{m_{f'_1}+m_{f_2}}
}{\widehat{E}_{f'_1}}-\frac{p_y^2+p_z^2-p_z|\vec{q}\,|
\frac{m_{f_2}}{m_{f'_1}+m_{f_2}}}{\widehat{E}_{f_1}}
\right)\nonumber\\%
V^{(1^+,S_{q\bar{q}}=0)\,3}_{\lambda=0}(|\vec{q}\,|)&=&i\sqrt3
\sqrt{2m_I2E_F(-\vec{q}\,)}\ \int\,d^3p\ \frac{1}{4\pi|\vec{p}\,|}
\left(\hat{\phi}^{(M_F(1^+,S_{q\bar{q}}=0))}_{f'_1,\,f_2}(|\vec{p}\,|)\right)^*\,
\hat{\phi}^{(M_I(0^-))}_{f_1,\,f_2}\left(\bigg|\,\vec{p}-\frac{m_{f_2}}{m_{f'_1}+m_{f_2}}
|\vec{q}\,|
\vec{k} \bigg|\right)\nonumber\\
&&\hspace{3cm}\sqrt{\frac{\widehat{E}_{f'_1}\widehat{E}_{f1}}{4E_{f'_1}
E_{f_1}}}\
p_z\,\Bigg(\frac{\frac{m_{f_2}}{m_{f'_1}+m_{f_2}}\,|\vec{q}\,|-p_z}{\widehat{E}_{f_1}}+
\frac{-\frac{m_{f'_1}}{m_{f'_1}+m_{f_2}}\,|\vec{q}\,|-p_z}{\widehat{E}_{f'_1}}
\Bigg)\nonumber\\%
V^{(1^+,S_{q\bar{q}}=1)\,3}_{\lambda=0}(|\vec{q}\,|)&=&-i\sqrt{\frac{3}{2}}
\sqrt{2m_I2E_F(-\vec{q}\,)}\ \int\,d^3p\ \frac{1}{4\pi|\vec{p}\,|}
\left(\hat{\phi}^{(M_F(1^+,S_{q\bar{q}}=1))}_{f'_1,\,f_2}(|\vec{p}\,|)\right)^*\,
\hat{\phi}^{(M_I(0^-))}_{f_1,\,f_2}\left(\bigg|\,\vec{p}-\frac{m_{f_2}}{m_{f'_1}+m_{f_2}}
|\vec{q}\,|
\vec{k} \bigg|\right)\nonumber\\
&&\hspace{3cm}\sqrt{\frac{\widehat{E}_{f'_1}\widehat{E}_{f1}}{4E_{f'_1}
E_{f_1}}}
\ (p_x^2+p_y^2)\,\left(\frac{1}{\widehat{E}_{f_1}}-\frac{1}{\widehat{E}_{f'_1}}
\right)\nonumber\\
\end{eqnarray}
and similarly
\begin{eqnarray}
A^{(1^+,S_{q\bar{q}}=0)\,1}_{\lambda=-1}(|\vec{q}\,|)&=&-i\sqrt{\frac{3}{2}}
\sqrt{2m_I2E_F(-\vec{q}\,)}\ \int\,d^3p\ \frac{1}{4\pi|\vec{p}\,|}
\left(\hat{\phi}^{(M_F(1^+,S_{q\bar{q}}=0))}_{f'_1,\,f_2}(|\vec{p}\,|)\right)^*\,
\hat{\phi}^{(M_I(0^-))}_{f_1,\,f_2}\left(\bigg|\,\vec{p}-\frac{m_{f_2}}{m_{f'_1}+m_{f_2}}
|\vec{q}\,|
\vec{k} \bigg|\right)\nonumber\\
&&\hspace{3cm}\sqrt{\frac{\widehat{E}_{f'_1}\widehat{E}_{f1}}{4E_{f'_1}
E_{f_1}}}
\ \frac{p_y^2|\vec{q}\,|}{\widehat{E}_{f_1}\widehat{E}_{f'_1}}
\nonumber\\%
A^{(1^+,S_{q\bar{q}}=1)1}_{\lambda=-1}(|\vec{q}\,|)&=&i\frac{\sqrt3}{2}\sqrt{2m_I2E_F(-\vec{q}\,)}\ \int\,d^3p\
\frac{1}{4\pi|\vec{p}\,|}
\left(\hat{\phi}^{(M_F (1^+,S_{q\bar{q}}=1))}_{f'_1,\,f_2}(|\vec{p}\,|)
\right)^*\,
\hat{\phi}^{(M_I(0^-))}_{f_1,\,f_2}\left(\bigg|\,\vec{p}-\frac{m_{f_2}}{m_{f'_1}+m_{f_2}}
|\vec{q}\,|
\vec{k} \bigg|\right)
 \nonumber\\
&&\hspace{3cm}\sqrt{\frac{\widehat{E}_{f'_1}\widehat{E}_{f1}}{4E_{f'_1}
E_{f_1}}}\bigg\{\
p_z\bigg(  
1-\frac{(-\frac{m_{f'_1}}{m_{f'_1}+m_{f_2}}\,|\vec{q}\,|\vec{k}-\vec{p}\,)
\cdot(\frac{m_{f_2}}{m_{f'_1}+m_{f_2}}\,|\vec{q}\,|\vec{k}-\vec{p}\,)}{\widehat{E}_{f'_1}\widehat{E}_{f_1}}
\bigg)\nonumber\\ 
&&\hspace{5.cm}+\frac{m_{f_2}-m_{f'_1}}{m_{f'_1}+m_{f_2}}
\frac{p_x^2|\vec{q}\,|}{\widehat{E}_{f'_1}\widehat{E}_{f1}}
\bigg\}\nonumber\\
\end{eqnarray}

\begin{itemize}
\item {\large Case $J^\pi= 2^-$}
\end{itemize}

\begin{eqnarray}
V^{(2^-)\,0}_{T\lambda=0}(|\vec{q}\,|)&=&i\,\sqrt{\frac{15}{2}}
\sqrt{2m_I2E_F(-\vec{q}\,)}\ \int\,d^3p\ \frac{1}{4\pi|\vec{p}\,|^2}
\left(\hat{\phi}^{(M_F(2^-))}_{f'_1,\,f_2}(|\vec{p}\,|)\right)^*\,
\hat{\phi}^{(M_I(0^-))}_{f_1,\,f_2}\left(\bigg|\,\vec{p}-\frac{m_{f_2}}{m_{f'_1}+m_{f_2}}
|\vec{q}\,|
\vec{k} \bigg|\right)\nonumber\\
&&\hspace{3cm}\sqrt{\frac{\widehat{E}_{f'_1}\widehat{E}_{f1}}{4E_{f'_1}
E_{f_1}}}\
\frac{p_z\left(p_x^2+p_y^2\right)|\vec{q}\,|}
{\widehat{E}_{f'_1}\widehat{E}_{f1}}\nonumber\\
V^{(2^-)\,1}_{T\lambda=+1}(|\vec{q}\,|)&=&i\,\frac{\sqrt5}{2}
\sqrt{2m_I2E_F(-\vec{q}\,)}\ \int\,d^3p\ \frac{1}{4\pi|\vec{p}\,|^2}
\left(\hat{\phi}^{(M_F(2^-))}_{f'_1,\,f_2}(|\vec{p}\,|)\right)^*\,
\hat{\phi}^{(M_I(0^-))}_{f_1,\,f_2}\left(\bigg|\,\vec{p}-\frac{m_{f_2}}{m_{f'_1}+m_{f_2}}
|\vec{q}\,|
\vec{k} \bigg|\right)\nonumber\\
&&\hspace{3cm}\sqrt{\frac{\widehat{E}_{f'_1}\widehat{E}_{f1}}{4E_{f'_1}
E_{f_1}}}
\Bigg\{\left(p_z^2-p_x^2\right)
\left(\ \frac{-p_z-\frac{m_{f'_1}}{m_{f'_1}+m_{f_2}}|\vec{q}\,|}
{\widehat{E}_{f'_1}}
-\frac{-p_z+\frac{m_{f_2}}{m_{f'_1}+m_{f_2}}|\vec{q}\,|}
{\widehat{E}_{f_1}}
\right)\nonumber\\
&&\hspace{5cm}-p_zp_y^2\left(\frac{1}{\widehat{E}_{f'_1}}-\frac{1}{\widehat{E}_{f_1}}
\right)\Bigg\}\nonumber\\
V^{(2^-)\,3}_{T\lambda=0}(|\vec{q}\,|)&=&i\,\sqrt{\frac{15}{2}}
\sqrt{2m_I2E_F(-\vec{q}\,)}\ \int\,d^3p\ \frac{1}{4\pi|\vec{p}\,|^2}
\left(\hat{\phi}^{(M_F(2^-))}_{f'_1,\,f_2}(|\vec{p}\,|)\right)^*\,
\hat{\phi}^{(M_I(0^-))}_{f_1,\,f_2}\left(\bigg|\,\vec{p}-\frac{m_{f_2}}{m_{f'_1}+m_{f_2}}
|\vec{q}\,|
\vec{k} \bigg|\right)\nonumber\\
&&\hspace{3cm}\sqrt{\frac{\widehat{E}_{f'_1}\widehat{E}_{f1}}{4E_{f'_1}
E_{f_1}}}\ p_z\left(p_x^2+p_y^2\right)
\left(
\frac{1}
{\widehat{E}_{f'_1}}-\frac{1}
{\widehat{E}_{f_1}}
\right)\nonumber\\
\end{eqnarray}
and similarly
\begin{eqnarray}
A^{(2^-)\,1}_{T\lambda=+1}(|\vec{q}\,|)&=&i\,\frac{\sqrt5}{2}
\sqrt{2m_I2E_F(-\vec{q}\,)}\ \int\,d^3p\ \frac{1}{4\pi|\vec{p}\,|^2}
\left(\hat{\phi}^{(M_F(2^-))}_{f'_1,\,f_2}(|\vec{p}\,|)\right)^*\,
\hat{\phi}^{(M_I(0^-))}_{f_1,\,f_2}\left(\bigg|\,\vec{p}-\frac{m_{f_2}}{m_{f'_1}+m_{f_2}}
|\vec{q}\,|
\vec{k} \bigg|\right)\nonumber\\
&&\hspace{3cm}\sqrt{\frac{\widehat{E}_{f'_1}\widehat{E}_{f1}}{4E_{f'_1}
E_{f_1}}}
\Bigg\{\left(p_z^2-p_y^2\right)
\left(1-
\frac{(-\frac{m_{f'_1}}{m_{f'_1}+m_{f_2}}\,|\vec{q}\,|\vec{k}-\vec{p}\,)
\cdot(\frac{m_{f_2}}{m_{f'_1}+m_{f_2}}\,|\vec{q}\,|\vec k-\vec{p}\,)}{\widehat{E}_{f'_1}\widehat{E}_{f1}}
\right)
\nonumber\\
&&\hspace{5cm}-p_zp_x^2|\vec{q}\,|\,\frac{m_{f'_1}-m_{f_2}}{m_{f'_1}+m_{f_2}}
\,\frac{1}
{\widehat{E}_{f'_1}\widehat{E}_{f_1}}\ 
\Bigg\}\nonumber\\
\end{eqnarray}

\begin{itemize}
\item  {\large  Case $J^\pi= 2^+$}
\end{itemize}

\begin{eqnarray}
V^{(2^+)\,1}_{T\lambda=+1}(|\vec{q}\,|)&=&i\,\frac{\sqrt3}{2}
\sqrt{2m_I2E_F(-\vec{q}\,)}\ \int\,d^3p\ \frac{1}{4\pi|\vec{p}\,|}
\left(\hat{\phi}^{(M_F(2^+))}_{f'_1,\,f_2}(|\vec{p}\,|)\right)^*\,
\hat{\phi}^{(M_I(0^-))}_{f_1,\,f_2}\left(\bigg|\,\vec{p}-\frac{m_{f_2}}{m_{f'_1}+m_{f_2}}
|\vec{q}\,|
\vec{k} \bigg|\right)\nonumber\\
&&\hspace{3cm}\sqrt{\frac{\widehat{E}_{f'_1}\widehat{E}_{f1}}{4E_{f'_1}
E_{f_1}}}
\left(\frac{p_y^2-p_z^2-p_z|\vec{q}\,|\frac{m_{f'_1}}{m_{f'_1}+m_{f_2}}}
{\widehat{E}_{f'_1}}
-\frac{p_y^2-p_z^2+p_z|\vec{q}\,|\frac{m_{f_2}}{m_{f'_1}+m_{f_2}}}
{\widehat{E}_{f_1}}
\right)\nonumber\\
\end{eqnarray}
and similarly

\begin{eqnarray}
A^{(2^+)\,0}_{T\lambda=0}(|\vec{q}\,|)&=&\frac{-i}{\sqrt2}
\sqrt{2m_I2E_F(-\vec{q}\,)}\ \int\,d^3p\ \frac{1}{4\pi|\vec{p}\,|}
\left(\hat{\phi}^{(M_F(2^+))}_{f'_1,\,f_2}(|\vec{p}\,|)\right)^*\,
\hat{\phi}^{(M_I(0^-))}_{f_1,\,f_2}\left(\bigg|\,\vec{p}-\frac{m_{f_2}}{m_{f'_1}+m_{f_2}}
|\vec{q}\,|
\vec{k} \bigg|\right)\nonumber\\
&&\hspace{2cm}\sqrt{\frac{\widehat{E}_{f'_1}\widehat{E}_{f1}}{4E_{f'_1}
E_{f_1}}}
\left(\frac{p_x^2+p_y^2-2p_z^2-2p_z|\vec{q}\,|\frac{m_{f'_1}}{m_{f'_1}+m_{f_2}}}
{\widehat{E}_{f'_1}}
+\frac{p_x^2+p_y^2-2p_z^2+2p_z|\vec{q}\,|\frac{m_{f_2}}{m_{f'_1}+m_{f_2}}}
{\widehat{E}_{f_1}}
\right)\nonumber\\
A^{(2^+)\,1}_{T\lambda=+1}(|\vec{q}\,|)&=&i\,\frac{\sqrt3}{2}
\sqrt{2m_I2E_F(-\vec{q}\,)}\ \int\,d^3p\ \frac{1}{4\pi|\vec{p}\,|}
\left(\hat{\phi}^{(M_F(2^+))}_{f'_1,\,f_2}(|\vec{p}\,|)\right)^*\,
\hat{\phi}^{(M_I(0^-))}_{f_1,\,f_2}\left(\bigg|\,\vec{p}-\frac{m_{f_2}}{m_{f'_1}+m_{f_2}}
|\vec{q}\,|
\vec{k} \bigg|\right)\nonumber\\
&&\hspace{3cm}\sqrt{\frac{\widehat{E}_{f'_1}\widehat{E}_{f1}}{4E_{f'_1}
E_{f_1}}}\Bigg\{\ \,p_z\,\left(1-
\frac{(-\frac{m_{f'_1}}{m_{f'_1}+m_{f_2}}\,|\vec{q}\,|\vec{k}-\vec{p}\,)
\cdot(\frac{m_{f_2}}{m_{f'_1}+m_{f_2}}\,|\vec{q}\,|\vec k-\vec{p}\,)}{\widehat{E}_{f'_1}\widehat{E}_{f1}}
\right)\nonumber\\
&&\hspace{5.cm}
+\frac{4p_zp_x^2-p_x^2|\vec{q}\,|
\frac{m_{f_2}-m_{f'_1}}{m_{f'_1}+m_{f_2}}}
{\widehat{E}_{f'_1}\widehat{E}_{f1}}\Bigg\}\nonumber\\
A^{(2^+)\,3}_{T\lambda=0}(|\vec{q}\,|)&=&-i\,\sqrt2
\sqrt{2m_I2E_F(-\vec{q}\,)}\ \int\,d^3p\ \frac{1}{4\pi|\vec{p}\,|}
\left(\hat{\phi}^{(M_F(2^+))}_{f'_1,\,f_2}(|\vec{p}\,|)\right)^*\,
\hat{\phi}^{(M_I(0^-))}_{f_1,\,f_2}\left(\bigg|\,\vec{p}-\frac{m_{f_2}}{m_{f'_1}+m_{f_2}}
|\vec{q}\,|
\vec{k} \bigg|\right)\nonumber\\
&&\hspace{2cm}\sqrt{\frac{\widehat{E}_{f'_1}\widehat{E}_{f1}}{4E_{f'_1}
E_{f_1}}}\Bigg\{\ \,p_z\,\left(1-
\frac{(-\frac{m_{f'_1}}{m_{f'_1}+m_{f_2}}\,|\vec{q}\,|\vec{k}-\vec{p}\,)
\cdot(\frac{m_{f_2}}{m_{f'_1}+m_{f_2}}\,|\vec{q}\,|\vec k-\vec{p}\,)}{\widehat{E}_{f'_1}\widehat{E}_{f1}}
\right)\nonumber\\
&&\hspace{4.cm}
+\frac{1}
{\widehat{E}_{f'_1}\widehat{E}_{f1}}
\Bigg[
 \ \,2p_z(-\frac{m_{f'_1}}{m_{f'_1}+m_{f_2}}\,|\vec{q}\,|-p_z)
\cdot (\frac{m_{f_2}}{m_{f'_1}+m_{f_2}}\,|\vec{q}\,|-p_z)\nonumber\\
&&\hspace{5.75cm}+\left(p_x^2+p_y^2\right)\bigg(
-p_z+\frac{m_{f_2}-m_{f'_1}}{2(m_{f'_1}+m_{f_2})}|\vec{q}\,|
\bigg)
\Bigg]
\ \Bigg\}\nonumber\\
\end{eqnarray}


%
%
%
%
\section{Transitions involving antiquarks} 
\label{app:sign}
When it is the antiquark that suffers the decay the expressions are modified
as described below for a general transition between
a pseudoscalar
meson $M_I$ at rest with quark content $q_{f_1}\overline{q}_{f_2}$ and a final 
$M_F$ meson with total angular momentum and parity $J^\pi=0^-,0^+,1^-,1^+,2^-,2^+$,
three-momentum $-|\vec{q}\,|\vec{k}$ and  quark content
$q_{f_1}\overline{q}_{f'_2}$. 
We have
\begin{eqnarray}
{\cal V}^\mu(|\vec{q}\,|)-{\cal A}^\mu(|\vec{q}\,|)&=&\sqrt{2m_{I}2E_{F}(-\vec{q\,})}\
{}_{\stackrel{}{\stackrel{}{NR}}}
\left\langle\, M_F(J^\pi),\,\lambda\,
-|\vec{q}\,|\,\vec{k}\,\bigg|\, J^{f_2\,f'_2\, \mu}(0)\,
\bigg| \, M_I(0^-),\,\vec{0}\right\rangle_{NR}\nonumber\\
&=&-\sqrt{2m_{I}2E_{F}(-\vec{q\,})}\ \int\,d^3p\sum_{s'_2}\sum_{s_1,s_2}
\left(\hat{\phi}^{(M_F(J^\pi),\lambda)}_{(s_1,f_1),\,(s'_2,f'_2)}(\vec{p}\,)\right)^*\ 
\hat{\phi}^{(M_I(0^-))}_{(s_1,f_1),\,(s_2,f_2)}(\vec{p}
+\frac{m_{f_1}}{m_{f_1}+m_{f'_2}}\,|\vec{q}\,|\vec{k})
\nonumber\\
&&\hspace{1cm}\frac{(-1)^{s_2-s'_2}}
{\sqrt{2E_{f'_2}\,2E_{f_2}}}\ \bar{v}_{s_2,f_2}(\frac{m_{f_1}}{m_{f_1}+m_{f'_2}}
|\vec{q}\,|\vec{k}+\vec{p}\,)\, 
\gamma^\mu(I-\gamma_5)\,v_{s'_2,f'_2}(-\frac{m_{f'_2}}{m_{f_1}+m_{f'_2}}
|\vec{q}\,|\vec{k}+\vec{p}\,)\nonumber\\
\end{eqnarray}
where ${\cal V}^\mu ({\cal A}^\mu)$ represent any of the $V^\mu\, (A^\mu)$,
$V^\mu_{(\lambda)}\, (A^\mu_{(\lambda)})$ or
$V^\mu_{T(\lambda)}\, (A^\mu_{T(\lambda)})$,
and $E_{f'_2},\,E_{f_2}$ are shorthand notation for
$E_{f'_2}(-\frac{m_{f'_2}}{m_{f_1}+m_{f'_2}}
|\vec{q}\,|\vec{k}+\vec{p}\,)$, $E_{f_2}(\frac{m_{f_1}}{m_{f_1}+m_{f'_2}}
|\vec{q}\,|\vec{k}+\vec{p}\,)$.
We can use now that
\begin{eqnarray}
v_{s,f}(\vec{p}\,)=(-1)^{(1/2)-s}\, {\cal C}\, \bar{u}^T_{s,f}(\vec{p}\,)
\hspace{.5cm};\hspace{.5cm}
\bar{v}_{s,f}(\vec{p}\,)=-(-1)^{(1/2)-s}\,  {u}^T_{s,f}(\vec{p}\,)\, {\cal C}^\dagger
\end{eqnarray}
where ${\cal C}$ is a matrix given in the Fermi--Dirac representation that we
use by
\begin{eqnarray}
{\cal C}=i\gamma^2\gamma^0
\end{eqnarray}
and that satisfies
\begin{eqnarray}
{\cal C}=-{\cal C}^{-1}=-{\cal C}^\dagger=-{\cal C}^T\hspace{.5cm};\hspace{.5cm}
{\cal C}\gamma_\mu^T{\cal C}^\dagger=-\gamma_\mu
\end{eqnarray}
Using the above information and making the change of variable
$\vec p\to -\vec p$\ \ we can rewrite
\begin{eqnarray}
{\cal V}^\mu(|\vec{q}\,|)-{\cal A}^\mu(|\vec{q}\,|)&=&
\sqrt{2m_{I}2E_{F}(-\vec{q\,})}\ \int\,d^3p\sum_{s'_2}\sum_{s_1,s_2}
\left(\hat{\phi}^{(M_F(J^\pi),\,\lambda)}_{(s_1,f_1),\,(s'_2,f'_2)}(-\vec{p}\,)
\right)^*\ 
\hat{\phi}^{(M_I(0^-))}_{(s_1,f_1),\,(s_2,f_2)}\bigg(-\bigg(\vec{p}
-\frac{m_{f_1}}{m_{f_1}+m_{f'_2}}\,|\vec{q}\,|\vec{k}\bigg)\bigg)
\nonumber\\
&&\hspace{1cm}\frac{1}
{\sqrt{2E_{f'_2}\,2E_{f_2}}}\ \bar{u}_{s'_2,f'_2}(-\frac{m_{f'_2}}{m_{f_1}+m_{f'_2}}
|\vec{q}\,|\vec{k}-\vec{p}\,)\, 
\gamma^\mu(-I-\gamma_5)\,u_{s_2,f_2}(\frac{m_{f_1}}{m_{f_1}+m_{f'_2}}
|\vec{q}\,|\vec{k}-\vec{p}\,)\nonumber\\
\end{eqnarray}%
By comparison with the corresponding expressions involving  quarks  we find 
that, apart from the changes in the masses involved, there is an
 extra minus sign for the vector part, and, due to Clebsch--Gordan
re-arrangements and the fact that $Y_{lm}(-\vec{p}\,)=(-1)^l\,Y_{lm}(\vec{p}\,)$,
a global sign given by $(-1)^{l_I+s_I-l_F-s_F}$ where $l_I,\,s_I$\,($l_F,\,s_F$) are 
the orbital and spin angular momenta of the initial (final) meson.

In any case this implies a change of sign in the relative phase between 
 vector and  axial contributions, which in its term produces 
a sign change in the tensor helicity
components combination $H_P$ due to the fact that 
${\cal H}_{+1\,+1}$ goes into ${\cal H}_{-1\,-1}$ and
vice versa. All other tensor helicity
components combinations defined in Eq.(\ref{eq:combinaciones}) keep their signs.\\

A simple way of anticipating the above result is the following: the current for 
$\bar q_{f_2}$ decay into $\bar q_{f'_2}$ is
\begin{eqnarray}
\overline{\Psi}_{f_2}(0)\gamma^\mu (I-\gamma_5)\Psi_{f'_2}(0)
\end{eqnarray} 
But for antiquarks the fields that play the similar role as the $\Psi$ fields
play for quarks are the charge conjugate ones $ \Psi^{\cal C}$. In terms of the
latter the above current is written as
\begin{eqnarray}
\overline{\Psi}_{f_2}(0)\gamma^\mu (I-\gamma_5)\Psi_{f'_2}(0)=
\overline{\Psi}_{f'_2}^{\cal C}(0)\gamma^\mu (-I-\gamma_5)\Psi_{f_2}^{\cal C}(0)
\end{eqnarray} 
Now this is similar to the current for  quark decay but with an extra minus
sign in the vector part. Whatever other changes might come from reorderings
in the wave functions we will have an extra relative sign between the vector and
axial part.

\section{Expresions for the helicity components of the hadron tensor}
\label{app:hcht}
In this appendix we give the expressions for the non--zero helicity components 
${\cal H}_{rs}$ of the
hadron tensor, as defined in Eq.(\ref{eq:hcht}), corresponding to a $B_c^-\to
c\bar c$ transition. The different cases correspond
to the ones discussed in the main text.
\begin{itemize}
\item {\large Case $0^-\to 0^-,0^+$}
\begin{eqnarray}
{\cal H}_{t\,t}({P}_{B_c},{P}_{c\bar{c}})&=&
\left(\frac{m^2_{B_c}-m^2_{c\bar c}}{\sqrt{q^2}}\,F_+(q^2)+\sqrt{q^2}\,
F_-(q^2)
\right)^2\nonumber\\
{\cal H}_{t\,0}({P}_{B_c},{P}_{c\bar{c}})&=&
{\cal H}_{0\,t}({P}_{B_c},{P}_{c\bar{c}})=
\lambda^{1/2}(q^2,m^2_{B_c},m^2_{c\bar c})\left(
\frac{m^2_{B_c}-m^2_{c\bar c}}{{q^2}}\,F_+^2(q^2)+
F_+(q^2)F_-(q^2)\right)\nonumber\\
{\cal H}_{0\,0}({P}_{B_c},{P}_{c\bar{c}})&=&
\frac{\lambda(q^2,m^2_{B_c},m^2_{c\bar c})}{q^2}\, F^2_+(q^2)
\end{eqnarray}
\item {\large Case $0^-\to 1^-,1^+$}.
\begin{eqnarray}
{\cal H}_{t\,t}({P}_{B_c},{P}_{c\bar{c}})&=&
\frac{\lambda(q^2,m^2_{B_c},m^2_{c\bar c})}{4m^2_{c\bar c}{q^2}}\,
\left((m_{B_c}-m_{c\bar c})\left(A_0(q^2)-A_+(q^2)\right)
-\frac{q^2}{m_{B_c}+m_{c\bar c}}A_-(q^2)\right)^2\nonumber\\
{\cal H}_{t\,0}({P}_{B_c},{P}_{c\bar{c}})&=&
{\cal H}_{0\,t}({P}_{B_c},{P}_{c\bar{c}})\nonumber\\
&=&\frac{\lambda^{1/2}(q^2,m^2_{B_c},m^2_{c\bar c})}{2m_{c\bar c}\sqrt{q^2}}\,
\left[(m_{B_c}-m_{c\bar c})\left(A_0(q^2)-A_+(q^2)\right)
-\frac{q^2}{m_{B_c}+m_{c\bar c}}A_-(q^2)\right]\nonumber\\
&&\times\ \left[(m_{B_c}-m_{c\bar c})\frac{m^2_{B_c}-q^2-m^2_{c\bar c}}
{2m_{c\bar c}\sqrt{q^2}}
\, A_0(q^2)-\frac{\lambda(q^2,m^2_{B_c},m^2_{c\bar c})}
{2m_{c\bar c}\,\sqrt{q^2}}\,\frac{A_+(q^2)}
{m_{B_c}+m_{c\bar c}}\right]\nonumber\\
{\cal H}_{+1\,+1}({P}_{B_c},{P}_{c\bar{c}})&=&\
\left(\frac{\lambda^{1/2}(q^2,m^2_{B_c},m^2_{c\bar c})}{m_{B_c}+m_{c\bar c}}\,
V(q^2)+(m_{B_c}-m_{c\bar c})\,A_0(q^2)\right) ^2\nonumber\\
{\cal H}_{-1\,-1}({P}_{B_c},{P}_{c\bar{c}})&=&\
\left(-\frac{\lambda^{1/2}(q^2,m^2_{B_c},m^2_{c\bar c})}{m_{B_c}+m_{c\bar c}}\,
V(q^2)+(m_{B_c}-m_{c\bar c})\,A_0(q^2)\right) ^2\nonumber\\
{\cal H}_{0\,0}({P}_{B_c},{P}_{c\bar{c}})&=&\
\left( (m_{B_c}-m_{c\bar c})\frac{m^2_{B_c}-q^2-m^2_{c\bar c}}{2m_{c\bar c}\sqrt{q^2}}\,
A_0(q^2)-\frac{\lambda(q^2,m^2_{B_c},m^2_{c\bar c})}{2m_{c\bar
c}\sqrt{q^2}}\,\frac{A_+(q^2)}{m_{B_c}+m_{c\bar c}}
\right)^2
\end{eqnarray}
\end{itemize}

For a $B_c^-\to B$ transition (with $B$ representing any of the 
$B=\overline B^0,\,\overline B^{*0},\,B^-,\,B^{*-}$), where it is the 
$\bar c$ antiquark that decays, we have
to change the mass of the final meson in the expressions above and 
take into account the changes in the form factors that derive from the
discussions  in appendix~\ref{app:sign}.
\begin{itemize}
\item {\large Case $0^-\to 2^-,2^+$}.
\begin{eqnarray}
{\cal H}_{t\,t}({P}_{B_c},{P}_{c\bar{c}})&=&
\frac{\lambda^{2}(q^2,m^2_{B_c},m^2_{c\bar c})}{24\,m^4_{c\bar c}\,q^2}\,
\left(\, T_1(q^2)+(m^2_{B_c}-m^2_{c\bar c})\,T_2(q^2)+q^2\,T_3(q^2)\,
\right)^2\nonumber\\
{\cal H}_{t\,0}({P}_{B_c},{P}_{c\bar{c}})&=&
{\cal H}_{0\,t}({P}_{B_c},{P}_{c\bar{c}})\nonumber\\
&=&\frac{\lambda^{3/2}(q^2,m^2_{B_c},m^2_{c\bar c})}{24\,m^4_{c\bar c}\,q^2}\,
\left(\, T_1(q^2)+(m^2_{B_c}-m^2_{c\bar c})\,T_2(q^2)+q^2\,T_3(q^2)\
\right)\nonumber\\
&&\hspace{2.55cm}\times\,\left(\,(m^2_{B_c}-q^2-m^2_{c\bar c})\,T_1(q^2)+\lambda(q^2,m^2_{B_c},m^2_{c\bar c})\,T_2(q^2)\,
\right)
\nonumber\\
{\cal H}_{+1\,+1}({P}_{B_c},{P}_{c\bar{c}})&=&
\frac{\lambda(q^2,m^2_{B_c},m^2_{c\bar c})}{8\,m^2_{c\bar c}}\,
\left(\,T_1(q^2)-\lambda^{1/2}(q^2,m^2_{B_c},m^2_{c\bar c})\,T_4(q^2)\,
\right)^2\nonumber\\
{\cal H}_{-1\,-1}({P}_{B_c},{P}_{c\bar{c}})&=&
\frac{\lambda(q^2,m^2_{B_c},m^2_{c\bar c})}{8\,m^2_{c\bar c}}\,
\left(\,T_1(q^2)+\lambda^{1/2}(q^2,m^2_{B_c},m^2_{c\bar c})\,T_4(q^2)\,
\right)^2\nonumber\\
{\cal H}_{0\,0}({P}_{B_c},{P}_{c\bar{c}})&=&
\frac{\lambda(q^2,m^2_{B_c},m^2_{c\bar c})}{24\,m^4_{c\bar c}\,q^2}\,
\left(\,(m^2_{B_c}-q^2-m^2_{c\bar c})\,T_1(q^2)+\lambda(q^2,m^2_{B_c},m^2_{c\bar c})\,T_2(q^2)\,
\right)^2\nonumber\\
\end{eqnarray}
\end{itemize}

\end{document}